**Paolo Zanna e Costantino Sigismondi**

# DÚNGAL, LETTERATO E ASTRONOMO
## Note di stilistica e di astronomia sulla Lettera a Carlo Magno circa le eclissi di sole dell'810

**Abstract**

Dúngal's letter to Charlemagne on the double solar eclipse in the year 810 is extremely interesting both for its form and for its subject matter. Part I of the present study deals with the *epistula* as a literary work (genre, language, sources), dealing in turn with vocabulary, morphology and syntax, rythmical prose and rhetorical figures, literary and Biblical references. If we compare it with Dúngal's other works, the letter is cast in the canonical *oratio* structure similar to his later *Responsa contra Claudium* (ed. Zanna, Firenze 2002) and it is likewise based on lengthy quotations drawns from Macrobius' commentary *In Somnium Scipionis*. A possible echo of Vitruvius' astronomical presentation in book IX of his *De architectura* is suggested. Finally, we attempt to define how the author's *persona* as *famulus et orator* and *reclusus* at St-Denis relates to Charlemagne and to abbot Waldo in his pursuit of Chirstian wisdom based on the Bible rahter than of scholarship *per se* based on academic research.
Part II is a thorough technical discussion of the astronomical issue presented to the Irish scholar by the Emperor, i.e. The frequency of solar eclipses and their visibility in the two emispheres in the year 810 (see maps provided). It assesses Dúngal's case in terms of eclipse theory and reviews previous comments on his letter to Charlemagne by astronomer Ismael Bullialdus (1605-1694). It also introduces us to first-hand knowledge of eclipses in history and nowadays, providing a glimpse into the complex problems tackled by the Irishman and his sources in Late Antique and in the Early Middle Ages. Ample footnotes to the Italian translation of Dúngal's work are an essential guiding tool for Latinists unfamiliar with astronomy.

## *PARTE I Dúngal latinista e letterato*
## *di Paolo Zanna*

### 0 Introduzione

### Elementi dell'analisi

A ulteriore integrazione di una serie di studi su Dúngal,[1] maestro alla scuola di Lotario a Pavia,[2] che lasciò i propri codici alla biblioteca dell'abbazia di san Colombano a Bobbio,[3] già *reclusus* a Saint-Denis,[4] presentiamo qui il testo[5] e una traduzione italiana a fronte della lettera a Carlomagno sulla duplice eclisse di Sole dell'810.
L'esame della lettera sarà diviso in quattro parti riguardanti:

(1) genere e struttura
(2) lingua e stile
(3) *clausulae* ritmiche
(4) fonti e modalità di citazione

Nelle conclusioni, riassumeremo gli aspetti salienti circa l'autore e il suo *modus operandi* come *famulus et orator* di Carlo Magno.

### 1 Genere e struttura della lettera: *oratio* et *fabula*

Nel rispondere alle richieste dell'imperatore, Dúngal segue la struttura compositiva dell'*oratio* messa in luce anche nella mia introduzione ai *Responsa* dungaliani.[6] Il testo comprende *exordium, narratio, argumentatio* ed *epilogus*. La numerazione dei paragrafi qui introdotta nel testo della *Patrologia* consente di distinguere più agevolmente la scansione delle sezioni.

L'intera composizione appare caratterizzata da un mirato equilibrio tra i corrispondenti in dialogo e le parti del testo stesso. L'*exordium* in preghiera (§ 1) è speculare rispetto all'*epilogus* che chiude lo scritto dungaliano (§ 40a-b); la *narratio*, che descrive l'occasione della lettera (§§ 2-3) si riferisce alla richiesta imperiale inoltrata dall'abate Waldo (*Waldoni abbati direxistis epistolam*, *ut per illam me interrogaret*), cui corrisponde la consegna del *responsum* dungaliano allo stesso Waldo (*per illum vobis remitto*, § 41).

Nell'*exordium* e nella *narratio* Dúngal si presenta nei termini consueti della sua corrispondenza successiva e dei *Responsa*: *vester fidelis famulus et orator* (§ 2), *volontarius* (§ 3). Tale caratterizzazione è identica a quella dell'abate Waldo al § 41 *fideli famulo Waldoni*, *vobis*

[1] «"A cavallo" tra teologia e retorica. Dúngal e il decoro di un latinista irlandese sul Continente», *Acme* LIV, Fascicolo 1, Gennaio-Aprile 2001, p. 33-57; *Responsa contra Claudium*, a new edition by Paolo Zanna, Firenze, Sismel, Edizioni del Galluzzo, 2002; «I carmi di Dúngal per Ildoardo e Baldo con un bilancio della sua opera in poesia e in prosa», *Archivum Bobiense* 25, p. 143-186 ; «La Bibbia in Dúngal», in: *Biblical Studies in the Early Middle Ages Gargnano, 24-27 June 2001* (in stampa); e «*Dungali Poetria*. Il carme *Quisquis es hunc cernens:* Dúngal, Aldelmo, Alcuino», letto al *IV Congreso Internacional de Latín Medieval*, *Poesia Latina Medieval (siglos IV- XV),* 12-15 settembre 2002 (in stampa).
[2] Mirella FERRARI, '*In Papia conveniant ad Dungalum*', *Italia medioevale e umanistica* 15 (1972), p. 1-52.
[3] Per l'elenco completo di questi testi cfr. *Responsa contra Claudium*, cit. alla nota 1, appendice 1, p. 297-299.
[4] Mirella FERRARI, «Dungal», in *Dizionario biografico degli italiani* 42, Roma, 1993, p. 11-14.
[5] *PL* 105, col. 447-458; sui due testimoni v. Mirella FERRARI, '*In Papia conveniant ad Dungalum*', cit. n. 2, pp. 4-5: il Berlinese lat. 177 (Phillips 1784) con il testo inframmezzato da estratti di Beda e con un passo di Isidoro in calce e il cod. c-9/69 (già LXXXII + LXII) della Biblioteca Capitolare di Monza, in cui la lettera fa parte di una raccolta di computistica divisa in 140 capitoli,
[6] Cfr. Paolo ZANNA, *Responsa contra Claudium*, cit. alla nota 1, p. CIV-CVI.

*fidelis*. Anche a Carlo Magno sono attribuiti epiteti *standard* ripetuti (v. § 2 di quest'introduzione stilistica e cfr. § 1, 39, 40a, 41). La ripetizione lessicale che rappresenta un tratto stilistico dell'opuscolo e dell'intera opera dungaliana (come illustrato sotto) risponde pertanto anche a precisi criteri e obiettivi strutturali. Da notare in particolare l'uso dell'apostrofe all'imperatore (vedi sotto § 2.2).

Il § 4 presenta la *propositio* dungaliana, ossia l'impegno profuso dall'Irlandese nel rispondere alla sollecitazione di Carlo Magno. L'intento non pare, però, tanto quello di dar conto della ricerca condotta sui *leves compendosiosque libellos qui inter manus sunt*, bensí quello di dichiarare la propria inadeguatezza al compito da svolgere (a causa del *torpor obtunsi cordis et tardus sensus lento conamine pigroque nisu reptans et movens*), nonché di fare appello alla *clementia* dell'imperatore. Secondo il già descritto principio di equilibratura dell'*oratio*, Dúngal ripete lo stesso appello nell'*epilogus* (*poenitentiam clementer imponatis*, § 41).

Nell'ambito alla *propositio* è opportuno, però, soffermarsi sull'inciso dungaliano:

> *ne vulgari proverbio lupus in fabula, pavido stupidoque silentio reprimi videar, utcunque respondebo*

Si tratta di una peculiarità interessante tanto per l'introduzione di un elemento faceto all'interno di una lettera indirizzata all'imperatore quanto, soprattutto, per l'allusione letteraria alla base del 'proverbio popolare'. L'espressione *lupus in fabula* è in origine terenziana, *Adelphoe* 537 (IV, 1, 21):

> quid nam est – lupus in fabula – pater est? – ipsust

Il commento dello scoliaste Eugrafio è il seguente:

> "silentii indictio est in hoc proverbio atque eiusmodi silentii ut in ipso verbo ve ipsa sillaba conticescat, quia lupum ridisse homines dicimus qui repente obtumuerunt; quod fere his evenit quos prior viderit lupus, ut ***cum cogitatione in qua fuerint etiam verbis et voce careant***"[7]

Qualunque sia l'origine dell'allusione di Dúngal, l'elemento popolare (*vulgari proverbio*) introdotto nella sua *captatio benevolentiae* avrà probabilmente fatto sorridere l'imperatore (vedi anche *infra* § 2.1.4).[8]

Nella successiva *argumentatio*, ossia la vera e propria trattazione del 'problema' da parte *orator* (§§ 5 ss): l'autore non fa altro che affidarsi a un'*auctoritas*.[9] La sua fonte non solo apre il

---

[7] Eugraphius, 14, 22, in Patrick McGlynn, *Lexicon Terentianum*, 2 vols., London, Glasgow, Bombay, 1963-1967, I, 1963, p. 339.

[8] Dal punto di vista della teoria linguistica (vedi § 2.1.1), il rimando dungaliano al proverbio popolare è esplicito e pertanto comprensibile a Carlo Magno, in quanto contenuto nel testo (*ne vulgari proverbio lupus in fabula*), quantunque il suo contesto di riferimento sia implicito, ossia esterno al testo stesso (cfr. LECKIE-TARRY, *Language and Context*, cit. *infra* alla n. 16, cap. 8: «Theme and information structure», p. 132-157, «Implicitness/Explicitness», p. 132-133, che, a p. 132, cita R. Hasan, «Ways of Saying, Ways of Meaning», in R.P. FAWCETT, M.A.K. HALLIDAY, S.M. LAMB e A. MAKKAI (eds), *The Semiotics of Language and Culture: Language as Social Semiotics*, London, Frances Pinter, 1984), p. 105-162, p. 109: [in caso di riferimento esplicito] «the correct interpretation of the message requires no more than a listener who has an average working knowledge of hte language in question».

[9] Rimandando alla trattazione tecnica di Costantino Sigismondi sulla frequenza delle eclissi nella II parte dell'articolo, cito qui una sua valutazione generale sul valore storico e il contributo scientifico della *responsio* dungaliana: «L'astronomia nel IX secolo tanto nell'occidente cristiano che nell'oriente arabo è un'astronomia di tipo pratico, utile al calcolo delle date delle feste mobili e alla misura del tempo per la preghiera.(cfr. Stephen C. MCCLUSKEY, *Astronomies and Cultures in Early Medieval Europe*, Cambridge, Cambridge University Press, 1998). Con Dùngal abbiamo un'idea delle fonti a cui un erudito poteva attingere per chiarire problemi di astronomia computazionale che andasse al di là del calcolo della Pasqua. Le effemeridi che Dùngal consulta sono abbastanza accurate da permettergli di rispondere sull'argomento della doppia eclissi, mentre dal valore dell'anno cosmico di 15000 anni si intuisce che i valori dei periodi sinodici e siderali dei pianeti usati nel calcolo (di cui si dà solo il risultato) sono lontanissimi dalla precisione che era stata raggiunta nell'Almagesto. Il valore di questi dati è quindi solamente estetico. Nel MURATORI, *Antiquitates Italicae Medii Aevii,* Tomus III.col. 817-

discorso ma occupa i due terzi dell'intero testo (ben 27 dei 41 paragrafi indicati; vedi *infra* § 3 di quest'introduzione). I §§ 5-8 e 36-37 introducono le citazioni fornendo i nessi di collegamento logico nell'ambito dell'*argumentatio* tra le parti della trattazione teorica e la sua applicazione pratica al caso specifico sottoposto da Carlo Magno a Dúngal.

L'*epilogus* (§§ 38-41), cui si è già fatto riferimento, è strutturato in diverse sotto-sezioni. Dapprima, viene semplicemente indicata la conclusione dell'*argumentatio* (*hic sit finis dicendi*), cui segue la duplice motivazione addotta dall'autore: l'incapacità di approfondire il discorso (*proprii ingenioli exiguitas*) e l'impossibilità di ricorrere a ulteriori fonti (*Plinius et alii libri.... non habentur nobiscum*) (§ 38). A una duplice citazione biblica è affidata l'*excusatio* dungaliana per la propria inadeguatezza (*stulta mundi elegit Deus; Non est apud eum personarum acceptio*) seguita da un duplice ricorso alla sapienza dell'imperatore (*purissimae et clarissimae sapientiae lux*; *serenissimi splendoris radius*) (§ 39) su cui torneremo.[10]

Il § 40 della lettera costituisce una sottosezione dell'*epilogus* che merita di essere trattata separatamente. Estendendo l'invito alla preghiera benaugurale del § 1 a tutti i sudditi dell'imperatore, nuovamente chiamati ad intercedere per la sua *longaeva vita* (§ 1) *ut Dominus… suo populo donet et tribuat* ***multis annis*** *de tali et tanto principe et magistro gaudere* (§ 40a)*, in multos extendat annorum curriculos* (§ 40b).

---

824. edizione di Milano, tip. Palatina, 1740, è riportata la lista dei libri della biblioteca di Bobbio. La presenza nell'elenco dell'Astronomia di Boezio, lascia intendere che il redattore della lista sia stato Gerbert d'Aurillac quando era stato nominato abate, poiché in una sua lettera ne riporta la scoperta. In questa lista vi sono anche i libri di Dùngal, ed altri aggiuntisi posteriormente a lui. Il libro sulla musica di Agostino [*Aurelii Augustini De musica*, edizione critica a cura di Giovanni MARZI, Firenze, Sansoni, 1969] era a disposizione di Dùngal, ed anche il *librum* De *Computo* di Dionigi il Piccolo [cfr. H. MORDEK, *Dionysius Exiguus*, in *Lexikon des* Mittelalters III, München, Zürich, 1986, col. 1088-1092, bibliografia col. 1091-1092; . *Liber de Paschate*, edizione completa di W. Jan, del 1718, rist. in J.P. Migne, *Patrologia Latina* 67, col. 483-514; tavola pasquale e *argomenta* riediti in B. KRUSCH, *Studien zur christlich-mittelalterlichen Chronologie. Die Entstehung unserer heutigen Zeitrechnung*, Berlin, 1938, *Abhandlungen der Preussischen Akademie der Wissenschaften* 1937, *Phil-hist. Kl.* 8, pp. 63-81, citato in Georges LECLERCQ, «Dionysius Exiguus and the Introduction of the Christian Era», *Sacris Erudiri* 41 (2002), p. 165-246, p. 189, n. 63], così come *de Ysidori Libris. Expositio in Genesi I* [edizione critica a cura di Michael GORMAN in preparazione] *De diebus, & Septimanis, temporibus, & Signis, libros III.* Tra i libri del presbitero Benedetto (probabilmente posteriore a Dùngal perché elencato dopo) era Virgilio ed il *Computus* di Vittore [d'Aquitania, di cui Abbone di Fleury († 1004) fece un commentario, nuova ed. Alison M. PEDEN, *Abbo of Fleury and Ramsey. Commentary on the Calculus of Victorious of Aquitaine*, Oxford, Oxford University Press, 2002, recensione in ISIS, International Review Devoted to the History of Science and Its Cultural Influences 93, 5 (2002), p. 66], mentre tra i libri del presbitero Pietro erano il *Librum Martiani de Nuptiis Philologiae & Mercurii [...]* [*Le nozze di Filologia e Mercurio*, introduzione, traduzione, commento e appendici di Ilaria Ramelli, Milano, Bompiani, 2001] il *Librum Ciceronis de Senectute* [*Cato Maior de Senectute*, edited with introduction and commentary by J.G.F. Powell, Cambridge, Cambridge University Press, 1990] *in quo habetur Expositio in Somnio Scipionis* [Ambrosii Theodosii Macrobii Commentarii in Somnium Scipioni,s, edidit Iacobus WILLIS, Stutgardiae; Lipsiae, in aedibus B. G. Teubneri, 1994; Macrobe, *Commentaire au songe de Scipion*, texte établi, traduit et commenté par Mireille Armisen-Marchetti, 2 vol.,Paris, Les Belles Lettres, 2001-2003]*, et Boetii de Musica* [Anicii Torquati Severini Boetii *De institutione aritmetica libri duo: De institutione musica libri V / accedit Geometria quae dicitur Boetii*; e libris manu scriptis edidit Godfredus Friedlein, Frankfurt a.M., Minerva G.M.B.H., 1966, p. 175-371, cfr. Alison WHITE, «Boethius in the Medieval Quadrivium», in *Boethius. His Life, Thought and Influence*, edited by Margaret GIBSON, Oxford, Blackwell, 1981, p. 162-205, M. Bernhard, «Glosses on Boethius' *De institutione musica*», in *Music Theory and Its Sources: Antiquity and the Middle Ages*, Notre Dame, In, 1990, p. 136-149].... *quas non reperimus*. Del libro di Marziano Capella, in cui si esprime in termini semplici il sistema tolemaico, ce n'era un'altra copia poco dopo menzionata tra i libri *fratris Smaragdi.* Da questa lista si capisce come Gerberto abbia letto tutto il possibile, ma anche che Dùngal, quasi 200 anni prima, forse non aveva tutti quei testi a disposizione. Le sue citazioni esplicite restano limitate a Virgilio [v. *infra*, § 4.1.2], Macrobio e Cicerone [e forse Vitruvio]». Per un'introduzione alla computistica medievale europea, v. Paolo PICCARI, «Il «computus» nell'alto medioevo», in *Il Tempo nel Medioevo. Rappresentazioni storiche e concezioni filosofiche. Atti del Convegno internazionale di Roma, 26-28 novembre 1998*, a cura di Roberto CAPASSO – Paolo PICCARI, praef. Ludovico GATTO, Roma, Accademia Europea per gli Studi Storici – Società di Demodossalogia, 2000, p. 49-62; *Arno* BORST, *Computus: Zeit und Zahl in der Geschichte Europas*, Berlin, K. Wagenbach, 1991, spec. capp. VI e VII sui secoli VII-IX. Borst (p. 38 e n. 70) cita il contributo per via epistolare di Alcuino di York (*Epistula* 171, *Monumenta Germaniae Historica*, *Epistulae* IV (1895), pp. 281-3) a Carlo Magno in materia di *computus* e *calculatio* del moto solare e lunare. Su matematica e astronomia alla scuola palatina di Carlo, vedi, infine, Paul Leo BUTZER, «Mathematics and Astronomy at the Court of Charlemagne and Its Mediterranean Roots», *CRM, Cahiers de recherches médiévales (XIIIe-XIV siècles)* 5 (1998), p. 203-244.

[10] Vedi *infra* § 4.2.

*Una laus regia incorporata nell'epilogus*

Dúngal introduce, quindi, a questo punto nell'*epilogus* una sorta di *laus regia* che si chiude proprio con la tradizionale formula, *exaudi, exaudi, exaudi, Coriste* (§§ 40a-b).[11] Infatti, dopo aver additato Carlo Magno a modello per governanti, soldati ed ecclesiastici, per filosofi ed accademici, ciascuno nella sua area d'azione e di competenza (§ 40a), Dúngal prega che vengano moltiplicati i trionfi di Carlo, conservata la sua discendenza e rafforzata la sua salute perché possa vivere ancora per molti anni:

> *rogemus ut nostri optimi domini Augusti Caroli triumphos multiplicet, imperium dilatet, sacram conservet progeniem, sanitatem confirmet, vitam in multos extendat annorum curriculos. Exaudi, exaudi, exaudi, Christe.*

«Vita et victoria» sono l'augurio contenuto nella *laus* dedicata a papa Leone, Carlo Magno e figli[12] citata a esempio da Kantorowicz[13] dal codice parigino, Bibliothèque Nationale MS 13159:

| | |
|---|---|
| Exaudi Christe | Leoni summo pontifici<br>et universali pape vita |
| … | |
| Exaudi Christe | Carolo excellentissimo et a Deo coronato atque amgno et<br>pacifico Regi Francorum et Longobardorum ac patricio Romanorum vita et victoria |
| …. | |
| Exaudi Christe | Nobilissime proli regali vita |
| … | |

Quest'articolato *Fürstenspiegel* si espande, poi, fino all'ultimo conclusivo *non solum optime domine sed et piissime atque amantissime pater* della chiusa al § 41, con la triplice aggettivazione quasi a riecheggiare l'*incipit*, *in nomine Patri et Filii et Spiritus Sancti*.

## 2. Lingua e stile della lettera

La capacità di Dúngal quale *fidelis orator* di Carlo non si manifesta solo nel costruire la risposta alla sollecitazione di Waldo, ordinandone le componenti strutturali con abilità pari alla concisione e all'efficacia (cfr. § 3: *et utinam tam efficax quam voluntarius existerem, ut non solum velle, sed et compote voto assequi cupita valerem*). Essa si evidenza altresí nell'utilizzare la lingua in funzione degli obiettivi stabiliti non solo dal *dominus* ma dal *famulus* stesso, consulente dell'imperatore, studioso e suddito (cfr. Conclusioni, *infra*).

[11] Ernst H. Kantorowicz, *Laudes Regiae. A Study in Liturgical Acclamations and Medieval Ruler Worship*, Berkeley and Los Angeles, University of California Press, 1946.
[12] Cfr. *ibidem*, p. 14, n. 26, Hans Hirsch, «Der mittelalterliche Kaisergedenke in den liturgischen Gebeten, *Mitteilungen der Oesterreischen Instituts fur Gechichtsforschung* 44 (1930), p. 1ss.
[13] Ernst H. Kantorowicz, *Laudes Regiae*, cit. n. 11, p. 15.

Esamineremo, dunque, lessico, morfologia, sintassi (occorrenze e ricorrenze) e retorica dell'opera (figure retoriche e sintassi, con particolare riferimento all'uso del parallelismo e dell' iperbato[14]). Infine, illustreremo l'uso del *cursus* esaminando le *clausulae* ritmiche nel breve testo dungaliano.

## 2.1 Lessico e referenti semantici della lettera[15]

### *2.1.1 Spunti di ermeneutica[16]*

Lo studio del lessico della lettera di Dúngal a Carlo Magno sull'eclissi di Sole ne rivela alcune costanti semantiche, che mettono in evidenza le sottolineature presenti nel testo. Se, da un lato, ripetizione e *variatio* costituiscono un tratto stilistico significativo per cui rimandiamo al § 2.4, ci interessa qui individuare alcuni *campi semantici* di rilievo nello scritto dungaliano, a mo' di chiavi ermeneutiche suggerite dall'autore stesso circa l'*opera* e l'*attività* dello scrivente. Potremmo schematicamente suggerire che al livello puramente dichiarativo si lega un livello pragmatico,[17] come ulteriore introduzione al livello espressivo.

Livello espressivo (connotazione)
↑ ↑
Livello pragmatico (ripetizione)
↑ ↑
Livello dichiarativo (lemma)

Significativa per una lettura pragmatica della lettera di Dúngal a Carlo Magno la combinazione tra due nozioni sviluppate dai linguisti,'tenore personale' e 'tenore funzionale': il primo, infatti, "realizza la funzione Interpersonale del linguaggio, mentre il tenore funzionale riflette il ruolo che il linguaggio sta giocando nella situazione, o ciò per cui il linguaggio viene utilizzato nella situazione contingente (p. es. persuadere, esortare, disciplinare)".[18] In particolare, due sono gli ambiti in cui emergono tali nozioni nell'analisi qui proposta:

---

[14] Per gli stessi elementi nei *Responsa contra Claudium*, cfr. *Responsa contra Claudium*, cit. n. 1, p. LX-LXIX.. Johann Baptist HOFMANN, Anton SZANTYR, *Stilistica latina*, a cura di Alfonso TRAINA, traduzione di Camillo NERI, Aggiornamenti di Renato ONIGA, Revisione e indici di Bruna PIERI, Bologna, Patron, 2002, p. 91-267, parte III, 'Espressione e rappresentazione

[15] Cfr. ZANNA, *Responsa contra Claudium*,, p. LXXVII-LXXX.

[16] Cfr. Helen LECKIE-TARRY, *Language and Context*, *A Functional Linguistic Theory of Register*, edited by David Birch, London and New York, Frances Pinter, 1995.

[17] Cfr. J. Austin, *How to do things with words*, *The William James Lectures delivered at Harvard University in 1955*, edited by J.O. Urmson and Marina Sbisà, Oxford, Oxford University Press, 1978), tr. it. *Come fare cose con le parole*, a cura di Carlo Penco e Marina Sbisà, Genova, Marietti, 1998; cfr. John R. Searle *Speech Acts: an essay in the philosophy of* language, Cambridge, Cambridge University Press, 1977; tr. it. *Atti linguistici: saggio di filosofia del linguaggio*, Torino, Boringhieri, 1976; cfr. anche il volume *Gli atti linguistici*: *aspetti e problemi di filosofia del linguaggio*, Milano, Feltrinelli, 1995.

[18] Michael GREGORY and Suzanne CARROLL, *Language and Situation: Language Varieties in their Social Contexts*, London, Routledge and Kegan Paul, 1978, pp. 51-3, come riferito in Helen LECKIE-TARRY, cit. n. 12, p. 26: «Personal tenor realizes the Interpersonal function of language, while functional tenor reflects the role that language is playing in the situation, or what language is being used for in the situation (e.g. to persuade, exhort, discipline)».

(a) la caratterizzazione dell'occasione e della composizione dell'opera
la caratterizzazione della lettera in quanto testo letterario e non

Quanto al primo aspetto, attraverso la ripetizione di alcuni lemmi, l'attenzione del lettore / spettatore è attirata sul dialogo tra le *dramatis personae* della lettera, Dúngal, Carlo Magno e Waldo e sui loro reciproci ruoli e interazioni. Quanto al secondo, presenteremo una lettura dell'epistola a cavallo di generi e stili diversi (*oratio*, *fabula*, *encomium*).

All'elemento soggettivo, ossia alla rappresentazione del ruolo di ciascuna figura menzionata nella lettera, e a quello formale, ossia all'ideazione letteraria della lettera, si accompagna poi l'elemento oggettivo del contenuto della lettera stessa. In questa prima parte, l'analisi si limiterà alle considerazioni relative alla tipologia letteraria della lettera secondo le chiavi interpretative offerte dall'autore stesso. Nel merito della tematica specifica, le eclissi di Sole dell'810, come studiata e presentata da Dúngal, si addentrerà nella seconda parte di questo saggio uno storico dell' astronomia.

*2.1.2 La caratterizzazione di Carlomagno*

*Ubi maior minor cessat*. Le *virtutes*, la *paternità affettuosa* di Carlo sono illustrate attraverso una serie di sinonimi e paronomasie:

*gloriosissimo; gloria sine fine* (§ 1)
*Dilectissime* (§ 2); *beatissimae et clarissimae* [*serenitati*] (§ 3); *beatissime* (§ 28)
*serenissimo* (§ 1), *serenitati* (§ 3), *serenissimae et*
*piissimae longanimitatis* (§ 4); *piissime* (§§ 39, 41)
*sanctissimo et utilissimo praecepto* (§ 3)
*fortis, sapiens, et religiosus* (§ 40a)
*optimi domini; optime domine* (§§ 40-41)
*reverentissime atque dulcissime, amantissime* (§ 41)

Nei *Responsa contra Claudium* si trova la stessa abbondanza di superlativi nel ricordo di Carlo (§ 12):

> *dominus beatissimae memoriae Carolus ferventissimus vigilantissimusque catholicae fidei tutor et propugnator*[19]

Tre sono gli ambiti semantici richiamati dagli aggettivi usati nella lettera a Carlo Magno sulle eclissi di Sole:

- affetto (*dilectissime*, *dulcissime*, *amantissime*)
    - virtú umane e religiose (*fortis, sapiens*, *utilissimo praecepto*; *religiosus*, *sanctissimo, reverendissime, piissimae longanimitatis* (§ 4); *piissime* (§§ 39, 41); *optimi domini; optime domine (§§ 40-41)*
    - gloria (*gloriosissimo*; *beatissimae et clarissimae* [*serenitati*] (§ 3); *beatissime* (§ 28)

All'immagine pubblica di Carlo quale integerrimo *defensor fidei* dei *Responsa*, scritti circa quindici anni dopo, la lettera aggiunge dunque un elemento più personale descrivendo l'intenso rapporto tra l'imperatore e il suo corrispondente e le qualità morali e di comando (*fortis*, *utilis*) di *talis rex et talis princeps* (§ 40a).

[19] Paolo ZANNA, *Responsa contra Claudium*, cit. alla n. 1, p. 6.

### *2.1.3 La caratterizzazione dell'abate Waldo*

Vi è poi il ruolo di intermediazione di Waldo, abate di Saint-Denis (806-814), chiamato a rappresentare l'imperatore e a sollecitare la *perizia* contenuta nella lettera. La caratterizzazione di Waldo è duplice, rispetto all'imperatore

> *vestro fideli famulo; vobis fidelis* (§ 41),

e a Dúngal stesso:

> *mihi de hac re gravis et importunus exactor* (§ 41) [cfr. *exactionem* (§ 28) *urgentem exactionem* (§ 41)].

Dúngal sarebbe dunque sottoposto all'autorità dell'abate di Saint-Denis, forse tra le file dei suoi monaci, e a Saint-Denis sarebbe rimasto con ogni probabilità anche dopo la morte di Waldo sotto Ilduino, in cui onore compose un carme.[20]

### *2.1.4 L'autorappresentazione di Dúngal: definizioni*

Nelle autodefinizioni dungaliane contenute nell'*exordium* della lettera a Carlo Magno: *vester fidelis famulus et orator* (§ 2); *quasi sectator sapientiae* (§ 3) sono ravvisabili piani e toni diversi.

L'aggettivo *vester* qualifica Dúngal all'interno del suo rapporto familiare con l'imperatore. Tale definizione si colloca perciò sul piano orizzontale, personale, affettivo.

Il sostantivo *famulus* qualifica Dúngal all'interno di un rapporto di obbedienza al suo signore. Tale definizione si colloca perciò sul piano verticale, organico, istituzionale.

*Orator* è un termine tecnico che qualifica Dúngal all'interno di una categoria professionale. Tale definizione si colloca perciò sul piano traversale, comunitario, professionale.

Il binomio *sectator sapientiae* va interpretato soffermandosi su ciascuno dei componenti (a) e (b) nonché dando il dovuto valore all'avverbio *quasi* che lo precede.

(a) L'eco di *sectator* risuona verso la fine della lettera: *tantum studium erga astronomiae aut cujuscunque disciplinae assectationem* (§ 37). La ricerca è un compito che Dúngal indica come auspicabile per sé come per ogni studioso.

(b) La *sapientia* è poi un dono che Dio ha abbondantemente elargito a Carlo Magno (*sapiens*, § 40a), perché egli ne possa illuminare i vicini e i lontani, i filosofi (per definizione amanti della σοφια) e i maestri: *vos, piissime Auguste, quibus prae omnibus affluentiam* ***sapientiae****, sicut et caeterarum sanctarum virtutum Deus distribuit*; *vestrae purissimae et clarissimae* ***sapientiae*** *lux his qui prope sunt luceat, sed et his qui longe* (§ 39); ***philosophis et scholasticis*** *ad honeste de humanis philosophandum et* ***sapiendum*** (**§** 40a). La sapienza è definita come una delle virtù cui aspirare e di cui servire da modello.

[20] Cfr. M. FERRARI, «'*In Papia conveniant ad Dungalum*'», cit. n. 2, p. 6 e n. 4, *MGH*, *Poetae Latini Aevi Carolini* I (1884), 664-665, n. XXVII.

(c) L'avverbio *quasi* "in concetti attributivi, per mitigare l'improprietà dell'espressione troppo larga, troppo ardita" nel significato di "per così dire"[21], attribuisce una connotazione auto-ironica all'autodefinizione dungaliana.

Tale definizione si colloca perciò sul piano impersonale, paradigmatico, simbolico.

Alle tre autodefinizioni analizzate sopra corrispondono diversi toni espressivi: si va dall'informale *vester*, all'accondiscendente *famulus*, al solenne *orator*, all'evocativo *quasi sectator sapientiae*. Si tratta di definizioni inclusive ma a gradi diversi: Dúngal sceglie di rappresentarsi tanto identificandosi nei referenti indicati *per via affermativa* (*vester famulus et orator*) quanto rileggendovisi per *via negativa* (***quasi** sectator sapientiae*).

Infine la stessa forma di auto-nascondimento si trova nel § 39, dove Dúngal si annovera tra i *reclusos* che beneficiano del «raggio del Vostro serenissimo splendore… anche se solo attraverso le fessure e le fenditure».

Potremmo riassumere le varie implicazioni semantiche dell'autodefinizione dungaliana in questa lettera a Carlo Magno nella seguente tabella:

| **L'AUTODEFINIZIONE DI DUNGAL DI FRONTE A CARLO MAGNO – ANALISI** | | | | |
|---|---|---|---|---|
| DEFINIZIONE | *Vester* | *Famulus* | *Orator* | *Quasi sectator sapientiae, reclusos* |
| PIANI | orizzontale, | verticale, | traversale, | globale |
| | personale | organico | comunitario, | impersonale |
| | affettivo | sociale | Professionale | simbolico |
| TONI | informale | Servile | Ufficiale | Ironico, allusivo |
| IDENTIFICAZ | Dichiarata | | | proiettiva[22] |

Per chiarire meglio la posizione personale e sociale di Dúngal nella lettera occorre esaminare altri aspetti dell'ideazione e stesura della lettera a Carlomagno.

*2.1.5 Il responso di Dúngal: azione e coscienza di uno studioso*

Le indagini dungaliane sono descritte in termini di impegno obbediente e di coscienza del limite insito nella ricerca:

*interrogarer*, *respondendo* (§ 3); *respondere; respondebo*; ***responsio*** (§ 4)
***relatu*** (§ 2)

*proferendo / faterer*
*quid scirem et quid sentirem*
***exceptum scriberetur***
*scriptumque vobis deferretur* (§ 3)

[21] Ferruccio CALONGHI, *Dizionario Latino-Italiano*, 3° edizione interamente rifusa ed aggiornata del Dizionario *Georges-Calonghi*, Torino, Rosenberg & Sellier, 1964, s.v., p. 2282.

[22] Ovvero «che trasferisce su una realtà o una situazione esterna il proprio mondo interiore», Salvatore Battaglia, *Grande Dizionario della Lingua Italiana*, 21 voll., Torino, Utet, 1964-2002, vol. XIV, s.v., p. 569.

I riferimenti terminologici elencati sopra riguardano due ambiti:
a) il genere letterario della lettera-trattato astronomico in questione (*responsio*)
b) il rapporto tra scritto e orale in tale contesto di ricerca (*relatu*, *proferendo, faterer*, *exceptum scriberetur*, *scriptum*).

Se al primo punto abbiamo dedicato il § 1, qui concentreremo l'attenzione sul secondo.

Al § 3 Dúngal descrive, infatti, il processo di composizione e definizione del testo. Dúngal è chiamato a verificare le affermazioni di altri (*plurium relatu*) in un dialogo aperto con Carlo Magno (*interrogarer*) e con questi ultimi (*proferendo et rispondendo*) e a curare la stesura e la consegna della nota scritta (*exceptum*, *scriptum*).

Se Dúngal è chiamato a rispondere a tale richiesta imperiale, sembra poi possibile isolare un campo semantico molto significativo concernente lo studio diligente della letteratura specialistica (nel caso specifico in ambito astronomico) considerando la forte sottolineatura della *peritia* e *subtilitas* e della *diligentia* necessarie:

*peritia* (§§ 2 ablativo, nominativo 4)
*omnium disciplinarum* ***peritissimi*** (§ 28)
*conamine* (§ 4); *acumine* (§ 4)
*industriam studiumve* (§ 4)
*nisu* (ablativo § 4); *nisus* (participio passato, § 37): *nitor* (§ 40a)
*subtilissime* (§ 28); *subtilioribus… motibus* (§ 29)
*diligentiores libri* (§ 4) *diligentissime*, *diligentia* (§ 28); *diligentem observationem* (§ 30)

Suggestivo l'accostamento con una lettera di Cicerone che citeremo anche più avanti per un altro rimando lessicale. Il grande avvocato così scrive al fratello Quinto:

> *valent pueri,* ***studiose*** *discunt,* ***diligenter docentur***[23]

Nel contesto degli studi si sottolinea la competenza multidisciplinare degli *antiqui philosophi* (§§ 2, 28), specie di quelli egiziani: *omnium philosophiae disciplinarum parentes* (§ 7), *omnium disciplinarum* ***peritissimi*** (§ 28) e si raccomanda *tantum studium erga astronomiae aut cujuscunque disciplinae assectationem* (§ 37)

Proprio per definire la precisa collocazione disciplinare dell'opera dungaliana più marcata ancora appare la concentrazione del lessico proprio della *dialettica* e della *retorica*: *argumentum, ratio, sermo, expressio*:

***talibus argumentis*** (§ 30)

***de ratione*** (§ 2); *rationis investigatio* (§ 4)
*credibilibus rationibus*, *ratio* (§ 8)
*naturas,* ***rationes***, *causas et origines* (§ 28)

*exercitatiori sermone et enucleatiori expressione* (§ 4)
*compositiores et diligentiores libri* (§ 4)

In Cicerone, *exercitatus* è un aggettivo particolarmente legato alla formazione umana e intellettuale del *sapiens*:

[23] *Ep.* CXLIX, *Ad Quintum fratrem* (*Q. fr.* III, 3), 1, ed. *Cicéron, Correspondance*, 10 vols, Paris, Belles Lettres, 1934-1991), vol. III, texte établi et traduit par L.A. Constans, Paris, Belles Lettres, 1936, p. 100.

*earum rerum scientiam non doctis hominibus ac sapientibus, sed in illo genere exercitatis concedendam putant* (*De re publica* I, 11)

*in quibus (rebus sapiens) versatus exercitatusque sit* (*Academicorum libri* II, 110)[24]

*quis exercitatior*? (*in M. Antonium orationes Philippicae* VI, 17)[25]

*consuetudine beneficientiae paratiores erunt et* ***exercitatiores*** *ad bene de multis promerendum* (*De officiis* II, 53)

Significativo appare poi l'abbinamento *exercitatus sermo – enucleata expressio* di chiara impronta retorica. Sembra rilevante la lezione ciceroniana circa il *genus dicendi enucleatum* (Cic. *Or*. 91), in cui si abbina chiarezza ed eleganza:

*modo id* ***eleganter enucleateque*** *faciat* (*Or*. 28, 5).
*Enucleate… et polite* (*Brutus* 115)[26]

Eppure nel *De finibus* Cicero distingue tra la chiarezza nell'esprimere le cose di minor conto (*minora*) e l'ornamento confacente a quelle di maggior importanza (*grandia*)

*Qui grandia* ***ornate*** *vellent,* ***enucleate*** *minora* ***dicere*** (*De finibus*, IV, 6)[27]

E nel contesto già citato delle *Familiares*, Cicerone si esprime in toni molto vicini alla sensibilità dell'epistola di Dúngal:

*de via nihil praetermisi quadam epistula quin* ***enucleate*** *ad te perscriberem*.[28]

La chiarezza espositiva è vista come il risultato di due operazioni sensoriali e mentali: *intentio e observatio*

*sagacissima intentione* , *intentissime observaverunt* (§ 28); *sagacem observationem* (§ 30)
*Platonis observatio* (§ 8); *tanta explorationis et observationis diligentia intentus* (§ 37)
*plenius et eruditius* (§ 4); *plenissime exploraverunt* (§ 28)
*elimatae et defaecatae mentis*, *perspicacissima et purgatissima* ***sensus*** *acie* (§ 28)

Si tratta cioè di operazioni connesse al campo del sensibile (*perspicacissima sensus acies*) e dell'intelletto (*mentis*, ***sagacissima*** *intentione*, ***sagacem*** *observationem*). Comune obiettivo di entrambe è il perfezionamento conoscitivo (*plenius et eruditius*, *pienissime*, *elimatae et defaecatae mentis*, *purgatissima acie*). Strumento comune e correlato (anche linguisticamente a *intentio*) è l'estrema attenzione e/o diligenza (*intensissime, diligentia intentus*).

---

[24] *Cfr.* H. MERGUET, *Lexicon zu den philosophischen Schriften Cicero's mit Angabe sämtlicher Stellen*, Erster Band, Hildesheim, Georg Olms Verlagsbuchhandlung, 1961, p. 911, *s.v. exercitatus*

[25] *Cfr.* H. MERGUET, *Lexicon zu den Reden des Cicero*, Jena, Verlag von Hermann Dufft, 1877), p. 265.

[26] *Cfr.* H. MERGUET, *Handlexicon zu Cicero*, Hildesheim, Georg Olms Verlagsbuchhandlung, 1962, p. 227, *s.v. enucleate*.

[27] *Cfr.* H. MERGUET, *Lexicon zu den philosophischen Schriften Cicero's mit Angabe sämtlicher Stellen*, Erster Band, Hildesheim, Georg Olms Verlagsbuchhandlung, 1961, p. 815, *s.v. enucleate*.

[28] *Ep.* CXLIX, *Ad Quintum fratrem* (*Q. fr.* III, 3), 1, ed. cit., p. 101.

## 2.2 Morfologia e sintassi

Elenchiamo di seguito alcune strutture morfologico-sintattiche ricorrenti nella lettera a Carlo Magno come nei *Responsa contra Claudium*.[29]

2.2.1 ***Abbinamenti morfologici***[30]

2 nomi

§ 1 *donis et exercitiis*
§ 2 *famulus et orator*
§ 4 *investigatio et peritia*
*industriam studiumve per fragilitatem infirmitatemque*
§ 28 *causas et origines*, *ortus et obitus; cursus et recursus, accessus et recessus*
§ 37 explor*ationis et observationis;*
*otiositate et curiositate*
§ 32 *rimas et juncturas*
§ 40a *Dominus et Salvator;*
*principe et magistro*
*operum et virtutum*

2 aggettivi (specie al grado superlativo[31])

§ 1 *nobilis honestisque*
§ 2 *usitatae et certae*
§ 3 *beatissimae et clarissimae*
*sanctissimo et utilissimo*
*pronus et alacris*
*effecta et adimpleta*
§ 4 *compositiores et diligentiores*
*leves compendiososque*
*pavido stupidoque*
*serenissimae et piissimae*
§ 6 ***proprias et speciales***
**§** 28 ***diutissima et meditatissima***
**§** 29 *nobiliores et faciliores*
§ 39 *purissimae et clarissimae*
§ 40a *attentis et assiduis*
*summis…. et eccellentibus*
§ 41 *gravis et importunus*
*piissime et amantissime*

2 avverbi

§ 4 ***proprie et specialiter*** (*iterum* al § 28)
*plenius et eruditius*

[29] Cfr. anche Paolo ZANNA, *Responsa contra Claudium*, cit. n. 1, p. LIV-LXXVII.
[30] Cfr. *ibidem*, p. LXIX-LXXII.
[31] Come già fatto notare in riferimento alla *persona* di Carlo Magno, al § 2.1.2.

§ 28 *subtilissime et instantissime, accuratissime et efficacissime*; *diligentissime et intentissime*
§ 29 *certissime et studiosissime*
§ 40a *reventerque atque orthodoxe*
§ 41 *reverentissime atque dulcissime*

2 verbi

§ 4 *reptans et movens*
§ 6 *nominata et numerata*
§ 28 *scierunt et praescierunt*; *inventa et deprehensa*
§ 39 *instruere et dirigere*
§ 40a *rogare et postulare*;
*donet et tribuat*
*sentiendum et credendum*
§ 40b clamemus…. et rogemus

*2.2.2 Espressioni del latino volgare*[32]

Rimandando al paragrafo successivo per le altre particolarità stilistiche che caratterizzano l'assetto sintattico della prosa dungaliana (ad es. parallelismo, epanalessi, iperbato) rileviamo un tratto del latino volgare prima parlato e poi scritto in una proposizione dichiarativa introdotta da *quod* in luogo di un'infinitiva:[33]

§ 2: audivi …. quod….

**2.3 Stile: figure retoriche**[34]

*2.3.1* ***Figure di suono***

***Allitterazione***[35]

§ 2 *elementorum effectum*
§ 3 *parere praecepto*

*Rima e omeoteleuto*[36]

§ 3 *sanctissimo et utilissimo*
§ 3 *existerem… valerem*
§ 4 *donaturam… castigaturam*
§ 5 *ferendo et continendo*

---

[32] Cfr. ad es. Dag NORBERG, *Manuale di latino medievale*, a cura di Massimo Oldoni, Cava dei Tirreni, 2002 e il mio saggio sulle epistole dungaliane, «"A cavallo" tra teologia e retorica. Dúngal e il decoro di un latinista irlandese sul Continente», cit. n. 1.
[33] Cfr. Dag NORBERG, cit. alla nota precedente p. 130: "La proposizione infinitiva, caratteristica del latino classico, fu ben presto sostituita nella lingua parlata da una dichiarativa introdotta da *quod* o *quia* (talvolta anche da *quoniam*). Questo tipo di proposizione si è estesa nei testi letterari di un'epoca più tarda sotto l'influenza della Bibbia, dove i traduttori si sono rifatti alla costruzione greca corrispondente."
[34] Cfr. ***ibidem***, p. LXII-LXXV. Cfr. HOFMANN, SZANTYR, *Stilistica latina*, cit. n. 9.
[35] Cfr. HOFMANN, SZANTYR, *Stilistica latina*, cit. n. 21, p. 29-35, 284-288.
[36] ***Ibidem*****,** p. 29-40, 288.

§ 40a *rogare et postulare*

*Espressione e rappresentazione*
*Lessico*

*Paronomasia*

*serenissimo* (§ 1), *serenitati* (§ 3), *serenissimae et piissimae longanimitatis (§ 4)*

*Litote*[37]

§ 1 *non immemor*

Nel carme dedicato a Baldo di Salisburgo (vv. 3 e 6) Dúngal stesso scrive *inmemor haud sum, non obliviscor.*[38]

*Sineddoche*

*Apices* (§ 4)

*Ripetizione*[39]

§ 2-3 *quasi*
§ 3 *quid*

§ 3 e 28 *indubitatissime*

§ 28 *diligentissime*
§ 30 *diligentem*
§ 28 *exploraverunt*, *explorando*
§ 30 *explorantes*

§ 40b, 41 *optimi domini, optime domine*
§ 4 e 41 *clementiam, clementer*

*Epanalessi*[40]

§ 6 *duo Coluri*…duo

*Costruzione del periodo*

*Parallelismo*[41]

---

[37] *Ibidem*, p. 151-154, 314-315.
[38] Cfr. Paolo ZANNA, «I carmi di Dúngal», cit. n. 1, p. 166.
[39] Cfr. HOFMANN, SZANTYR, *Stilistica latina*, cit. n. 21, p. 225-227, 325-326.
[40] *Ibidem*, p. 208-209, 324.
[41] *Ibidem*, p. 69-72, 293-294.

§ 29 *sagacem explorationem et diligentem observationem*
§ 40b *altissimis vocibus et devotissimis cordibus*
§ 41 *commonendo interrogaret et exigendo commoneret*
§§ 28 e 30 *intuentes et intuendo experientes = praesciebant et praesciendo praedicebant*

*Figure di collocazione*[42]

Iperbato[43]

Nella lettera a Carlo Magno troviamo la forma più semplice di iperbato riscontrata nelle altre opere dungaliane citate in nota, secondo tratti comuni alla tradizione insulare e continentale tardoantica e alto-medioevale[44]:

aggettivo – verbo – nome

§ 3 *utilissimo parere praecepto*
§ 4 *vestri continent apices*; *congrua reddatur… responsio*
§ 27 *duas haberi medias*
§ 36 *vestrae indicant litterae*
*prima initiante luna*,
*trigesima incipiente luna*
§ 40 *perfectum habetur exemplar*

Chiasmo[45]

§ 1 vita longaeva, fida salus

**3 *Clausulae* ritmiche**[46]

(a) *Frequenza*

In appendice elenchiamo i tipi di *clausulae* utilizzate da Dúngal:

*cursus planus* p 3p p. es. *illustrátur emíttit* (§9), 25 occorrenze
*cursus tardus* p 4pp p. es. *laténtis emisphaérii* (§ 27), 21 occorrenze
*cursus velox* pp 4p p. es. *véniam donatúram* (§ 4), 11 occorrenze
*cursus trispondaicus* con accenti tonici che distano tre sillabe; 3 tipi:
*in una singola parola*, *dísputatiónibus* (§ 4)
p 4p, p. es. *cleménter imponátis* (§ 41), 34 occorrenze (!)
pp 3p, p. es. *ratio comméndat* (§ 8), 10 occorrenze

---

42 *Ibidem*, cit. n. 8, p.11-27, 280-282.
43 *Ibidem*, 11-19; 280-28, In Dúngal, cfr. ZANNA, «I carmi di Dúngal per Ildoardo e Baldo», cit. n. 1, p. 151 e ZANNA, *Responsa contra Claudium*, cit. n. 1, p. LXII-LXIX., es. cit. anche in *idem*, «"A cavallo" tra teologia e retorica», cit. n. 1, p. 52: *suam intuetur effigem*.
44 Cfr. J.N. Adams, «A Type of Celtic Hyperbaton in Latin Prose», *Bulletin of the Board of Celtic Studies* 27 (1976-8), p. 207-12.
45 Cfr. HOFMANN, SZANTYR, *Stilistica latina*, cit. n. 8, p. 22-25, 281. In Dúngal, cfr. ZANNA, «I carmi di Dúngal per Ildoardo e Baldo», cit. n. 1, p. 151, 166.
46 Cfr. HOFMANN, SZANTYR, *Stilistica latina*, cit. n. 8, p. 53-62, 291-292; Tore Janson, *Prose Rhythm in Medieval Latin from the 9th to the 13th* Century, Stockholm, Göteborg, Uppsala, 1975, spec. pp. 13-14.

Come nei *Responsa*[47], da notare, dunque, la frequenza eccezionale del *cursus trispondaicus* che, come hanno fatto notare Neil Wright e Michael Winterbotton, è tratto caratteristico della prosa ritmica di Colombano e di altri autori irlandesi.[48]

(b) *Tecnica ritmica*

Una seconda osservazione riguarda un duplice stratagemma utilizzato da Dúngal nella formazione delle *clausulae*.

(1) come nei versi per Ildoardo e per Waldo[49], anche nelle *clausulae* troviamo la ripetizione di sintagmi identici: ad es. avverbio al superlativo e verbo corrispondente:

*intentissime observaverunt plenı́ssime exploravérunt; studiosíssime cognovérunt* (§§ 28-29)

(2) la ripetizione di abbinamenti morfologici identici nelle *clausulae* è accompagnata dalla *variatio* sinonimica:

es. *intentissime observaverunt, studiosíssime cognovérunt* (§§ 28, 29)

**4 Fonti della lettera a Carlo Magno, con un'ipotesi[50]**

Oltre ai confronti stilistici e lessicali istituiti sopra all'interno del *corpus* dungaliano stesso, il maestro irlandese inserisce nel testo della lettera una serie di citazioni da un singolo autore, Macrobio, lamentando la mancanza di altre fonti cui fa soltanto cenno, come il secondo libro della *Naturalis Historia* di Plinio (§ 38).[51] Vanno esaminati, dunque:

- il mosaico di citazioni da un singolo autore
- le fonti secondarie citate nella singola fonte
- le citazioni bibliche

In ultimo presenterò un'eco suggestiva del *De architectura* di Vitruvio.

*4.1 Macrobio*

Tutte le citazioni che abbiamo indicato in nota a piè di pagina nel testo latino provengono, come detto, dal *Commento al Somnium Scipionis* ciceroniano di Macrobio (420-440 circa)[52], con cui Dúngal dimostra una grande familiarità[53] (v. note a piè pagine a cura di Costantino Sigismondi).

*4.1.1 Tecniche di citazione*

---

[47] *Responsa contra Claudium*, p. LXXV-LXXVII.

[48] Ibid., p. LXXVI, per il rimando agli studi di Neil WRIGHT, «Columbanus' s *Epistulae*», in *Columbanus: The Latin Writings*, ed. Michael LAPIDGE, Woodbridge, Suffolk, Boydell, 1997. p. 29-92, p. 57 e di Michael WINTERBOTTOM, «Aldhelm's Prose Style and its Origins», *Anglo-Saxon England* 6 (1977), p. 39-77, p. 71-2 (Table 1).

[49] Cfr. Paolo ZANNA, «I carmi di Dúngal per Ildoardo e Baldo», cit. n. 1, rispettivamente p. 150, 164-5.

[50] Cfr. Mirella FERRARI, «'*In Papia conveniant ad Dungalum*'», cit. n. 2, p. 5-6, 8.

[51] Cfr. Mirella FERRARI, *ibid*., p. 6, sull'indisponibilità di copie di Plinio a Saint-Denis.

[52] Cfr. «Introduction», in Macrobio, *In Somnium Scipionis* I, 14, 23, ed. M. Armisen-Marchetti, cit. n. 9, vol. I, pp. XVI-XVIII (datazione).

[53] Cfr. Mirella FERRARI, «'*In Papia conveniant ad Dungalum*'», cit. n. 2 p. 8.

Per quanto riguarda le citazioni del commento stesso, dal punto di vista tecnico si possono distinguere tre modalità di riuso di quest'opera da parte di Dúngal:

(1) la citazione completa
(2) la modifica testuale-contestuale
(3) l'alternanza fra lacerti di citazioni e interventi dungaliani

Alle diverse tipologie di re-impiego della fonte corrispondono probabilmente finalità complementari che integrano l'immagine di Dúngal letterato e maestro.

Al primo tipo di citazione, la citazione completa e continua, è riconducibile la maggior parte delle citazioni (§§ 9-10, 12-19, 22). Dúngal lascia, per cosí dire, la parola alla propria fonte, dove ritiene che essa sia pienamente comprensibile. Questa preoccupazione per la massima chiarezza si manifesta nell'uso frequente di *hoc est* (§§ 5 – ripetuto 3 volte – 7 – ripetuto 2 volte)

Nel secondo tipo precisato sopra, una leggera modifica testuale-contestuale serve a chiarire meglio il contenuto o ad adattarlo al discorso dungaliano

§§ 10, 20 Macrobio: *duo lumina / utrumque lumen* = Dúngal *solem et lunam, sol et luna*
§ 12 Macrobio *consideremus enim signorum ordinem* = Dúngal *considerato signorum ordine*

Al § 23, Dúngal rende al tempo presente un futuro in Macrobio: *mediam regionem tenebit numero, spatio non tenebit*" Dúngal: *mediam regionem tenet numero, sed non spatio*, in cui si semplifica anche il chiasmo del testo originale; al § 25,

Nel terzo tipo di riuso testuale, Dúngal crea dei collegamenti tra parti diverse del testo macrobiano (v. §§ 11 e 19). Al § 6 Dúngal riprende il numerale *duo* della prima citazione (*duo Coluri sunt*), come se stesse ragionando "a voce alta". Al § 7 Dúngal esplicita il concetto riferito da Macrobio sulla posizione di Platone.

> Macrobio: *Plato Aegyptios, omnium philosophiae disciplinarum parentes secutus est, qui ita solem inter lunam et Mercurium locatum volunt*
> Dúngal: *Plato* ***vero a luna sursum secundum, hoc est inter septem a summo locum sextum solem tenere confirmat,*** *secutus Aegyptios, omnium philosophiae disciplinarum parentes, qui ita solem inter lunam et Mercurium locatum volunt.*

*4.1.2 Citazioni virgiliane nel testo di Macrobio*

In Macrobio troviamo anche due citazioni dalle *Georgiche*, forse una dall'*Eneide* (vedi note a pié pagina del testo latino). Virgilio compare una sola volta nei *Responsa contra Claudium* in una citazione dal *De civitate* agostiniano.[54] Quanto alle *Georgiche*, nell'epitaffio che compose per se stesso, la Ferrari ha evidenziato che «l'uso del non frequente vocabolo *squalorum* crea una connessione con il codice *Bernese* 363 (Berna, Burgerbibliothek), grande enciclopedia formata nella cerchia di Sedulio Scoto: dove il nome di Dúngal è annotato accanto alla spiegazione di Servio per la parola *squalentibus* (in *Georg*. IV, 91)».[55]

[54] *Aen*. 1,12, in *De civitate Dei*, X, 1, l. 59, Sancti Aurelii Augustini *De civitate Dei*, Ad fidem quartae editionis Teubnerianae quam A. MCMXXVIII-MCMXXIX, curaverunt Bernardus DOMBART et Alphonsus KALB, paucis emendatis mutatis additis, *Corpus Christianorum Series Latina* 47-48, Turnholti, Typographi Brepols editores Pontificii, 1955, vol. 47, p. 272, *cfr*. *Responsa contra Claudium*, ed. cit. n. 1, p. 76 e p. CVIII-CX.
[55] Mirella FERRARI, «Dungal», cit. n. 4, p. 13.

*4.2 Citazioni bibliche*

L'uso della Bibbia in Dúngal è stato discusso altrove.[56] Due citazioni paoline confermano il frequente ricorso a Paolo da parte di Dúngal in prosa e in poesia (cfr. ad as. preghiera-epigramma, v. 23[57]) ma sono importanti soprattutto per il contenuto. Agli *excerpta* paolini al § 39:

*Stulta enim mundi elegit Deus*:[58] et
*Non est apud eum personarum acceptio*[59]

segue la chiosa dungaliana:

*ut non solum vestrae purissimae et clarissimae sapientiae lux his qui prope sunt luceat, sed et his qui longe*

Andando a leggere l'intero brano della Prima Lettera ai Corinzi della prima citazione, si scoprono le sfumature di senso di quest'ultima:

«27Ma *Dio ha scelto ciò che nel mondo è stolto* [citazione dungaliana] **per confondere i sapienti**, Dio ha scelto ciò che nel mondo è debole per confondere i **forti**. 28Dio ha scelto ciò che nel mondo è ignobile e disprezzato per ridurre a nulla le cose che sono, 29**perché nessun uomo possa gloriarsi davanti a Dio.** 30Ed è per lui che voi siete in Cristo Gesù, il quale per opera di Dio è diventato per noi **sapienza**, giustizia, santificazione e redenzione, 31perché, come sta scritto,
***Chi si vanta, si vanti nel Signore***. [libera citazione di Ger 9, 22 ss.]»[60]

La **sapienza** di Carlo, "luce" per i vicini e i lontani è quasi contrapposta all'elezione divina "di ciò che nel mondo è stolto per confondere i sapienti" e dei deboli per confondere i forti. Oltre che *sapiens* e ***fortis*** (§ 40a), Carlo è perciò definito *religiosus*, in virtú di una superiore **sapienza** cristiana (vv. 30-31). Il Dúngal che cita la Bibbia è necessariamente il Dúngal teologo, *reclusus* a Saint-Denis. Nelle conclusioni vedremo come tale elemento "paolino" si fonda con il ruolo di Dúngal *orator*.

A conferma della stessa tesi circa l'origine della vera sapienza anche la citazione della Lettera di Giacomo (1, 17) al § 28:

*illo* ***omne datum optimum et omne donum perfectum*** *offerente*
offerte da Colui da cui proviene «ogni buon regalo e ogni dono perfetto»[61].

4.3 *Un'eco di Vitruvio?*

Suggestivo l'accostamento tra il testo del § 29 della lettera a Carlo Magno di Dúngal e

---

[56] «La Bibbia in Dúngal», negli atti del convegno: *The Study of the Bible in the Early Middle Ages, Gargnano, 24-27 June 2001,* in uscita; «Dungali Poetria. Il carme Quisquis es hunc cernens: Dúngal, Aldelmo, Alcuino», letto al *IV Congreso Internacional de Latín Medieval, Poesia Latina Medieval (siglos IV- XV), Santiago de Compostela, 12-15 settembre 2002 (*in uscita negli atti del convegno).
[57] Cfr. scheda 5 in «I carmi di Dúngal per Ildoardo e Baldo», cit. n. 1, p. 186.
[58] 1 Cor 1, 27
[59] Eph 6, 9; cfr. Rm 2, 11, Col 3, 25.
[60] *La Sacra Bibbia*, Edizione Ufficiale della C.E.I., Roma, UELCI, 1996, p. 1134.
[61] *La Sacra Bibbia*, Edizione Ufficiale della C.E.I., Roma, UELCI, 1996, p. 1208.

*solis et lunae cursus, qui vere notabiliores et* ***faciliores sunt ad cognoscendum***

e il passo relativo alle costellazioni boreali del IX libro del *De architectura* di Vitruvio (9, IV, 4):

**eorum** ***faciliores sunt*** *capitum vertices* ***ad cognoscendum***[62]

Data la prossimità tematica tra i due scritti – la seconda parte del libro IX del *De architectura* tratta schematicamente dell'universo; dei pianeti, della loro orbite di rivoluzione, della loro velocità di approssimazione e di allontanamento e della loro temperatura; delle costellazioni boreali e australi; della Luna e delle sue fasi; del Sole e del suo percorso attraverso lo Zodiaco nonché della durata variabile delle ore e dei giorni; della sfera celeste e di astronomia e meteorologia[63]- esplorando l'ipotesi di un nesso testuale, si può rilevare che il testimone più antico dell'opera vitruviana, l'Harleiano (Londra, British Museum, MS Harley 2767, saec. IX), è stato associato anche con Eginardo, biografo di Carlo Magno.[64]

## 5 Conclusioni: *Dungalus vester famulus et orator* nella lettera a Carlo Magno

Dal punto di vista letterario possiamo concludere che la lettera a Carlo Magno sulle eclissi di sole dell'810 rappresenta non tanto e non soltanto una perizia scrive (*hujus rationis investigatio et peritia*, § 4) di carattere scientifico su un problema astronomico in senso stretto (*proprie et specialiter*, *ibidem*) quanto, piuttosto, una drammatizzazione del ruolo della cultura e della ricerca 'filologica' delle *auctoritates* (*quid sentirem, et quid scirem, et quid sentirem proferendo et respondendo faterer, exceptum scriberetur, scriptumque vobis deferretur*, § 3, *libri compositiores et diligentiores quamvis mihi non suppetant*, § 4).[65]

A conclusione della *peritia*, Dúngal accompagna la consegna con un'*encomium* del 'suo' *amantissimus pater*:

*omnibus aequaliter omnium bonorum operum et virtutum et honestarum disciplinarum doctor praecipuus, et perfectum habetur exemplar rectoribus ad suos subjectos bene regendos, militibus ad suam exercendam legitime militiam, clericis ad universalis Christianae religionis ritum recte observandum, philosophis et scholasticis ad honeste de humanis philosophandum et sapiendum, reverenterque atque orthodoxe de divinis sentiendum et credendum. Quid plura de nostri domini Augusti Caroli summis virtutibus et excellentibus dicere nitor, cum licet multum elaborare velim totas referre non potero?* (§ 40a)

Tale visione non è però né adulatoria né statica. Carlo stesso, *talis rex et talis princeps*, nel momento in cui chiede l'intervento del suo *famulus et orator* Dúngal, è chiamato a esemplare e a trasmettere in prima persona tale modello articolato e dinamico di virtú di azione e di comando:

---

[62] Citato in *Thesaurus Linguae Latinae* vol. 6 pars prior [lettera *F*], Lipsia, Teubner, 1926, p. 60, righe 28-29: *Vitruve*, *De l'Architecture*, *livre IX*, texte établi, traduit et commenté par Jean Soubiran, Paris, Les Belles Lettres, 1969, p. 21.

[63] Cfr. *Vitruve*, *De l'Architecture*, *livre IX*, cit. n. 60, p. XII e p. XXXIV-LXXI.

[64] Precisamente da H. Degering, v. *Vitruve*, De *l'Architecture*, *livre I*, texte établi, traduit et commenté par Philippe Fleury, Paris, Les Belles Lettres, 1990, p. LVII, n. 144, che rimanda a P. Thielscher, «Vitruvius», *R.E.* A, 1 (1961), coll. 419-481, col. 471. F. Granger, «The Harleian Manuscript of Vitruvius (H) and the Codex Amiatinus», *Journal of Theological Studies* 32 (1931), p. 47-77, cit. *ibidem*, n. 146, opta per una datazione anteriore all'VIII secolo e un'origine inglese, contro cui Fleury cita L.W. Jones, «The Provenience (*sic*!) of the London Vitruvius», *Speculum* 7 (1932), p. 64-70.

[65] Come scrive Mirella FERRARI, «'*In Papia conveniant ad Dungalum*'», cit. n. 2, p. 33: «Per dirla in termini moderni_Dungal fu più filologo che filosofo; più si preoccupò di completare, emendare, accertare i testi che leggeva che di comporre commenti. Tale atteggiamento si rispecchia nella sua opera, ove i brani, ordinatamente stralciati da alcuni autori, si susseguono per pagine e pagine; sarebbe assurdo se non si intendesse come un'antologia laboriosamente raccolta e criticamente corretta»

*prae omnibus affluentiam sapientiae, sicut et caeterarum sanctarum virtutum Deus distribuit, rogo suppliciter ut in quo vobis de hac causa ignorare videar, aut aliter aestimare quam rectum est, instruere et dirigere dignemini* (§ 39)

Dúngal mostra una profonda consapevolezza dell'importanza della conoscenza e dell'ortodossia (*ignorare, aliter aestimare quam rectum est*, che riprende l'avvertimento "*si quid* ***minus aut aliter*** *dixero*" dell'*exordium*), per cui è necessaria la duplice azione correttiva della ricerca 'filologica' personale a lui affidata, da un lato (*Macrobius* ***igitur***, **§** 5), e quella dell'imperatore stesso, dall'altro (*instruere et dirigere dignemini*), *exemplar philosophis et scholasticis ad honeste de humanis* ***philosophandum et sapiendum****, reverenterque atque* ***orthodoxe*** *de divinis* ***sentiendum et credendum***. Gli avverbi *honeste*, *reverenterque atque orthodoxe* sono legati ciascuno nel proprio ambito come i binomi verbali *philosophandum et sapiendum, sentiendum et credendum*, allo stesso modo l'*affluentia sapientiae* di Carlo si accompagna inscindibilmente alle altre *sanctae virtutes* di Carlo, *bonorum operum et* ***virtutum et honestarum disciplinarum*** *doctor praecipuus*.

La destrezza formale di Dúngal, *famulus et orator*, si dimostra nel rendere prontamente e puntualmente conto (*quid sentirem, et quid scirem, et quid sentirem proferendo et respondendo faterer, exceptum scriberetur, scriptumque vobis deferretur*, § 3) dell'efficace e volenteroso svolgimento (*efficax quam volontarius*, § 3) di un compito improbo (*praeter industriam studiumve per fragilitatem infirmitatemque*, con la felice antitesi tra due coppie di nomi e il secondo binomio rimato, § 4): espressione di un impegno esemplare (*quis... sollicitus tantum studium erga astronomiae aut* ***cujuscunque disciplinae assectationem*** *adhibuerit*, § 37) da parte di chi è investito di tale ruolo dall'autorità stessa (*quasi* ***sectator*** *sapientiae*, § 3).

Le continue eco lessicali sulla *peritia* in ogni *disciplina* che risuonano nell'articolazione di questa sagace *oratio-responsio* (***Respondi*** *ergo… ex eorumdem auctoritate quemadmodum antiqui philosophi et scierunt et praescierunt****… omnium disciplinarum peritissimi… sagacissima*** *elimatae et defaecatae mentis intentione*, § 28) rivelano, dunque, come il *sermo* e l'*expressio* di Dúngal additino la lezione di chi è più esperto e più eloquente di lui (*physicos... quorum libri compositiores et diligentiores quamvis mihi non suppetant, quibus de his rebus et de talibus* ***exercitatiori sermone et enucleatiori expressione*** *tractaverunt*, § 4) proprio a colui che è preso a modello di tale ideale, ***honestarum disciplinarum*** *doctor praecipuus*. Oltre alla ripetizione lessicale, l'abile retorica di Dúngal fa leva su figure di suono (allitterazione, omeoteleuto) e di collocazione (iperbato, chiasmo) che animano e strutturano l'assetto dei suoi periodi, oltre che sulla prosa ritmica, come si potrà verificare leggendo il testo con l'ausilio della scheda delle *clausulae* adottate.

La sagacia dell'*orator* (*ne vulgari proverbio lupus in fabula, pavido stupidoque silentio reprimi videar, utcunque respondebo*) si dimostra altresì nella presentazione delle fonti letterarie scientifiche e letterarie (Macrobio e Virgilio), e, specialmente, di quelle bibliche. Abbiamo sottolineato, infatti, le sfumature presenti nelle citazioni paoline sul rapporto tra *stultitia* in questo mondo e *sapientia* cristiana.

Riassumendo, la lettera di Dúngal a Carlo Magno sembra raccogliere diversi generi letterari, *oratio*, *epistula, fabula, encomium*, rappresentando cosí un significativo esercizio retorico da parte del futuro maestro di scuola irlandese a Pavia. La coloritura biblica conferisce però al dialogo tra Dúngal e Carlo Magno e alla dissertazione astronomica contenuta nel corpo della lettera il sapore di una riflessione spirituale sull'impegno dell'intellettuale in rapporto con l'autorità umana e divina. Nel capitolo successivo che segue la già discussa citazione di 1 Cor 28: "Dio ha scelto ciò che nel mondo è stolto" e l'ingiunzione che chiude il capitolo primo "*Chi si vanti si vanti nel Signore*" (v. 30), Paolo esprime sentimenti e concetti che richiamano per allusione o contrasto atteggiamenti e ideali descritti e prospettati da Dúngal a Carlo Magno:

"[1]Anch'io, o fratelli, quando venni tra di voi, non mi presentai ad annunziarvi la testimonianza di Dio **con sublimità di parola o di sapienza** [2]Io ritenni infatti di **non sapere altro** in mezzo a voi **se non Cristo, e questi crocifisso.** [3]Io venni in mezzo a voi **in debolezza** e con **molto timore** e tribolazione, [4]e **la mia parola e il mio messaggio non si basarono su discorsi persuasivi, ma** sulla manifestazione dello **Spirito** e della **<u>sua</u> potenza**, perché la **vostra fede non fosse fondata sulla sapienza umana, ma sulla potenza di Dio"** (1 Cor 2, 1-5)

A conclusione di quest'analisi, infatti, basti citare l'ammissione stessa del *famulus et orator*, che si annovera tra i *reclusos* (§ 39) e che si augura di raggiungere lo scopo desiderato invocando la clemenza dell'imperatore e di Dio per cui ogni sforzo devoto e diligente è ritenuto equivalente a un progetto realizzato compiutamente:

*utinam tam efficax quam voluntarius existerem, ut non solum velle, sed et compote voto assequi cupita valerem*, ***licet apud summum Rectorem pronus et alacris affectus pro re effecta et adimpleta reputatur.***

perché da Lui viene ogni cosa buona e ogni dono perfetto:

*a quo* ***omne datum optimum est et omne donum perfectum*** (§ 28)
*Ad maiora semper, Dungale, et Caroli regis et noster Bobiensis amice!*[66]

[66] A completamento dell'analisi delle opere dungaliane mi riprometto di dedicare uno studio al citato carme commissionato da Ilduino, abate di Saint-Denis (814/15-840) per la nuova cappella costruita per i martiri Dionigi, Rustico ed Eleuterio, e il *titulus* dell'altare d'oro della Basilica di S. Ambrogio in Milano, attribuito da Mirella Ferrari a Dúngal (vedi Mirella FERRARI, «Dungal», cit. n. 4, p. 13).

# APPENDICE 1
# LA PROSA RITMICA DELLA LETTERA A CARLO MAGNO

***Analisi delle clausulae per tipologia di cursus***

(I) *cursus planus* p 3p *íllum dedúxit* (Janson, p. 13)

| | | § |
|---|---|---|
| (1) | *compértum dixístis* | 2 |
| (2) | *parére praecépto* | 3 |
| (3) | *cupíta valérem* | 3 |
| (4) | *réptans et-móvens* | 4 |
| (5) | *explicánda procédens* | 4 |
| (6) | *vidéntur infíxa* | 5 |
| (7) | *paralléli vocántur* | 5 |
| (8) | *deprehendísse vidétur* | 8 |
| (9) | *ésse infíxas* | 11 |
| (10) | *stélla recédens* | 19 |
| (11) | *quám movétur* | 20 |
| (12) | *lócum regréssus* | 20 |
| (13) | *stéllae perlústrant* | 21 |
| (14) | *evénire compróbant* | 27 |
| (15) | *vísum arcénte* | 27 |
| (16) | *míhi vidétur* | 28 |
| (17) | *fínis dicendi* | 38 |
| (18) | *pósse supplére* | 38 |
| (19) | *néque praésumam* | 38 |
| (20) | *discurréntes illústret* | 39 |
| (21) | *magístro gaudére* | 40a |
| (22) | *óre conclámant* | 40a |
| (23) | *Déo donánte* | 40a |
| (24) | *béne regéndos* | 40a |
| (25) | *sanitátem confírmet* | 40b |
| (26) | *abbáti mandástis* | 41 |
| (27) | *vóbis remítto* | 41 |

(II) *cursus tardus* p 4pp *resilíre tentáverit* (Janson, p. 14)

| | | § |
|---|---|---|
| (1) | *míhi non-súppetant* | 4 |
| (2) | *pósse credíderim* | 4 |
| (3) | *interrogátis respónsio* | 4 |
| (4) | *fúlta conváluit* | 8 |
| (5) | *stéllae non fáciunt* | 11 |
| (6) | *sólem pervénerit* | 19 |
| (7) | *prohibeátur accédere* | 19 |
| (8) | *recedéndo contígerit* | 19 |
| (9) | *perácta convérsio* | 20 |
| (10) | *sémper appáreat* | 27 |
| (11) | *laténtis emisphaérii* | 27 |
| (12) | *Orientáles non-séntiunt* | 27 |
| (13) | *cúrsum dirígeret* | 27 |
| (14) | *traditióne cognóvimus* | 27 |
| (15) | *memoráre non quíverit* | 38 |
| (16) | *rógo supplíciter* | 38 |
| (17) | *dónet et-tríbuat* | 40a |
| (18) | *clamémus ad-Dóminum* | 40b |
| (19) | *triúmphos multíplicet* | 40b |
| (20) | *consérvet progéniem* | 40b |
| (21) | *annórum currículos* | 40b |

(III) *cursus velox* pp 4p ***hóminem recepístis*** (Janson, p. 14)

| | | § |
|---|---|---|
| (1) | *glória sine-fíne* | 1 |
| (2) | *véniam donatúram* | 4 |
| (3) | *partículae temperátur* | 24 |
| (4) | *latitúdinem adipísci* | 24 |
| (5) | *líneis mensurátur* | 27 |
| (6) | *fíeri quam vidéri* | 27 |
| (7) | *voluísse compérimus* | 37 |
| (8) | *áliter aestimáre* | 39 |
| (9) | *grátias referátis* | 41 |

(III) *cursus trispondaicus*

I tipo: in parola singola es. *díssimulávero* (Janson, p. 14)
II tipo: p 4p *ágnos admittátis* (Janson, p. 14)
III tipo: pp 3p *brácchio exténto* (Janson, p. 15)

I tipo

| | | § |
|---|---|---|
| (1) | *disputationibus* | 4 |

II tipo p 4p es. *aethéream vocavérunt* (§ 9)

| | | § |
|---|---|---|
| (1) | *legisse memorástis* | 2 |
| (2) | *interrogárer quid-sentírem* | 3 |
| (3) | *excéptum scriberétur* | 3 |
| (4) | *vóbis deferrétur* | 3 |
| (5) | *adimpléta reputátur* | 3 |
| (6) | *expressióne tractavérunt* | 4 |
| (7) | *utcúnque respondébo* | 4 |
| (8) | *moderári perhibétur* | 19 |
| (9) | *consuéta revocátur* | 19 |
| (10) | *demensióne moderátur* | 19 |
| (11) | *lustrári confirmántes* | 27 |
| (12) | *ménse a-prióri* | 27 |
| (13) | *exactiónem litterárum* | 28 |
| (14) | *sciérunt et-praesciérunt* | 28 |
| (15) | *investigatióne quaesivérunt* | 28 |
| (16) | *stellárum intuéntes* | 28 |
| (17) | *cúrsus ignorárent* | 29 |
| (18) | *pártem coeúntes* | 29 |
| (19) | *eclípsin praesciébant* | 30 |
| (20) | *Sólis praesciébant* | 30 |
| (21) | *ánnos sequerétur* | 30 |
| (22) | *expérti praesignábant* | 30 |
| (23) | *arguméntis protendérunt* | 30 |
| (24) | *Sólis evenísse* | 36 |
| (25) | *Decémbre inchoánte* | 36 |
| (26) | *Sólis definítur* | 36 |
| (27) | *pósse perveníre* | 37 |
| (28) | *récte observándum* | 40a |
| (29) | *philosophándum et-sapiéndum* | 40a |
| (30) | *sentiéndum et-credéndum* | 40a |
| (31) | *Fránci dominántur* | 40a |
| (32) | *tális oriátur* | 40b |
| (33) | *exigéndo commonéret* | 41 |
| (34) | *cleménter imponátis* | 41 |

III tipo **pp 3p** es. *speciáliter studéntes* (§ 28)

| | | § |
|---|---|---|
| (1) | *dícitur Satúrni* | 7 |
| (2) | *rátio affírmat* | 11 |
| (3) | *biénnio perágat* | 21 |
| (4) | *ménsibus non-sínit* | 27 |
| (5) | *beatíssime Augúste* | 28 |
| (6) | *ácie praefíxa* | 28 |
| (7) | *errática lustráret* | 28 |
| (8) | *diligéntia inténtus* | 37 |
| (9) | *éxerens perfúndat* | 39 |
| (10) | *impérium dilátet* | 40b |

**Altri tipi di *clausolae***

1 2 *hóc scíre* (Janson, p. 15)

§

| | | |
|---|---|---|
| (1) | *est mírum* | 36 |
| (2) | *sít fáctus* | 36 |

1 3p *ét redémpta* (Janson, p. 15)

§

| | | |
|---|---|---|
| (1) | *his partibus* | 38 |
| (2) | *ét rogémus* | 40b |

1 3pp *quaé fáciat* (Janson, p. 15)

§

| | | |
|---|---|---|
| (1) | *súnt lúceat* | 39 |
| (2) | *nón pótero* | 40a |

1 4p *néc impetrávi* (Janson, p. 14)

§

| | | |
|---|---|---|
| (1) | *ád cognoscéndum* | 29 |
| (2) | *plús mirémini* | 30 |

1 4pp *ét consílio* (Janson, p. 14)

§

| | | |
|---|---|---|
| (1) | *quém reférimus* | 24 |
| (2) | *Déus distríbuit* | 39 |

p 1 *consecútus súm* (Janson, p. 15)

§

| | | |
|---|---|---|
| (1) | *repeténda ést* | 4 |
| (2) | *recépta ést* | 8 |
| (3) | *necésse est* | 27 |
| (4) | *credéndus ést* | 37 |
| (5) | *aequális ést* | 37 |
| (6) | *réctum ést* | 39 |
| (7) | *vísus ést* | 40a |

p 2p *vidétur díctum* (Janson, p. 15)

§

| | | |
|---|---|---|
| (1) | *continéndo díctus* | 5 |
| (2) | *lúna ténet* | 7 |
| (3) | *aliquándo súpra* | 27 |
| (4) | *cérni póssunt* | 27 |
| (5) | *vísum ésse* | 27 |
| (6) | *Respóndi érgo* | 28 |
| (7) | *deféctus Sólis* | 28 |
| (8) | *sígni ésset* | 28 |
| (9) | *deféctum Sólis* | 30 |
| (10) | *íta dícit* | 31 |
| (11) | *átque pándit* | 32 |
| (12) | *diuturnitáte nísus* | 37 |
| (13) | *distá fácit* | 37 |
| (14) | *elaboráre vélim* | 40a |
| (15) | *vólens sólvi* | 41 |

pp 1

§

| | | |
|---|---|---|
| (1) | *definítio ést* | 19 |

pp 2p *víneam nóstram* (Janson, p. 15)

§

| | | | |
|---|---|---|---|
| (1) | *discúrrere dícunt* | 27 | |
| (2) | *indubitatíssime scírent* | | 28 |
| (3) | *fíeret Sólis* | 29 | |
| (4) | *dícere nítor* | 40a | |
| | (5) *amantíssime Pater* | | 41 |

p 3pp *fínem fácio* (Janson, p. 15)

§

| | | | |
|---|---|---|---|
| (1) | | *sphaerárum éfficit* | 21 |
| (2) | | *númerum pértinet* | 24 |
| (3) | | *médius dícitur* | 24 |
| (4) | | *quándo fíeret* | 28 |
| | (5) | *Africáni réferens* | 31 |
| (6) | | *incarnatióne Dómini* | 36 |
| (7) | | *índicant lítterae* | 36 |
| (8) | | *lúna fíeri* | 36 |
| (9) | | *aetáte géniti* | 37 |
| (10) | | *sufficiénter árbitrer* | 38 |
| (11) | | *moderáte éxtitit* | 41 |
| | pp | 3pp | *plúribus áttulit* (Janson, p. 15) |

| | | § |
|---|---|---|
| (1) | *dífferam ígitur* | 3 |
| (2) | *speciáliter pértinet* | 4 |
| (3) | *cóntinent ápices* | 4 |
| (4) | *réprimi vídear* | 4 |
| (5) | *áliter díxero* | 4 |
| (6) | *acúmine praéditus* | 37 |
| (7) | *legítime milítiam* | 40a |
| (8) | *veráciter dícimus* | 40a |

| | | | |
|---|---|---|---|
| | pp | 4pp | *indúbia tutámine* (Janson p. 14) |

| | § |
|---|---|
| *volontárius exísterem* | 3 |
| *ópus expónere* | 6 |
| *éxtima et-ínfima* | 7 |
| *dirígere dignémini* | 39 |

| | | | |
|---|---|---|---|
| | p | 5p | *vestrarum largitione* (Janson, p. 14) |

| | | § |
|---|---|---|
| (1) | *praedícto superaddúntur* | 6 |
| (2) | *subjéctas dispertivérunt* | 6 |
| (3) | *térram occuliári* | 27 |
| (4) | *intuéndo experiéntes* | 28 |
| (5) | *matutínos Occidentáles* | 28 |
| (6) | *rogáre et-postuláre* | 40a |
| (7) | *commonéndo interrogáret* | 41 |

| | | | |
|---|---|---|---|
| | p | 5pp | *nulla confidéntia* (Janson, p. 14) |

| | | § |
|---|---|---|
| (1) | *ubíque conspiciátur* | 36 |
| (2) | *assectationem adhibúerit* | 37 |

| | | | |
|---|---|---|---|
| | pp | 5p | *efficáciter exaudiétis* (Janson, p. 14) |

| | | § |
|---|---|---|
| *lácteus interpretátur* | 5 | |
| *síngulas obtinuérunt* | 6 | |
| *intentíssime observavérunt* | 28 | |

| | |
|---|---|
| *plenìssime exploravérunt* | 28 |
| *studiosíssime cognovérunt* | 29 |
| *áudeam excogitáre* | 38 |

pp 5pp *misériam praelibáverim* (Janson, p. 14)

§

| | | |
|---|---|---|
| (1) | *mémorat in-Geórgicis* | 5 |
| (2) | *pénitus non appáreat* | 36 |

1 5pp *séu iniustítia* (Janson, p. 14)

§

*si apparúerit* 36

## *Parte II Eclissi di Luna e di Sole: effemeridi e frequenze con un metodo grafico di Costantino Sigismondi*[67]

**Abstract**
The frequency of solar and lunar eclipses can be inferred by a simple geometrical model: for lunar eclipses the 'cross section' is given by the amplitude of penumbral circle with the addition of the lunar radius in order to allow partial penumbral eclipses. For solar eclipses, in the approximation of plane Earth with a radius double of that one of the Moon, an eclipse can occur if the Moon at conjunction has an ecliptical latitude not larger than 90', this is the limiting case of partial eclipse visible only at poles. With this model it is possible to predict fairly accurately the type of the eclipses and the zone of visibility, as well as the percentage of partial and total lunar eclipses. The Dungal's case of 810 eclipses series is discussed.

**Sommario:**
La frequenza delle eclissi solari e lunari è derivata a partire da un modello geometrico molto semplice. Per le eclissi di Luna (cfr. fig. 1 lato sinistro) nel cielo c'è il cerchio d'ombra e di penombra della Terra, che la Luna attraversa se la distanza del centro di questi cerchi è sufficientemente vicina al nodo dell'orbita lunare. Il rapporto tra le semi-ampiezze di ombra e penombra diminuite del semi-diametro lunare ed il diametro lunare determinano il rapporto tra numero di eclissi lunari totali, penombrali e parziali.
Nelle eclissi di Sole (cfr. fig. 1 lato destro) viene costruito un simile cerchio fittizio nel cielo, nell'ipotesi di trovarsi all'equatore terrestre con il nodo dell'orbita lunare esattamente sull'equatore celeste. In quel caso l'eclissi è totale sull'equatore. La fascia di parzialità si estende un diametro lunare a Nord e uno a Sud. Se invece la Luna si trova un diametro a Nord dell'equatore l'eclissi sarà totale nella zona temperata a Nord, e la fascia di parzialità si estende dall'equatore al polo Nord, che si trova alla distanza di un altro diametro lunare verso Nord. Con la Luna 62' a Nord della posizione iniziale, l'eclissi sarà totale al polo Nord, mentre altri 31' verso Nord è il limite della parzialità visibile solo dal polo Nord. La situazione verso Sud è simmetrica.
Una retta orizzontale individua l'eclittica nelle due parti della figura 1, mentre una retta inclinata di 5°09' individua l'orbita lunare. La distanza tra il nodo dell'orbita e la posizione del Sole (o dell'anti-sole per l'eclissi di Luna) determina il tipo di eclissi. Si esamina in dettaglio il caso dell'anno 810, proposto da Dúngal in una lettera a Carlo Magno.

*********************************************************************************

**Eclissi di Sole a medie latitudini : probabilità di ripetersi.**
La lettera di Dúngal a Carlo Magno apre naturalmente un problema statistico legato con la geometria delle eclissi di Sole e di Luna: quali di queste siano più probabili. Infatti Dúngal tratta il problema di due eclissi solari nello stesso anno.
Ismael Bullialdus (1605-1694, astronomo che meritò la dedica di un cratere lunare nella carta di Giovanni Battista Riccioli, 1651) commenta il testo di Dúngal con le effemeridi da lui calcolate, che mostrano come una delle due eclissi non sia stata visibile da Aquisgrana, sede della corte di Carlo Magno, poiché accadde nell'emisfero meridionale. Bullialdo afferma con chiarezza che da un posto lontano dall'equatore non si possono vedere due eclissi di Sole nello spazio di in un solo anno.
Entriamo dunque nel dettaglio del problema delle eclissi, per capire meglio i termini del problema, sia tecnici che storici.
**Eclissi: storia e scienza.**

[67] **Docente incaricato di Storia dell'Astronomia all'Università di Roma "La Sapienza" e-mail sigismondi@icra.it**

Si noti come le eclissi solari e quelle lunari abbiano rivestito un certo ruolo nella storia dell'Astronomia e anche dell'affermazione delle teorie copernicane. La previsione di un'eclissi lunare servì a Colombo il 1 marzo 1504 come spauracchio nei confronti dei nativi delle isole centro americane, ma altresì come metodo di misura della longitudine, a partire dall'istante del mezzogiorno locale. Il calcolo dell'eclissi di Sole del 1629 a Pechino fatto dai padri Gesuiti era più accurato di quello fatto dai cinesi, grazie alle effemeridi pubblicate da Keplero nel 1627 (le *Tabulae Rudolphinae*) basate sui dati di Tycho. In tempi più recenti la teoria delle eclissi solari ha permesso di monitorare il tasso di variazione della rotazione terrestre, che è variabile lungo i secoli, così come si è utilizzata la forma della Luna come profilo utile alla misura del diametro solare, al fine di identificarne eventuali variazioni secolari.

**Eclissi: teoria e calcoli**

Per trattare compiutamente il problema delle eclissi occorre lavorare con gli elementi Besseliani[68] in cui si considera il cono d'ombra lunare o terrestre proiettato sul piano perpendicolare ad esso passante per il centro della Terra. Tuttavia è possibile fare dei calcoli semplificati per verificare con modelli geometrici meno complessi la loro statistica.

Le eclissi di Sole sono più frequenti di quelle di Luna[69] Durante un Saros (18 anni 11 giorni 1/3, oppure 10 giorni a seconda del numero di anni bisestili intercalari) avvengono in media 84 eclissi, metà (circa) di Sole e metà di Luna[70] Poiché le modalità dell'eclissi di Luna sono indipendenti dalla posizione dell'osservatore (contrariamente con quanto avviene per le eclissi solari) le eclissi di Luna sono più frequenti di quelle di Sole che sono in genere piuttosto rare per uno stesso luogo[71] .

Queste tre affermazioni sembrano in contrasto tra loro. Vediamo quantitativamente perché non lo sono. Giannone si riferisce alla visibilità da uno stesso luogo: il problema sollevato dall'epistola di Dúngal . Un eclissi di Sole non è visibile contemporaneamente da tutto un emisfero, ma solo da una fascia limitata, mentre l'eclissi di Luna (*defectus Lunae*) è visibile da ovunque sia visibile la Luna.

L'eclissi di Sole è tecnicamente un'occultazione: il Sole non perde il suo splendore, ma viene occultato dalla Luna, mentre la Luna lo perde entrando nell'ombra della Terra, poiché essa brilla di luce riflessa. Perciò si parlava propriamente di *defectus Lunae*, come nella targa a ricordo[72] della visita di Copernico a Roma nel 1500, quando ebbe l'occasione di osservare un tale fenomeno. Meno corretto è l'uso dello stesso termine per il Sole. Romano invece afferma che le eclissi di Sole sono più frequenti di quelle di Luna rispetto a tutta la Terra, non ad un punto particolare, ma poi dice che in 19 anni ne avvengono quasi un ugual numero. Le due affermazioni non sono in contraddizione se quel quasi corrisponde ad uno sbilanciamento verso un maggior numero di eclissi di Sole, oppure se vengono incluse nel computo anche le eclissi lunari di penombra, che sono gli equivalenti lunari delle eclissi solari parziali, ma che da Terra sono meno evidenti di quelle lunari totali o parziali. La zona di penombra è un concetto ben preciso in astronomia e corrisponde con la regione da cui il Sole si vede parzialmente occultato o dalla Luna (eclissi parziale di Sole) o dalla Terra (eclissi penombrale di Luna). Nella fase parziale di un'eclissi di Luna, la parte in luce della Luna vede il Sole parzialmente occultato dalla Terra, ed è dunque in penombra, non in luce piena (cfr. figura 2).

---

[68] C. Barbieri, Lezioni di Astronomia, Zanichelli, Bologna (1999) p. 231-233

[69] G. Romano, *Introduzione all'Astronomia*, Muzzio (Padova) 1993. p. 79

[70] G. Romano, op. cit., 85.

[71] P. Giannone, Elementi di Astronomia, Sistema, Roma (1978), p. 107.

[72] Situata nella Facoltà di Giurisprudenza dell'Università di Roma "La Sapienza", nell'ambiente antistante l'aula Falcone-Borsellino, al primo piano. Fu realizzata nel 1873, in occasione del 400 anniversario della nascita di Copernico. *Nicolao Copernico Quod astrorum circuitus legesque dum divinitus meditatur veteres dissipaturus errores mathematicam in hoc archigymnasio tradiderit caelique rationem deficiente luna A. MD speculatus romanae sapientiae deecus maximum astronomorum maximus pepererit quadringentesimo natali die doctores et alumni honoris deferunt monimentumXI kal. Mart. MDCCCLXXIII.*

Nella figura 2, con angoli esagerati rispetto alla realtà, è pure indicata la fase di Luna piena normale, dove si vuole evidenziare come la Luna piena non sia esattamente a 180° dal Sole e quindi non possa essere pienamente illuminata da esso. La più perfetta Luna piena è proprio durante l'eclissi, ma la Terra si frappone tra il Sole e la Luna e questa non può essere illuminata in pieno.
La fase di totalità è rappresentata col colore rosso, tipico, mentre quelle di parzialità con un bordo rosso (dove la luce del Sole non arriva più) ed il resto rosaceo (dove il Sole è gia parzialmente occultato dalla Terra), la fase penombrale è giallastra per indicare che lì il Sole è quasi tutto visibile, mentre fuori dalla zona di penombra la luce del Sole arriva senza ostacoli.
Quantitativamente il cono d'ombra terrestre ha un raggio di 42'(circa 10/14 del raggio terrestre a 60 raggi terrestri di distanza), la Luna un raggio di circa 15', mentre la penombra –alla distanza media Terra-Luna- si estende per un raggio di circa 15' in più rispetto all'ombra. Infatti il Sole appare come un disco di raggio circa 15', di conseguenza anche la penombra si estende per lo stesso spazio.
Allora per avere un'eclissi totale di Luna il centro della Luna deve cadere entro 42'-15'= 27' dal centro del cono d'ombra della Terra. Quindi il nodo dell'orbita lunare (dove viene intersecata l'eclittica) e la sizigie (essere in opposizione al Sole) non devono distare più di 27'/tan(5° 09')=5°.
Questa condizione diventa 7.8° per avere un'eclissi lunare parziale e 10.5° per averne una penombrale (completa). I calcoli sono riferiti alle distanze medie Terra-Luna e Terra-Sole, quindi ai diametri angolari medi di Luna e Sole: queste cifre variano in un ambito compreso tra 9.5° e 12.1° rispettivamente con una penombra estesa tra 75'/2+15' (9.7°) e 90'/2+15' (11.1°)[73] .
Può essere d'aiuto anche la figura relativa all'eclissi del 28 ottobre 2004 (fig. 4), in cui si vede bene il rapporto tra ombra, penombra e diametro lunare.

**Misura diretta del diametro dell'ombra terrestre**
E' possibile ricavarlo anche da una sola foto dell'eclissi. Su una foto che ho ripreso attorno alle 2UT del 28 ottobre 2004 (figura 5), corrispondenti alle 22 EDT della figura 4, ho sovrapposto i cerchi in modo da misurare poi i rapporti dei diametri.
Il rapporto tra i diametri risulta circa 1:2.28. Con 1831 secondi d'arco di diametro per la Luna, ciò corrisponde a 70' di diametro per l'ombra terrestre, un valore consistente col minimo diametro di 75' per l'ombra della Terra. Con foto a maggior risoluzione si possono fare misure più accurate.
**Calcolo delle dimensioni dell'ombra della Terra**
Poiché il diametro del Sole è circa 1/100 della sua distanza dalla Terra, anche la forma dell'ombra della Terra sarà 100 volte il suo diametro. Se la Luna ha una distanza media di 60 raggi terrestri, cioè 30 diametri, la sezione dell'ombra della Terra alla distanza Terra-Luna sarà il 70% del diametro terrestre, cioè 2.57 volte il diametro lunare. ABE e CDE sono triangoli simili e CE<<CA.
**Eclissi di Sole**
Tornando alla parte destra della figura 1, perché si abbia un'eclissi di Sole sulla Terra, anche parziale, è necessario che almeno parte della figura della Luna si sovrapponga al Sole. Se immaginiamo di avere un'eclissi totale all'equatore, con Sole di 16' di raggio e la Luna di circa 15.5' di raggio, ambedue allo zenith, si ha che spostandosi verso Nord di metà del raggio terrestre (valore prossimo al diametro lunare) la Luna sembrerà abbassarsi rispetto al Sole di 31'. Analogamente andando verso Sud della stessa quantità la Luna si alzerà di 31'. Dunque la condizione di totalità per l'equatore corrisponde ad una fascia di parzialità estesa 31' sia a Nord che a Sud. Inoltre se il centro della Luna passasse 31' a Nord dell'equatore celeste la fascia di totalità sarebbe spostata di mezzo raggio terrestre verso Nord, e lambirebbe il polo Nord. Situazione simmetrica per il Sud. Alzandosi di altri 31', la fascia di totalità sarebbe centrata sul polo Nord, mentre con altri 31' sarebbe la fase parziale a lambire il polo Nord che sarebbe l'unico luogo da cui l'eclissi sarebbe visibile.

[73] Barbieri, op. cit., 229.

Dunque si ha un'eclissi (parziale o totale almeno su un punto della Terra) se il centro della Luna dista al più 90' dal centro del Sole, in caso di Sole all'equatore celeste. Questa condizione si modifica leggermente quando il Sole è ai tropici, perpendicolare quindi a regioni già 23.5° a Nord o a Sud dell'equatore, ma la differenza rispetto al caso trattato è abbastanza piccola.
Applicando gli stessi ragionamenti geometrici nel caso dell'eclisse di Luna abbiamo che la longitudine eclittica del Sole non deve differire dal nodo per più di 90'/tan(5° 09')= 16.6°.
I valori estremi sono compresi tra 15.4° e 18.5°[74], se si tiene conto di tutti i valori precisi degli angoli in gioco.

**Periodi delle eclissi**

Per aversi un'eclissi è necessario che la Luna attraversi l'eclittica, l'orbita del Sole nel cielo che si chiama così proprio perché su quel percorso possono avvenire le eclissi. Dunque la Luna deve trovarsi ad attraversare i nodi della sua orbita, e questo accade 2 volte al mese. Ma contemporaneamente deve essere in sizigie, cioè a 180° (opposizione) dal Sole oppure a 0° (congiunzione). Abbiamo visto che quando il Sole si trova entro ±17° circa dal nodo lunare e si verifica un novilunio, c'è un'eclissi solare. Mentre se il Sole si trova entro ±10.5° dal nodo lunare e si verifica un plenilunio c'è un'eclissi di Luna.
Poiché in un mese il Sole percorre circa 30°, se al principio del mese si trova a -17° dal nodo può aversi un'eclissi solare parziale, poi dopo 15 giorni si troverà a 2° dal nodo e si verifica un'eclissi di Luna, e dopo altri 15 giorni sarà a +17° per un'altra eclissi parziale di Sole visibile dalla parte opposta del globo. Dunque in quella stagione delle eclissi si saranno avute 3 eclissi, 2 di Sole ed una di Luna. Dopo sei mesi lunari (177 giorni) il Sole avrà percorso altri 177°, rispetto al novilunio si troverà a -20° (177°-17°=160°) dal nodo (180°, il quale però nel frattempo si è spostato retrogradando e viene raggiunto dopo 173 giorni, 4 giorni prima del novilunio, e si trova a 170°) pertanto ci sarà una eclissi poiché il Sole dista -10° dal nodo –questa eclissi sarà nuovamente visibile nella parte opposta del globo rispetto alla prima eclissi dell'anno, poiché il nodo in questione è quello opposto alla stagione delle eclissi precedente-; rispetto al plenilunio sarà a +5° dal nodo (eclissi di Luna), mentre rispetto al successivo novilunio sarà a +20° (non c'è possibilità di eclissi di Sole). Al compimento di altri 6 mesi lunari dalla prima configurazione di novilunio abbiamo (177°+20°-10°=187°)+7° rispetto al nodo (180°) che produce ancora un'eclissi di Sole –sempre dalla parte opposta del globo rispetto alla prima eclissi della prima stagione-. Totale 4 eclissi di Sole di cui e 2 di Luna in un anno solare occorse in tre stagioni delle eclissi . L'alternanza fa sì che se la prima eclissi era visibile nell'emisfero Sud, le altre sono al Nord, al Nord e a medie latitudini Nord.

| Eclissi di Sole/Stagione | Regione di visibilità |
|---|---|
| 1/I | Sud-estremo |
| 2/I | Nord-estremo |
| 1/II | Nord-medie latitudini |
| 1/III | Nord-medie latitudini |

Abbiamo visto quindi anche la funzione dell'anno draconitico di 346 giorni, detto anche anno delle eclissi, che è il periodo in capo al quale la Luna ha attraversato per 12 volte lo stesso nodo dell'orbita. In un anno le eclissi si presentano raccolte in due o tre gruppi a distanza di poco meno di 6 mesi l'uno dall'altro; le modalità sono tuttavia di volta in volta diverse. Nell'esempio precedente abbiamo visto un anno solare con 6 eclissi: 4 eclissi di Sole e 2 di Luna, come nel caso dell'810. Delle 5 eclissi solari interessavano tutte le regioni circumpolari, data la distanza dai nodi prossima ai valori estremi. Quindi un anno con 5 eclissi solari è facile che siano eclissi visibili dalle regioni polari. Il 1935 fu un anno con 7 eclissi, e le cinque solari (5/1 (Antartide) e 3/2 (Groenlandia); 30/6 (Polo Nord) e 30/7 (Antartide) e 25/12 (Antartide) erano tutte visibili dai poli, in modo alternato per ognuna delle stagioni delle eclissi, il 19/1 e il 16/7 si ebbero due eclissi totali di Luna.

[74] Barbieri, op. cit., p.229.

Le combinazioni in un anno raggiungono un minimo di 2 eclissi tutte solari, oppure 3+4 o 4+3 solari+lunari.

**Frequenza delle eclissi lunari e solari**

La maggiore frequenza delle eclissi solari rispetto alle lunari si ricava dalla maggior "sezione d'urto" dell'eclissi di Sole rispetto a quella di Luna (90' di raggio per avere un'eclissi solare almeno parziale contro 89.5' di raggio per il verificarsi di una parzialità di un'eclissi lunare penombrale, come risulta dagli schemi di figura 1). La frequenza risulta circa 1:1 (89.5':90'), come il rapporto tra i raggi dei cerchi massimi disegnati nella prima figura; la Luna infatti è vincolata a muoversi su una linea inclinata di 5°09' rispetto all'eclittica e l'intersezione con il diametro verticale dei due cerchi massimi in figura 1 determina il tipo di eclissi.

C. Barbieri afferma che se ne possono avere *al massimo* 7 in un anno giuliano (5 solari + 2 lunari o 4 solari e 3 lunari). In un Saros (18 anni 10.3 giorni o 11.3) se ne hanno 84 metà solari e metà circa lunari come afferma G. Romano, dato che conferma il rapporto 1:1, se includiamo anche le eclissi lunari penombrali. Un altro testo preso dal web riporta "Durante un ciclo di Saros avvengono circa 70 eclissi, di cui in genere 29 lunari e 41 solari; di queste ultime solitamente 10 sono totali e 31 parziali [questa affermazione risulta errata da quanto segue]. Ogni anno si verificano in media quattro eclissi, con un minimo di due e un massimo sette. Alla fine del XX secolo sono avvenute, nei cento anni, 375 eclissi: 228 solari e 147 lunari." Il sito web dell'osservatorio di Bologna riporta invece "Durante un saros avvengono in media 43 eclissi solari (di cui 12 totali e 16 anulari) e 43 lunari (di cui 13 totali e 15 parziali)". E ancora risulta che dal 1991 al 2000 abbiamo avuto 22 eclissi di Sole e 14 lunari[75]. Questi testi non sono in contraddizione tra loro, poiché eccetto Romano, tutti escludono dal computo le eclissi lunari penombrali.

Con il nostro modello basta considerare che si ha un'eclissi totale lunare se il bordo della Luna cade tutto dentro l'ombra, dunque il centro sia entro 27' a Nord o a Sud di latitudine eclitticale. I successivi 31' sono per le circostanze di un'eclissi parziale, ed i rimanenti 31'.5 per una parziale di penombra, su un totale di 89.5' di "sezione d'urto lineare" ciò dà luogo alle percentuali di 30.2% per le totali; 34.6% per le parziali e 35.2% per le penombrali, e su 43 eclissi per Saros si ottengono proprio le cifre su indicate di 13 totali e 15 parziali. Analogamente su 44 eclissi (1991-2000) ne aspetteremmo 28 lunari escluse le penombrali, ma il periodo considerato è inferiore al Saros. Su un secolo intero, il XX, il numero di eclissi lunari (escluse le penombrali) è proprio 147 cioè il 34.6%+30.2% di quelle solari (228).

Per quanto concerne l'affermazione sulla probabilità relativa di eclissi di Sole totali rispetto alle parziali possiamo dire quanto segue. Consideriamo innanzitutto nelle stessa classe delle eclissi totali anche le anulari, trascurando il problema dell'intersezione del cono d'ombra lunare con la superficie terrestre.

Un'eclissi di Sole è solo parziale quando la totalità, anche sotto forma di anularità, non è visibile da nessuna parte della Terra. Il nostro cerchio in figura 1 è per 1/3 occupato dalla fascia di parzialità, che interessa solo le regioni terrestri comprese tra le medie latitudini ed i poli. Era un'eclissi di questo genere quella del 25 dicembre 2000, visibile dal Nord America. Dunque le eclissi parziali sono 1/3 ovvero il 33.3% del totale.

Dal sito del catalogo delle eclissi solari della NASA troviamo che la frequenza su 6000 anni (dal 2000 a.C. al 4000 d.C.) risulta il 35.3% di eclissi parziali, ed il 64.3% tra totali (26.6%) anulari (32.9%) ed ibride (5.2%), in ottimo accordo con il modello geometrico qui sviluppato. Su 41 eclissi solari ne aspettiamo quindi 14 parziali ed il resto, 27, tra totali (11) anulari (14) ed ibride (2); concludiamo pertanto che sul sito precedentemente citato sotto la voce eclissi parziali sono state erroneamente conteggiate anche le anulari e le ibride (anulari-totali) che sono centrali.

**Eclissi di Sole a medie latitudini**

---

[75] Ferreri Pellequer, Piccola Guida del Cielo Piemme (Casale Monferrato) 1991, p.104-105.

Da un punto di vista probabilistico non c'è una latitudine terrestre privilegiata, od una fascia, sulla quale avvengano un numero maggiore di eclissi rispetto ad un'altra. Se consideriamo il polo Nord abbiamo che il centro della Luna ha una tolleranza di 62' attorno alla posizione che genera un'eclissi totale per poter produrre un'eclissi parziale in quel luogo, ma questo vale per qualunque altro punto della Terra. Quindi nel corso delle stagioni delle eclissi che capitano in un anno abbiamo l'equiprobabilità che capiti in qualunque punto della Terra. E' la seconda eclissi, o le successive, che sono vincolate da dove capita la prima. Se, ad esempio, la prima eclissi è capitata con il Sole a -17° dal nodo ed interessava regioni meridionali antartiche, quella dopo capiterà a +13°, e riguarderà quelle settentrionali e così via. Il commento di Ismael Bullialdus a Dúngal si basa proprio su questa alternanza. Per aversi due eclissi di Sole nella stessa stagione delle eclissi, infatti, queste devono essere necessariamente con il Sole abbastanza distante dal nodo, sia a destra che a sinistra. Quindi è escluso che nella stessa stagione delle eclissi se ne siano viste due consecutive dallo stesso luogo. Un'altra possibilità potrebbe essere quella di due eclissi in stagioni successive. Infatti, se la prima eclissi è capitata proprio con il Sole sul nodo dell'orbita lunare, dopo 6 mesi lunari -177 giorni- il Sole sarà 177° gradi dopo, con il nodo che è retrogradato di quasi 10° da 180° a 170°, a +7°. C'è dunque una differenza di 7° che in termini di latitudine eclitticale corrisponde a 38' di spostamento verso Nord. E' possibile perciò aver osservato un'eclissi parziale con la Luna che copriva la parte Sud del Sole e dopo 6 mesi assisterne ad un'altra con la Luna che copre la parte Nord del Sole.
La congiunzione avviene in media dopo 177.18 giorni, quindi anche l'orario potrebbe permettere la visibilità dell'eclissi nella stessa zona. Tuttavia questo non fu il caso dell'810, in cui la sequenza delle eclissi fu 9/1 (totale dall'Antartide) 5/6 (anulare nel Sud Pacifico) 20/6 totale di Luna (con la Luna al centro dell'ombra ed il Sole al nodo dell'orbita) 5/7 (anulare al polo Nord) 30/11 (quasi totale ad Aquisgrana) e 14/12 totale di Luna.
Dalla sequenza si può verificare l'alternanza Nord-Sud nell'ambito della stessa stagione delle eclissi (caso della coppia di eclissi del 5/6 e del 5/7) e la possibilità di ripetizione nella stessa zona da una stagione all'altra; è il caso delle eclissi del 9/1 e 5/6/810 in cui dal luogo di latitudine 63° S e 135 W furono visibili due eclissi una totale ed una anulare. Nel primo caso quel luogo era al limite Nord della fascia di totalità, nel secondo caso al limite Sud. Un salto di due stagioni rimanendo nello stesso anno permetterebbe ancora l'accadere di un'eclissi visibile dalla stessa regione. Le mappe di visibilità delle varie eclissi dell'anno 810 sono state realizzate adoperando il programma Occult v. 3.1.0 di D. Herald (2004) e sono nelle figure 7-12..

| | |
|---|---|
| | |
| | |

In questo caso le 4 eclissi di Sole sono raggruppate diversamente rispetto al caso teorico trattato nel paragrafo dei periodi delle eclissi.

| Eclissi di Sole/Stagione | Regione di visibilità |
|---|---|
| 1/I | Sud-estremo |
| 1/II | Sud-estremo |
| 2/II | Nord-estremo |
| 1/III | Nord-medie latitudini |

**Uso della figura 1 con i cerchi concentrici per calcolare le regioni di visibilità delle eclissi**
Quando la Luna attraversa il nodo ascendente dell'orbita, descrive un angolo di 5° 09' con l'eclittica e nelle vicinanze del nodo percorre una linea praticamente retta, si può usare quindi la geometria piana. Quando è al nodo discendente la linea è percorsa in senso discendente. Se il Sole si trova a -10° gradi dalla longitudine del nodo è alla sua sinistra. Le sizigie (syn+zygon stesso giogo, legati assieme, Sole-Luna a 0° o 180°) avvengono quando i centri sono allineati perpendicolarmente all'orizzontale (stessa longitudine eclitticale). La distanza tra il centro del Sole e quello della figura (la latitudine eclitticale al momento della sizigie) determina la latitudine della Terra dove l'eclissi è centrale. La totalità o anularità dipende se la Luna in quel momento è vicino al perigeo (ed è quindi angolarmente più grande del Sole) o all'apogeo (dove non è mai più grande del Sole).
Ad esempio calcoliamo le circostanze dell'eclissi del 5 ottobre 2005, sapendo quelle dell'eclissi di Luna del 28/10/2004. La Luna è passata al nodo alle 21:30 UT del 27 ottobre, 5 ½ ore prima della congiunzione col Sole. Poiché la Luna in un'ora percorre il suo diametro, 31', in 5 ½ ore avrà percorso 2.84°. Il Sole, al momento dell'eclissi ha dunque superato di 2.84 giorni il nodo. La latitudine eclitticale della Luna al momento della totalità era 16' Nord. Il Sole tornerà al nodo (ascendente) dell'orbita lunare il 6 ottobre 2005 alle 21:45 UT (6.9), dopo un anno draconitico di 346.62 giorni. La congiunzione al novilunio occorrerà dopo 12 mesi lunari di 29.53 giorni l'uno, a partire dal novilunio di ottobre 2004 (il periodo medio di 29.53 giorni è tra uguali fasi della Luna) 354.36 giorni dopo il novilunio del 14/10/2004 alle 2:50UT (14.12), il 3 ottobre 2005 alle 11 e 28 UT (3.47). Il Sole, in quella occasione, si troverà 3.43° gradi prima del nodo. La traiettoria della Luna interseca l'orizzontale 3.43° gradi a Est di dove avviene l'eclissi. La distanza minima dal centro della figura risulta 18.5' a Nord rispetto al punto dove il Sole sarebbe perpendicolare quel giorno come da figura 13, cioè 4° a Sud dell'equatore. Infatti18.5' corrisponde ad 1.2 il diametro della Luna (che è in media 31' ovvero 3476 km), ovvero 2079 km a Nord della latitudine 4° S, che sono già 445 km a Sud dell'equatore, si arriva così a circa 14.4° Nord, corrispondenti a 1634 km a Nord dell'equatore, calcolati sul piano che passa per il centro della Terra. Per fare quest'ultimo calcolo sarà utile confrontare lo schema in figura 14, per il calcolo della latitudine del punto subsolare.

L'eclissi comincia più a Nord per terminare più a Sud di questa latitudine, e siccome la Terra ruota da Ovest verso Est la traiettoria dell'ombra sarà data dalla combinazione del moto dell'ombra della Luna su un piano fisso, quindi da Nord –Ovest a Sud-Est, con la Terra in rotazione che tende a seguire l'ombra con il risultato di allungare la durata dell'eclissi. Poiché la Luna si trova vicino al nodo discendente dell'orbita, a sua volta nei pressi del nodo discendente dell'orbita solare (nodo dell'eclittica, dove con un angolo di 23.5° l'eclittica passa da Nord a Sud dell'equatore celeste) la traiettoria dell'ombra della Luna risulterà inclinata rispetto ai paralleli terrestri di 28.5° da Nord-Ovest a Sud-Est. Il punto subsolare è quello che si trova più vicino alla Luna ed al Sole dal quale si sperimenta l'eclissi di maggiore magnitudine. La mattina del 3 ottobre 2005 vedremo un'eclissi che sarà anulare in Spagna, e parziale da noi, tempo permettendo, e questa poi scenderà verso l'Africa, come mostra anche il diagramma calcolato con il programma di effemeridi (figura 15). L'eclissi sarà anulare poiché in quel momento la Luna sarà presso l'apogeo con un diametro angolare di 1815 secondi d'arco, o 30.25', inferiore ai quasi 32' del Sole (corrispondenti a 1918 secondi d'arco).

**Referenze:**

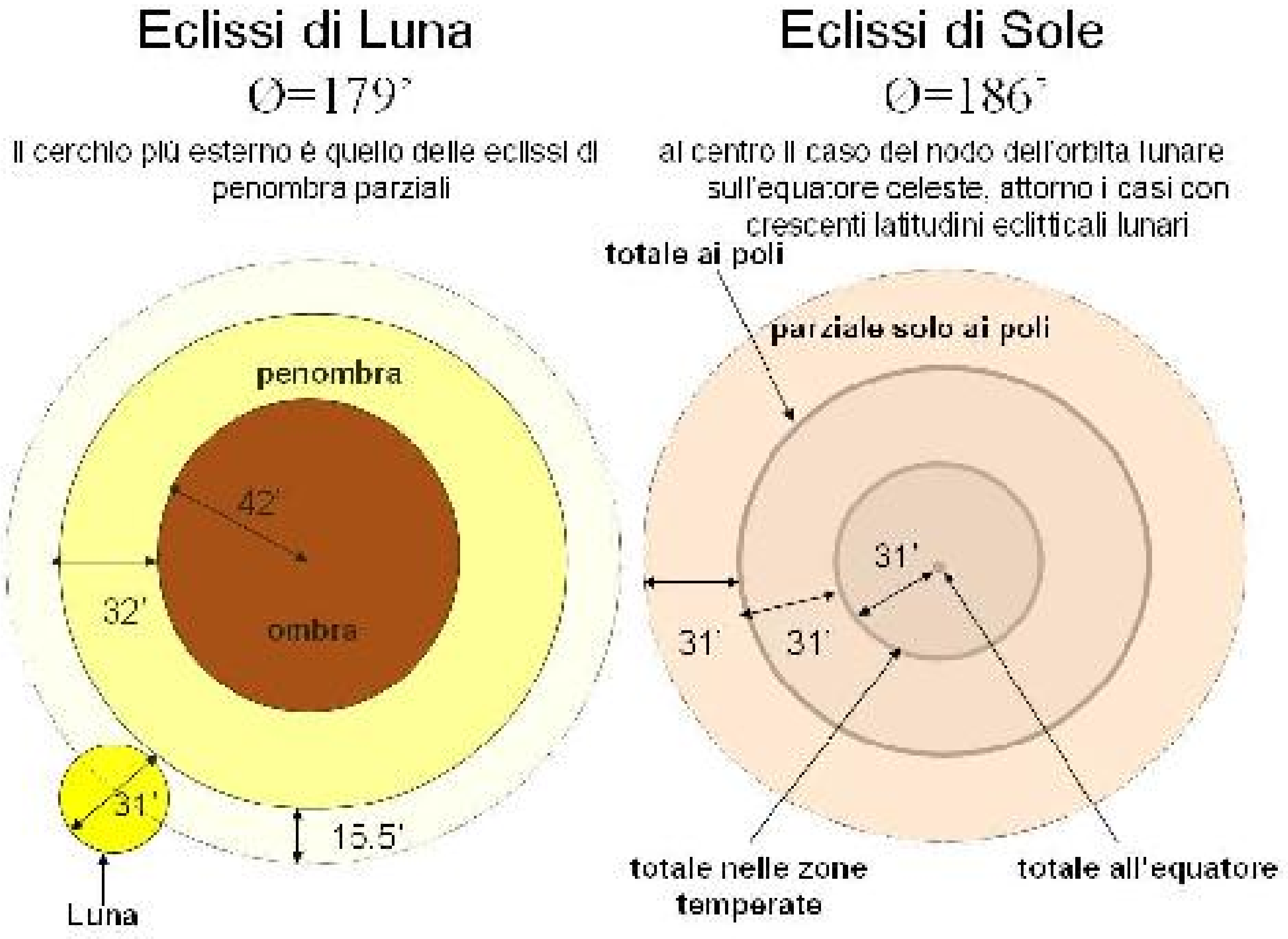

Figura 1. Lato sinistro: schema grafico per un'eclissi lunare. Lato destro: schema grafico per le eclissi di Sole. Il diametro angolare di queste figure va da 179 minuti d'arco per le eclissi di Luna a 186 per le eclissi di Sole, che sono numericamente più probabili, tenendo conto anche delle eclissi lunari di penombra.

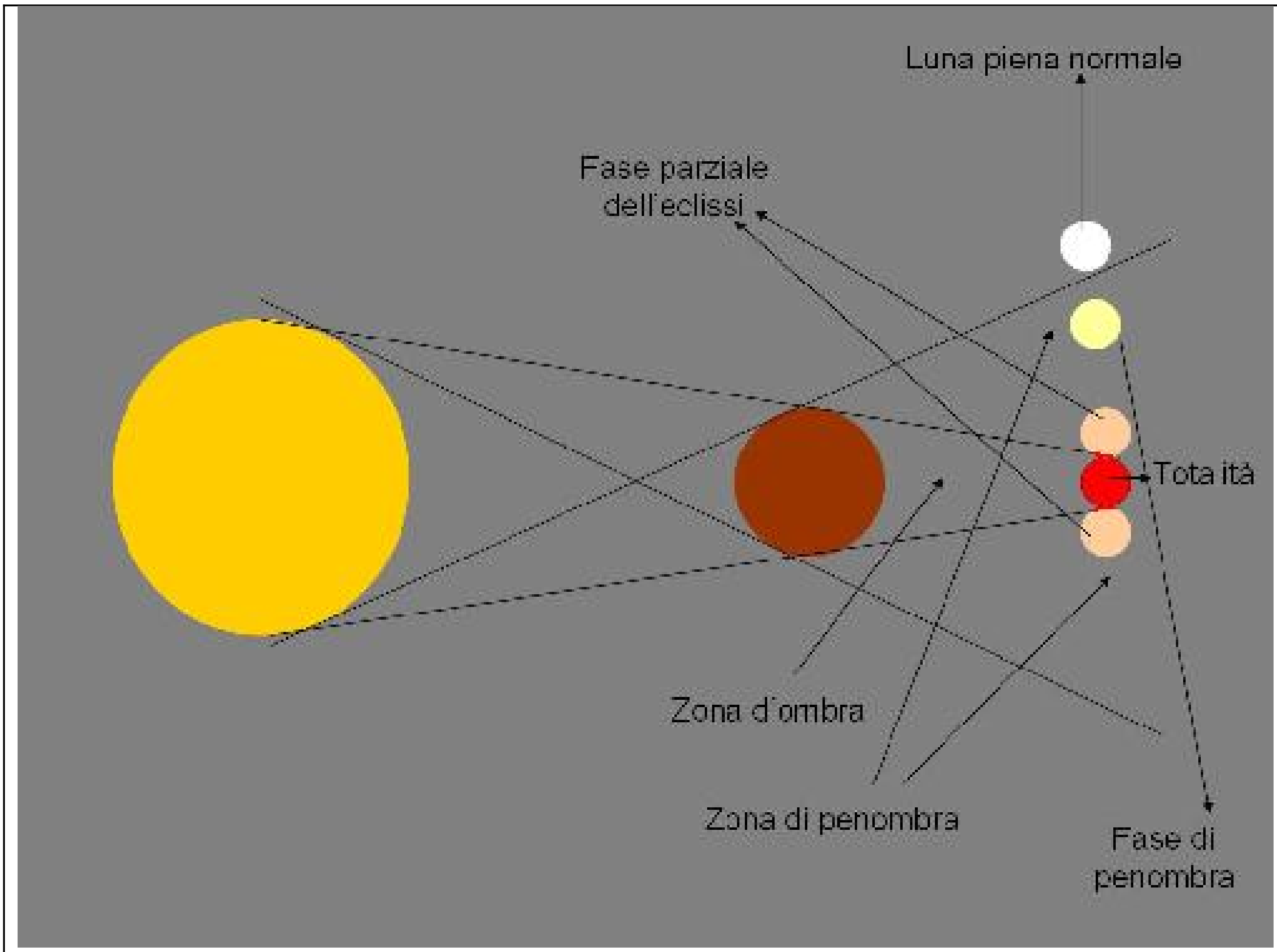

Figura 2. Schema per distinguere le varie fasi di un'eclissi di Luna totale. Quando l'eclissi è solo parziale o di penombra mancano le fasi centrali perché la traiettoria della Luna passa lontano dal centro del cerchio di figura 1, parte sinistra. Si noti anche la fase di Luna piena normale, ad un angolo esageratamente diverso da 180° rispetto al Sole. A rigori la vera Luna piena, in opposizione perfetta al Sole, finisce nel cono d'ombra della Terra.

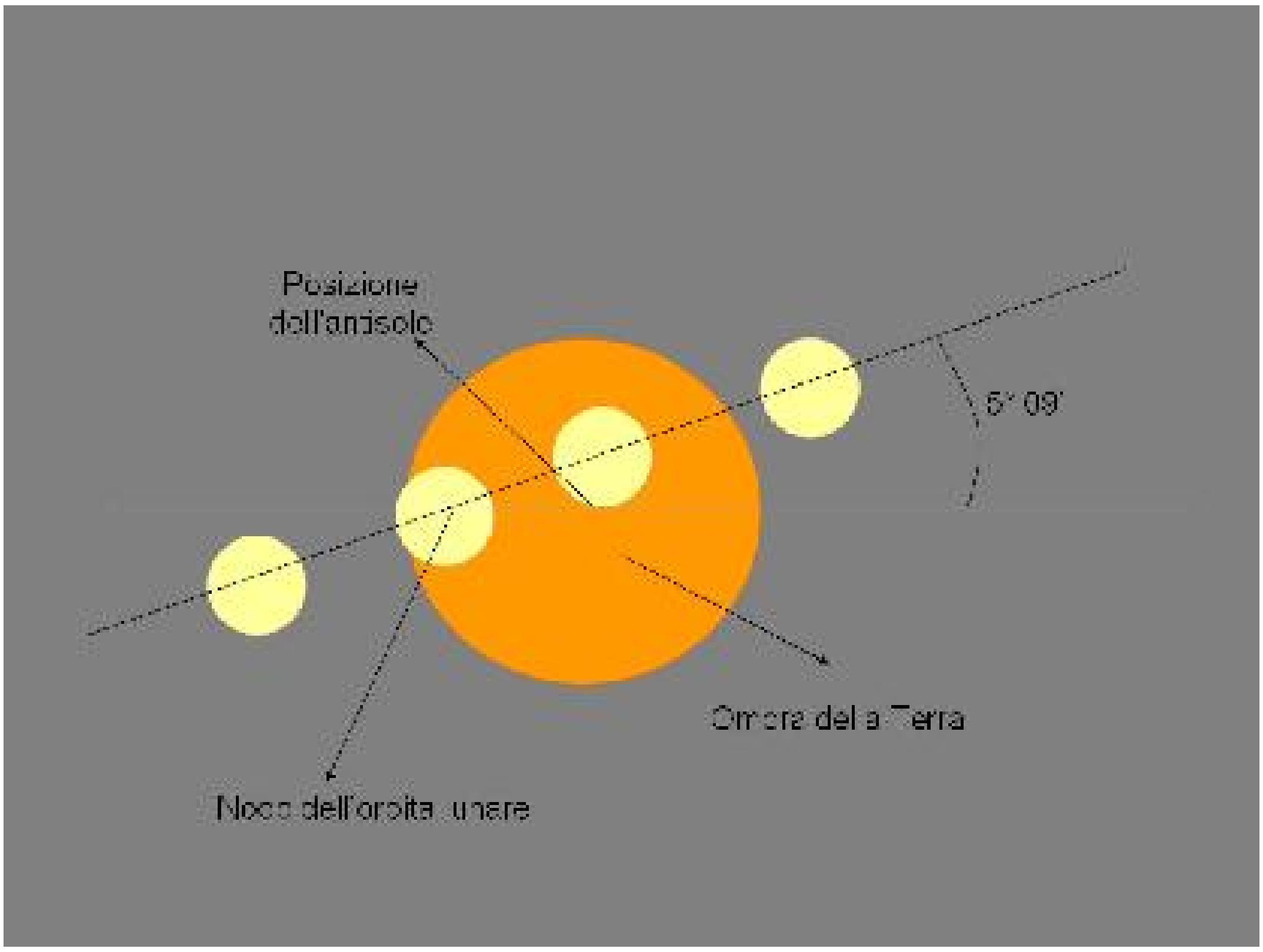

Figura 3. Eclissi di Luna e nodo dell'orbita lunare. La massima durata si ha quando il nodo dell'orbita coincide con la posizione dell'antisole.

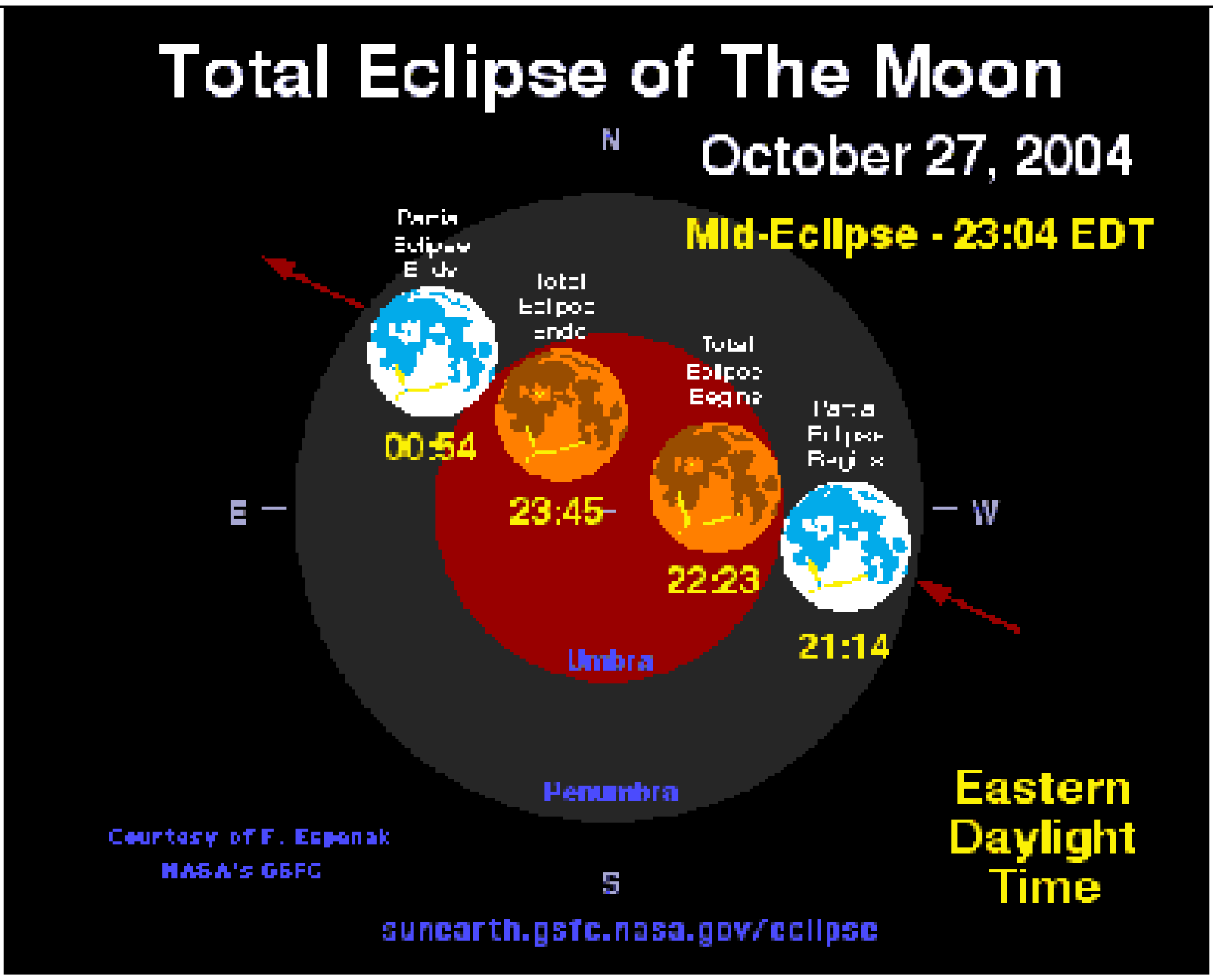

Figura 4. Schema dell’eclissi totale di Luna del 28 ottobre 2004.

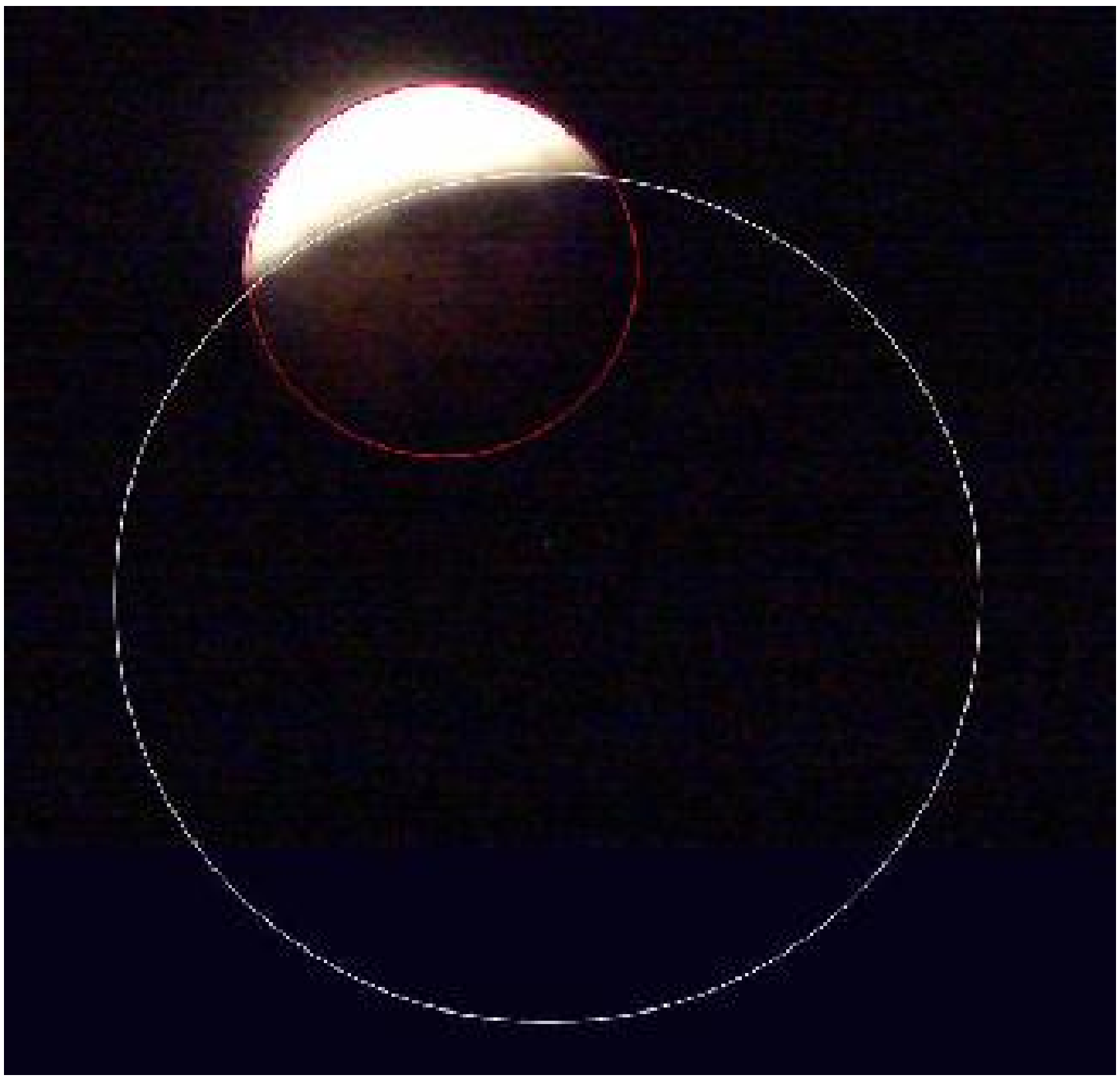

Figura 5. Eclissi di Luna del 28 ottobre 2004, fase parziale alle ore 2 del tempo di Greenwich (2 UT). Solo ad uno sguardo superficiale la Luna sembra quasi al primo quarto: in tutte le fasi lunari i corni (illuminati o oscuri) della Luna sono diametralmente opposti (a 180° di differenza di *angolo di posizione* rispetto al Nord), mentre nelle fasi di un'eclissi i corni sono il risultato di un'intersezione tra due cerchi di diverso raggio di curvatura e perciò la differenza tra i loro *angoli di posizione* varia in continuazione.

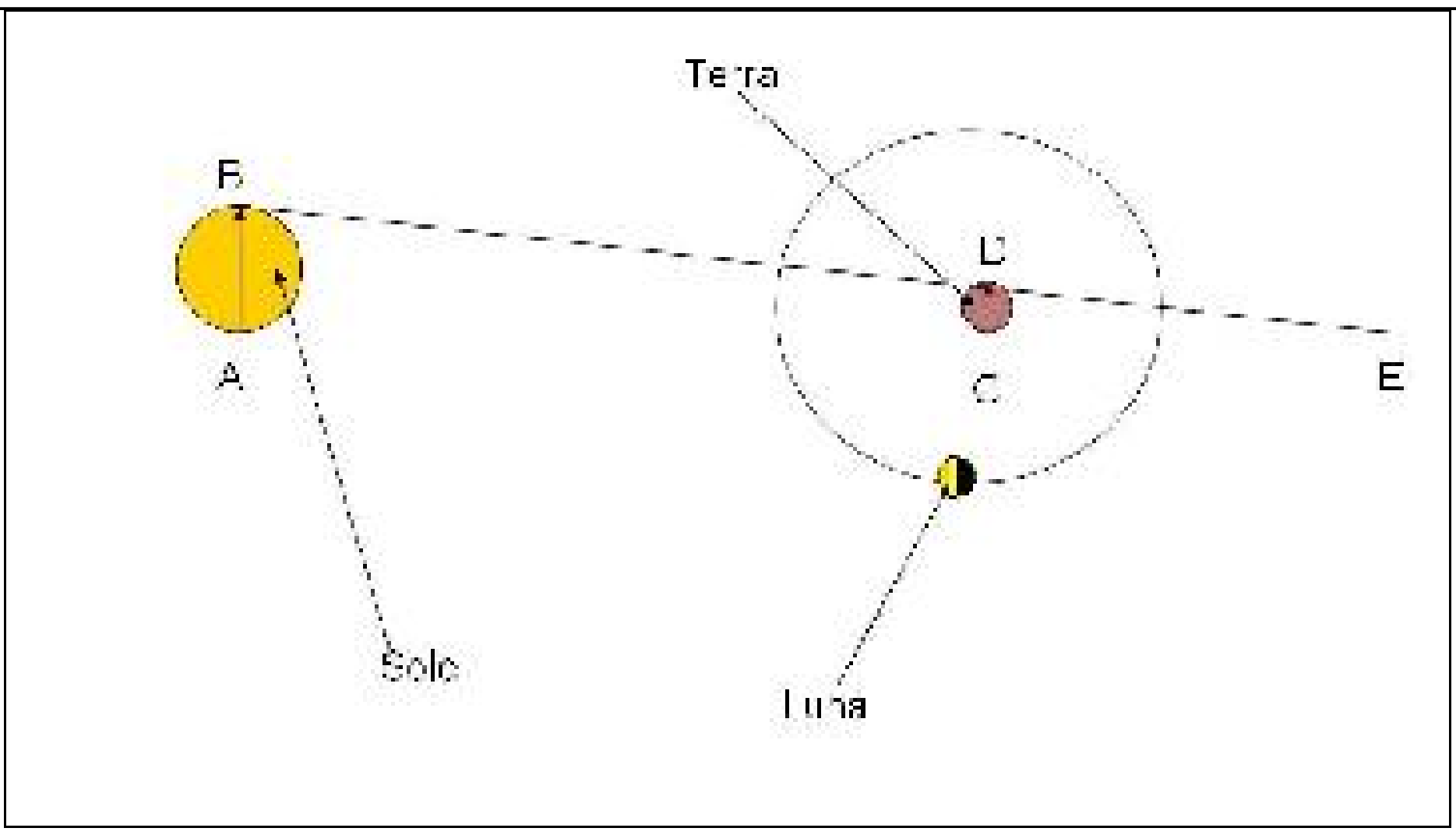

Figura 6. Schema geometrico del cono d'ombra della Terra.

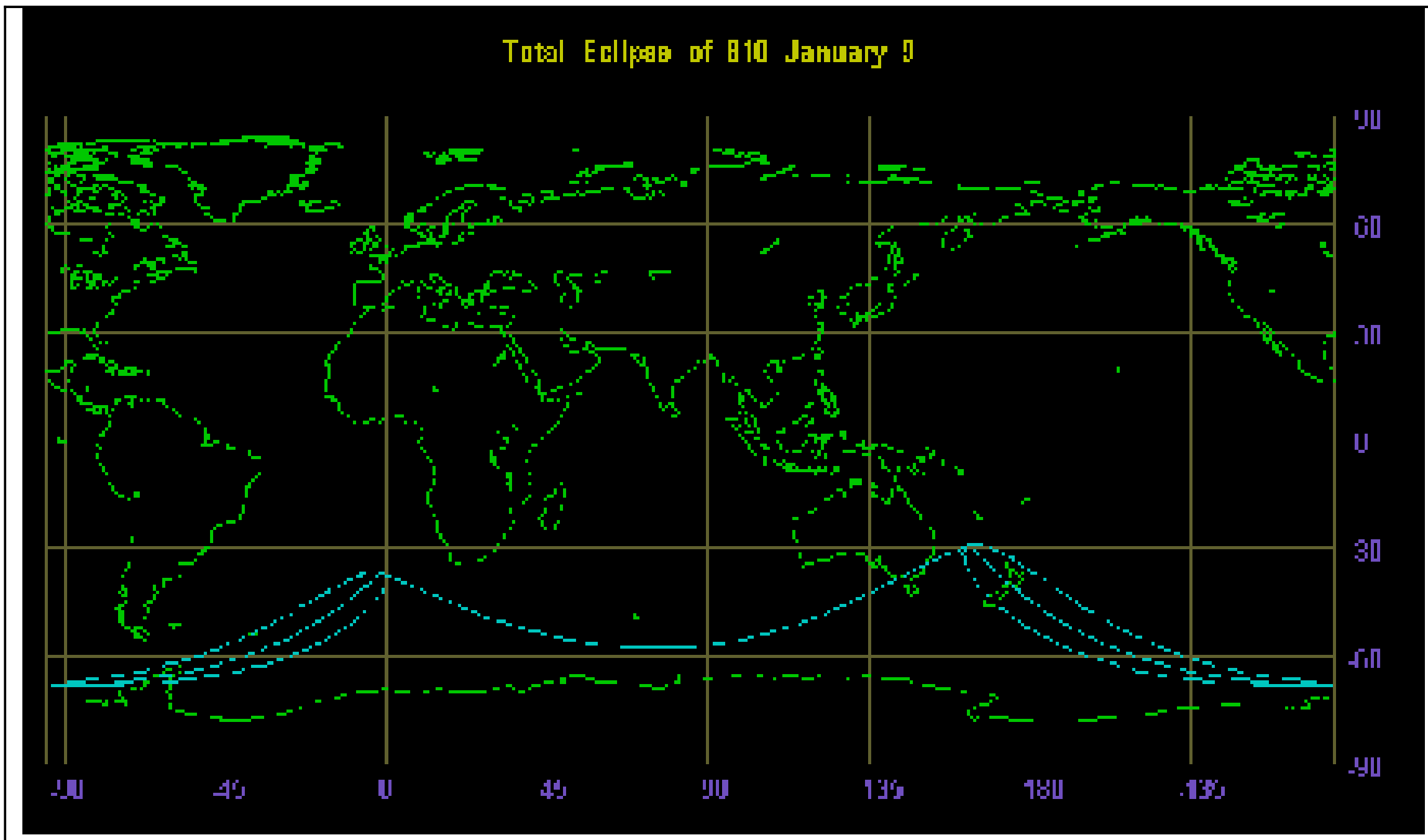

Fig. 7 Eclissi totale di Sole del 9 gennaio 810. Visibile dall'Antartide,

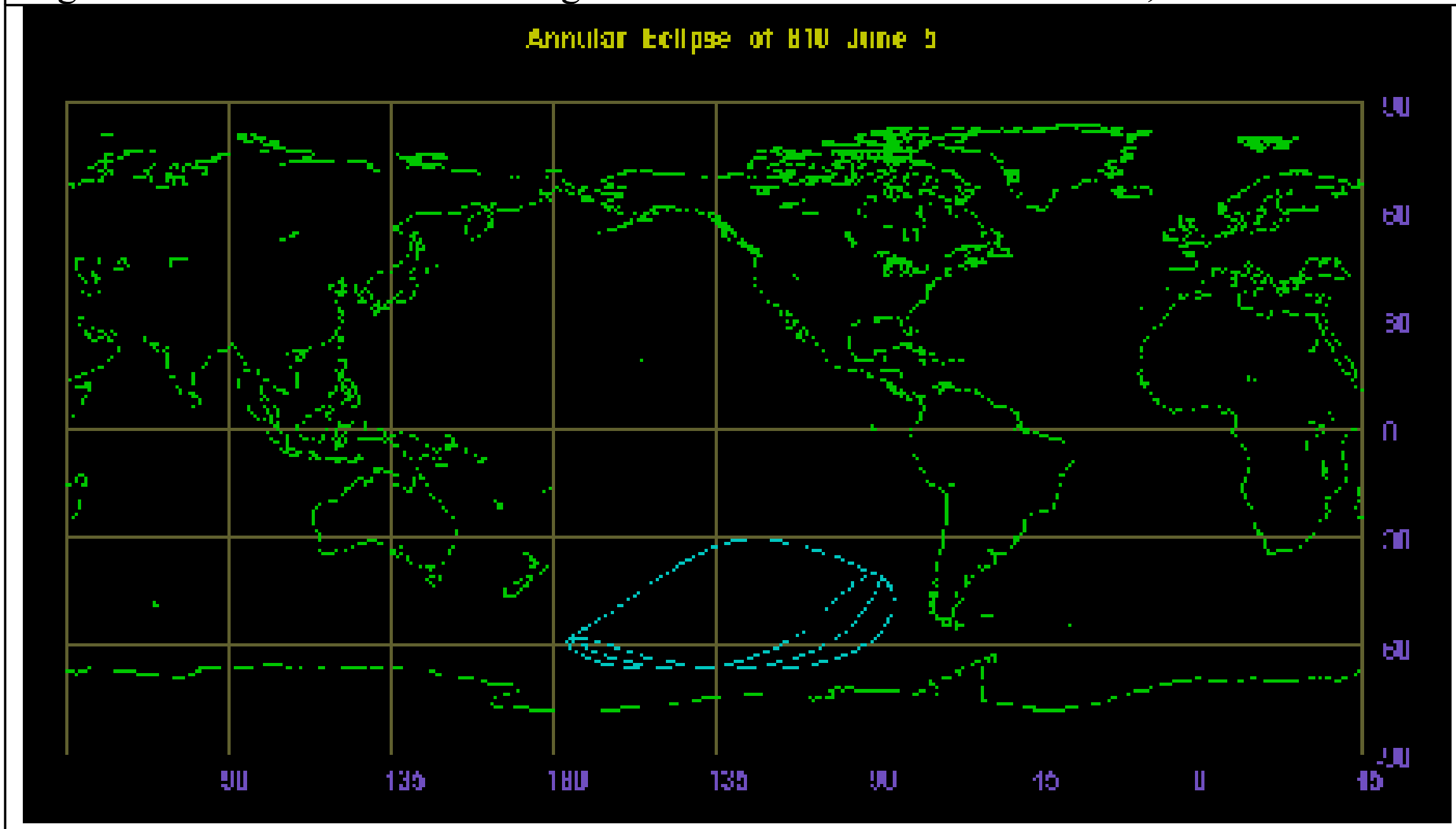

Fig. 8 Eclissi anulare di Sole del 5 giugno 810, visibile dal Sud Pacifico.

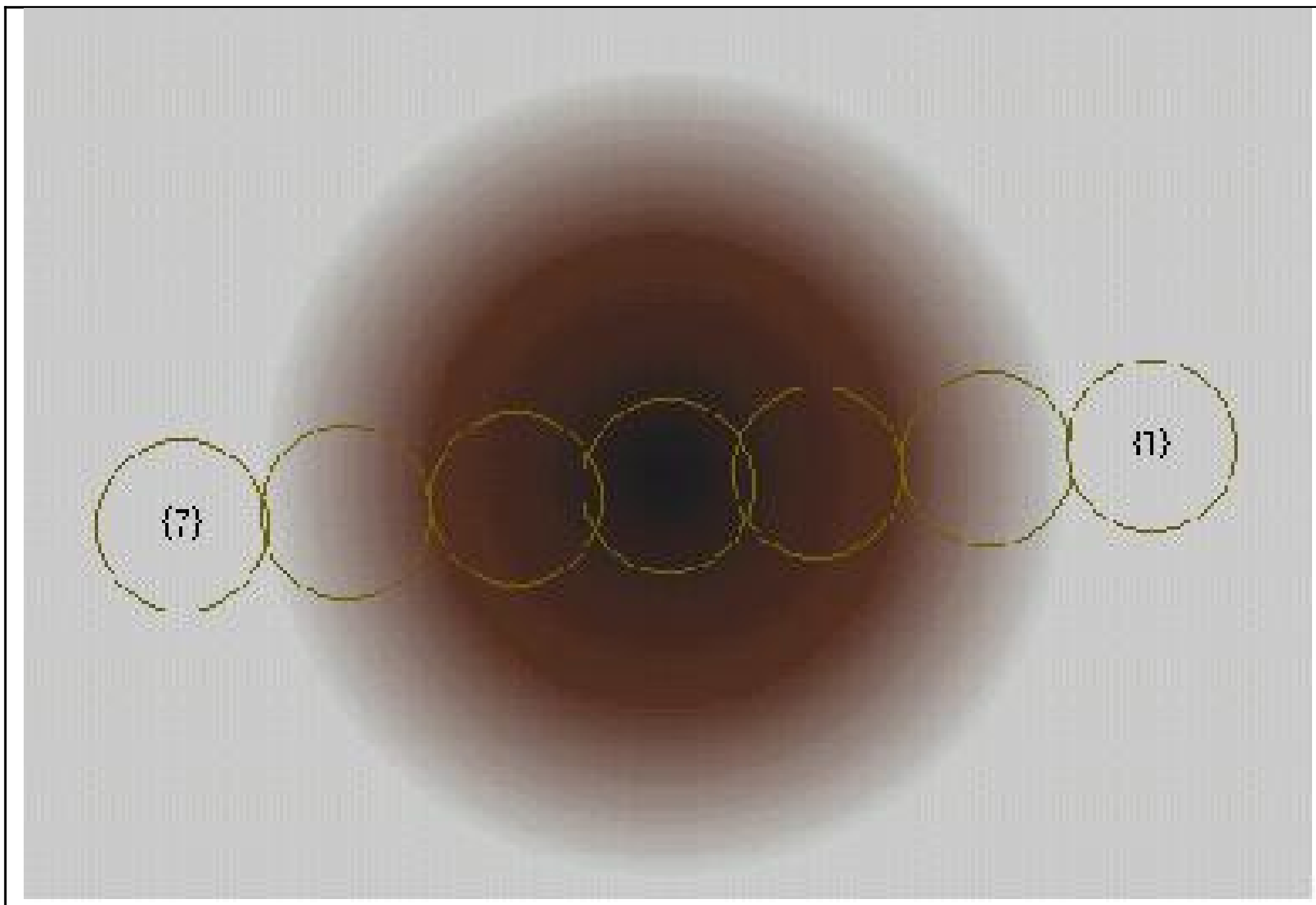

Fig. 9 Eclissi di Luna del 20 giugno 810, visibile di sera in Europa

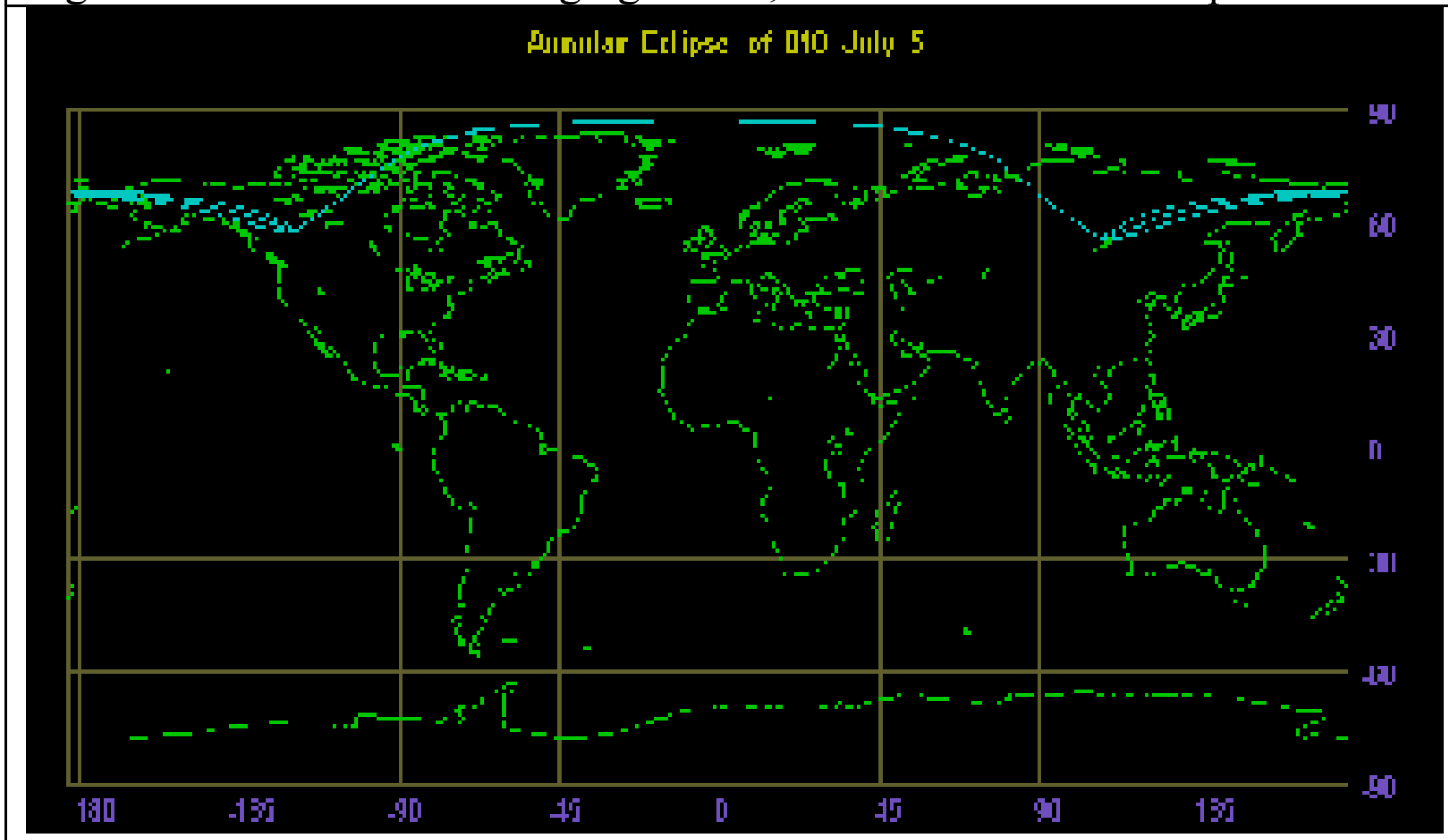

Fig. 10 Eclissi anulare di Sole del 5 luglio 810, visibile dalle regioni artiche.

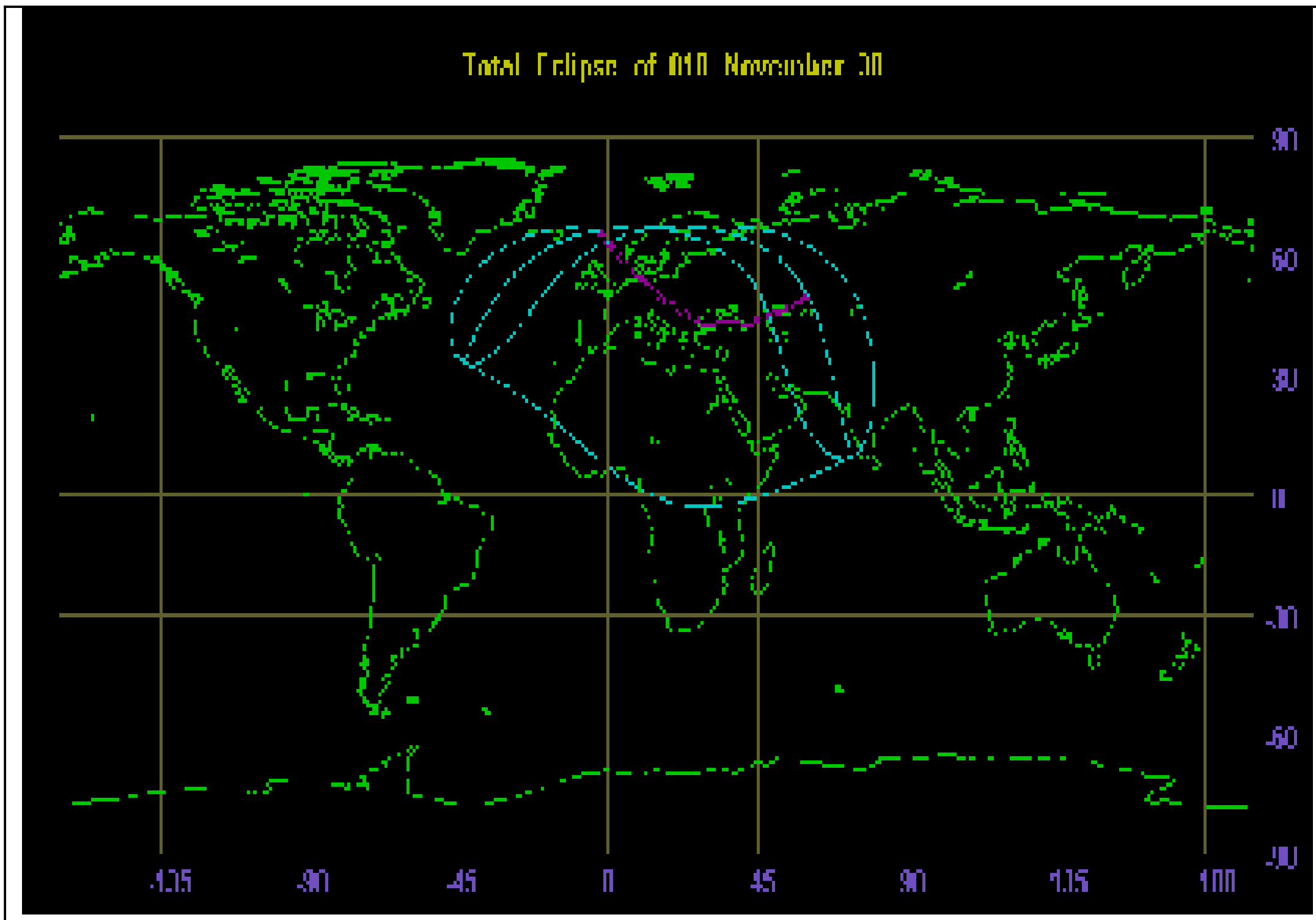

Fig. 11 Eclissi totale di Sole del 30 novembre 810, ad Aquisgrana fu parziale, ma ben visbile se il tempo era sereno, poiché la fascia di totalità era molto prossima.

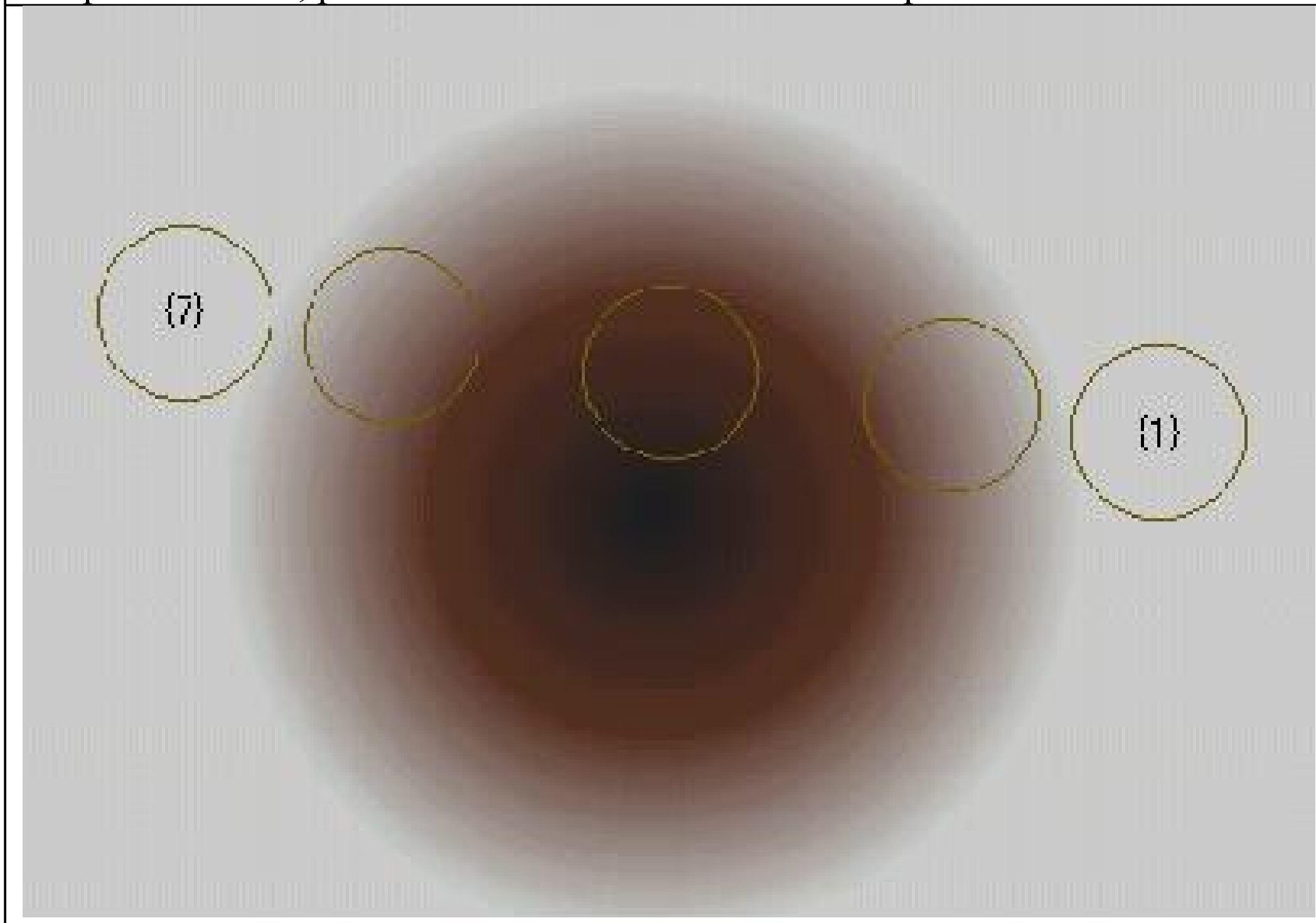

Fig. 12 Eclissi di Luna del 14 dicembre 810. Visibile di sera in Europa. Questa ha seguito di 15 giorni l'eclissi di Sole del 30 novembre. Ambedue erano visibili da Aquisgrana col tempo buono.

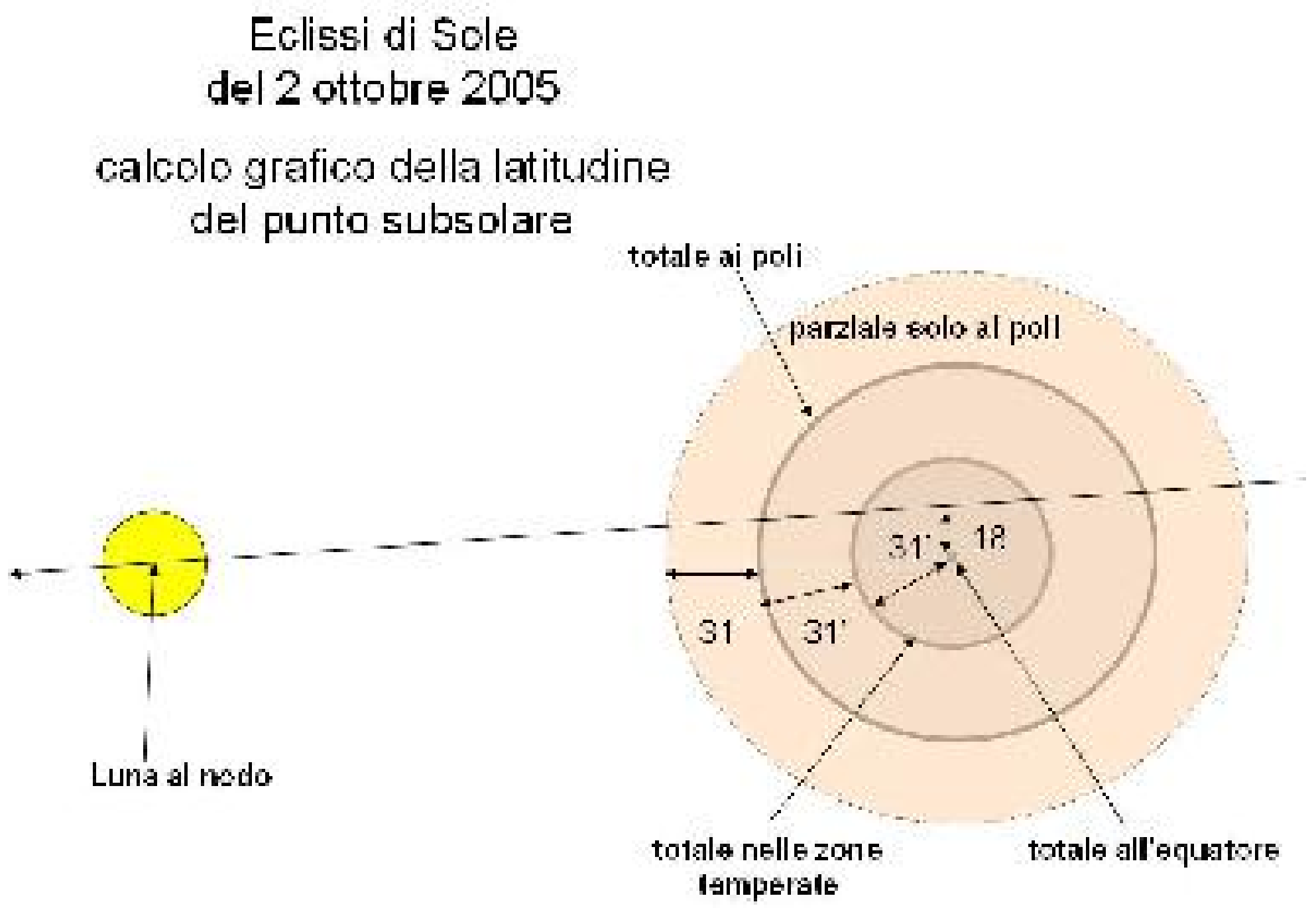

Figura 13. Calcolo grafico della latitudine del punto subsolare, e della posizione del nodo dell'orbita lunare per l'eclissi anulare di Sole del 2 ottobre 2005.

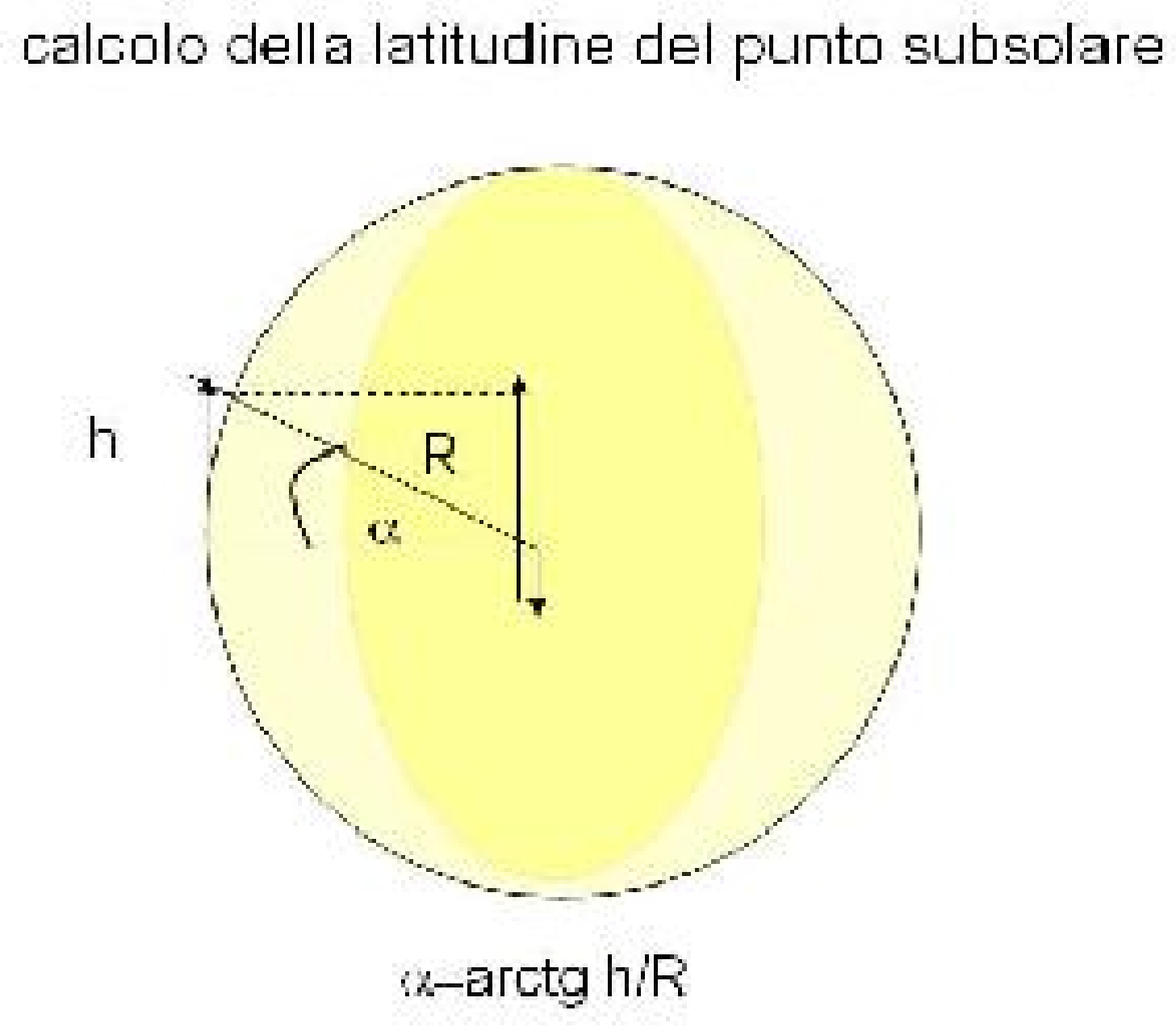

Figura 14, schema per il calcolo della latitudine del punto subsolare, dove cioè il Sole in eclissi è perpendicolare al suolo.

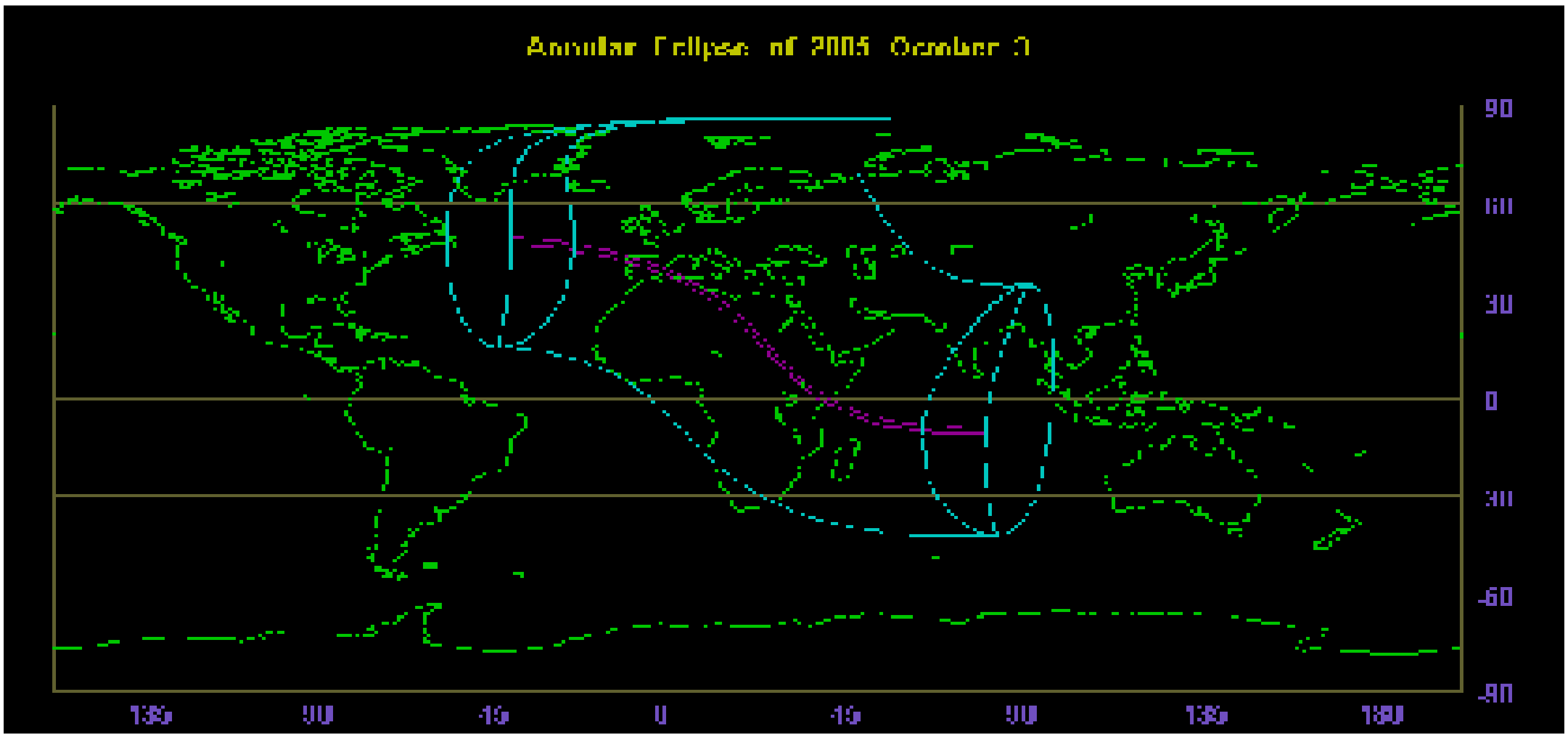

Figura 15. Eclissi anulare del 3 ottobre 2005, mappa realizzata con il programma Occult v. 3.1.0, di David Herald (2004). Il programma è disponibile gratuitamente sul web al sito dello IOTA (International Occultation Timing Association).

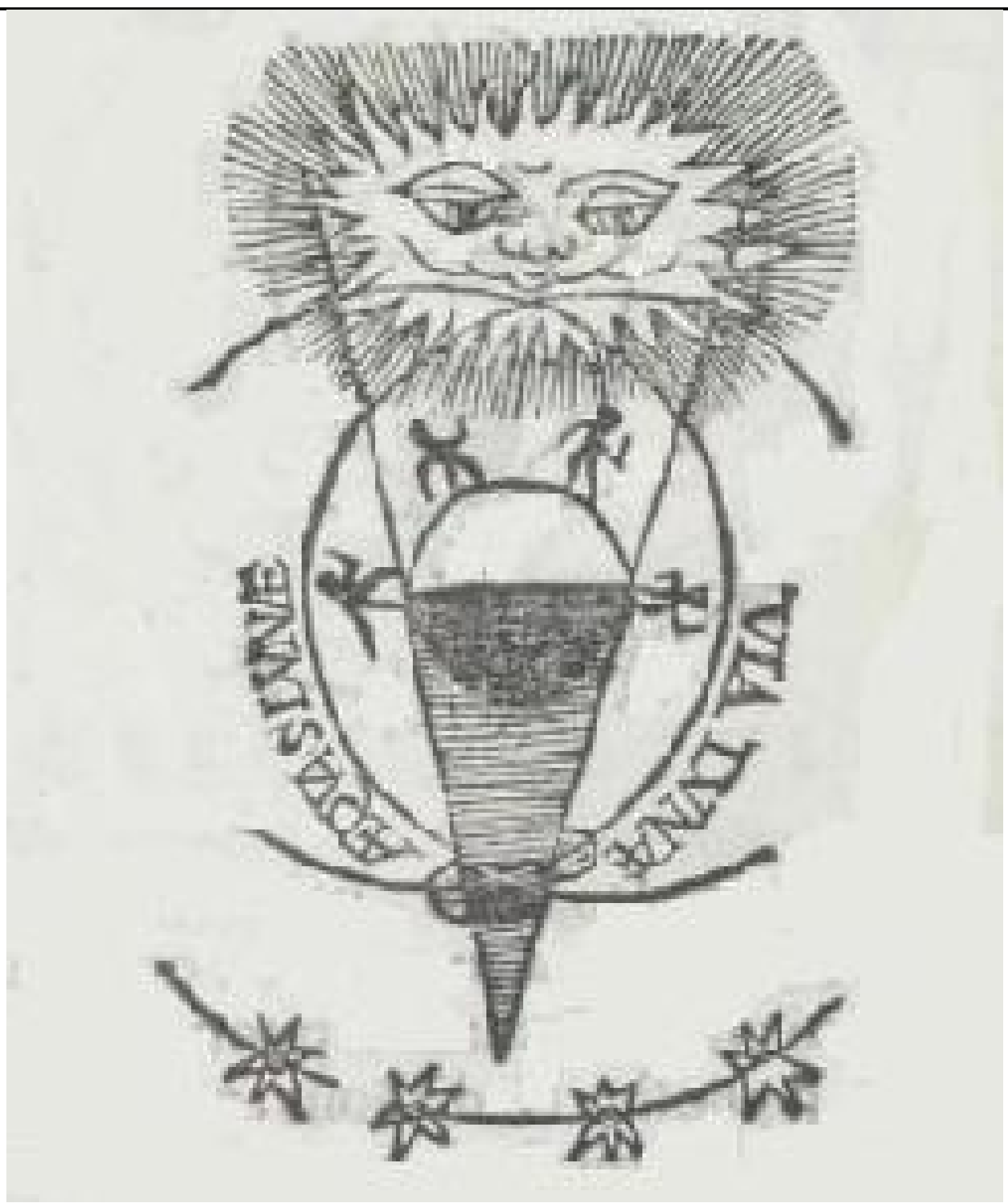

Schema del cono d’ombra terrestre dall’edizione del 1562 della Sfera del Sacrobosco (Vinetiis, apud Hieronimum Scotum).È accennata anche la sfera delle stelle fisse. Il sistema rappresentato è geocentrico.
Gentile concessione Biblioteca Universitaria Alessandrina, Roma.

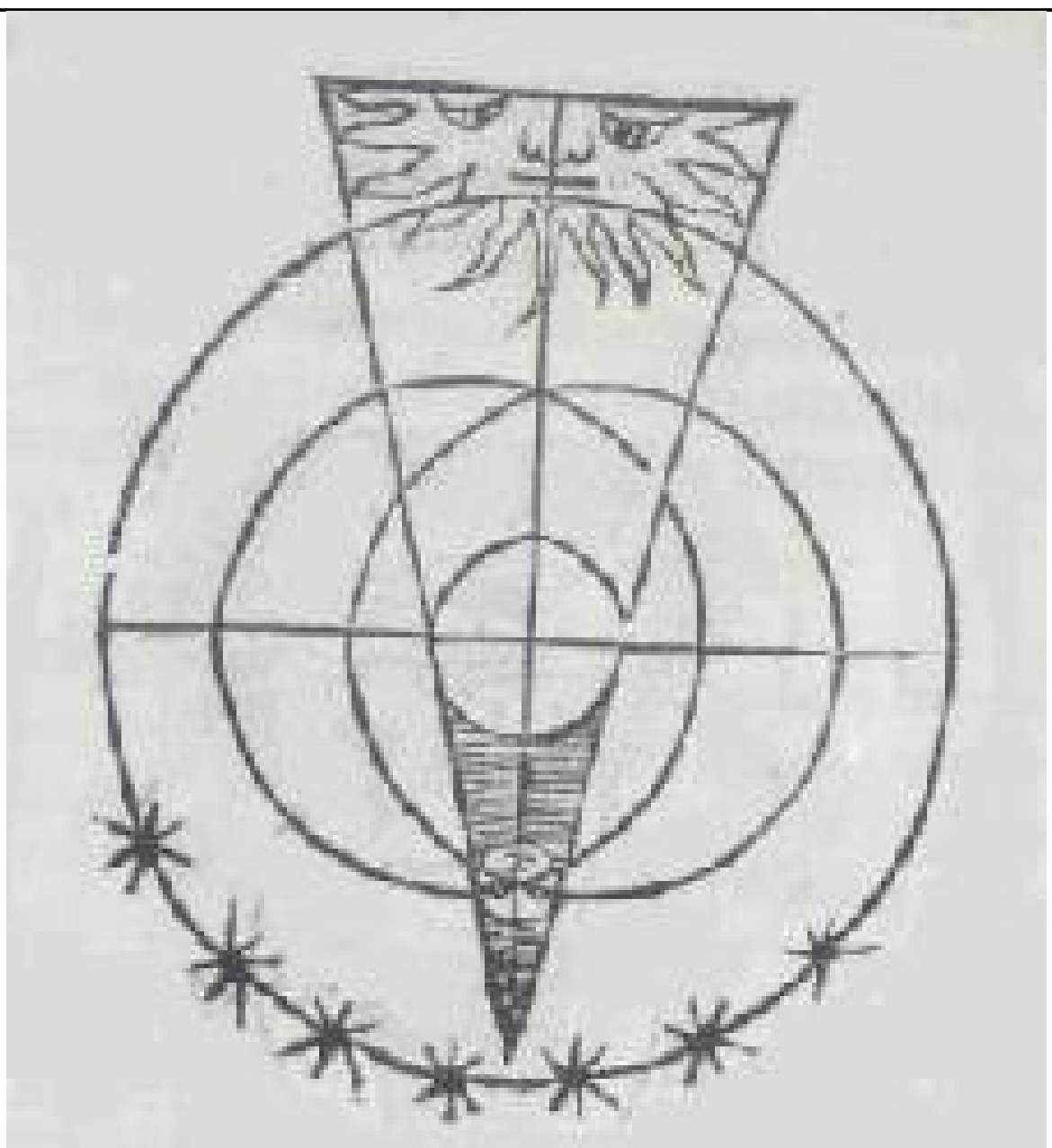

Schema di un’eclissi di Luna dall’edizione del 1562 della Sfera del Sacrobosco (1256) (Vinetiis, apud Hieronimum Scotum). Si vedono anche le stelle fisse. Il sistema rappresentato è geocentrico.
Gentile concessione Biblioteca Universitaria Alessandrina, Roma.

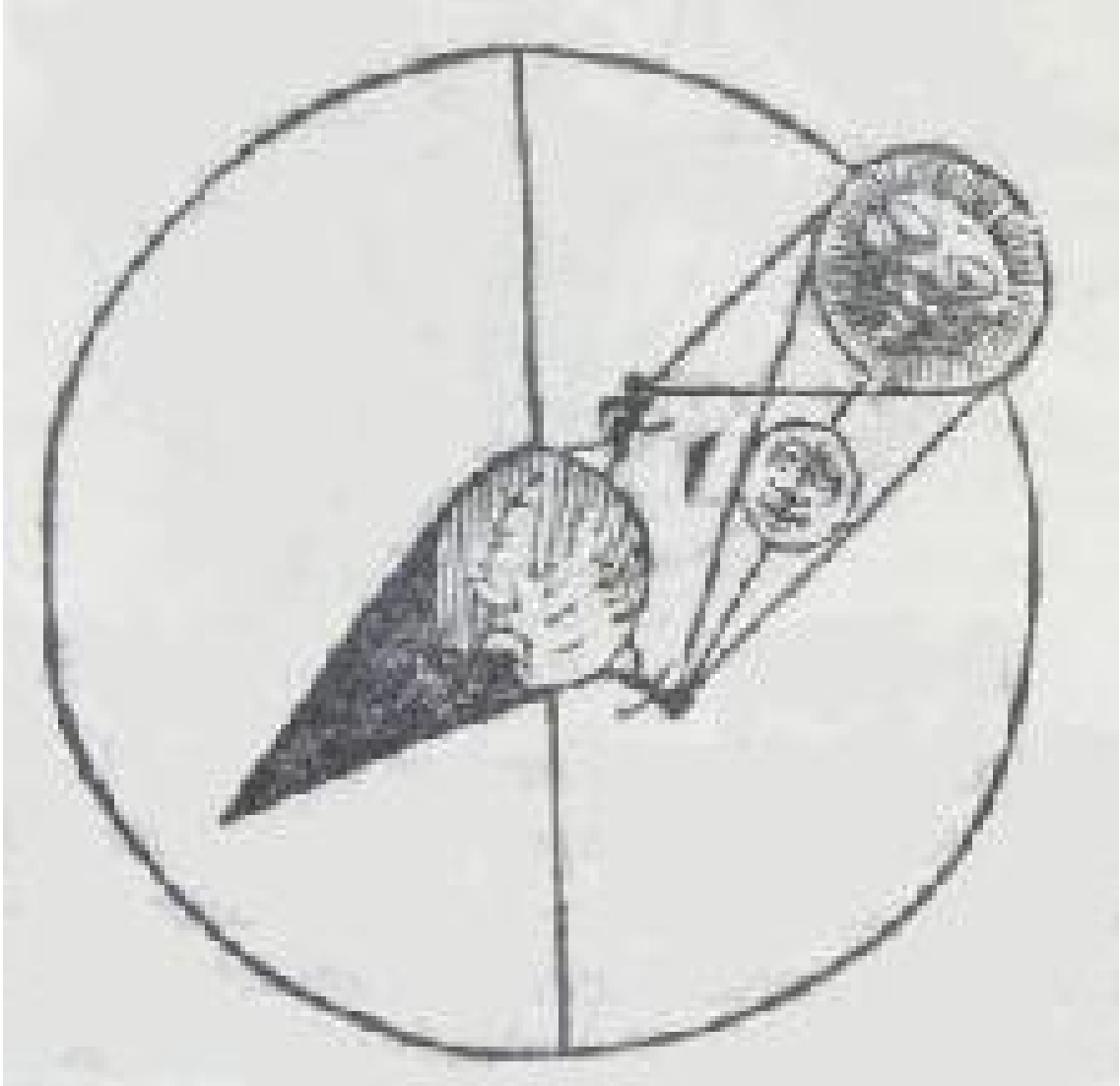

Caso di un’eclissi di Sole visibile da un punto della Terra e non da un altro.
Dall’edizione del 1562 della Sfera del Sacrobosco (Vinetiis, apud Hieronimum Scotum). Il sistema rappresentato è geocentrico, vedendosi chiaramente l’orbita del Sole.
Gentile concessione Biblioteca Universitaria Alessandrina, Roma.

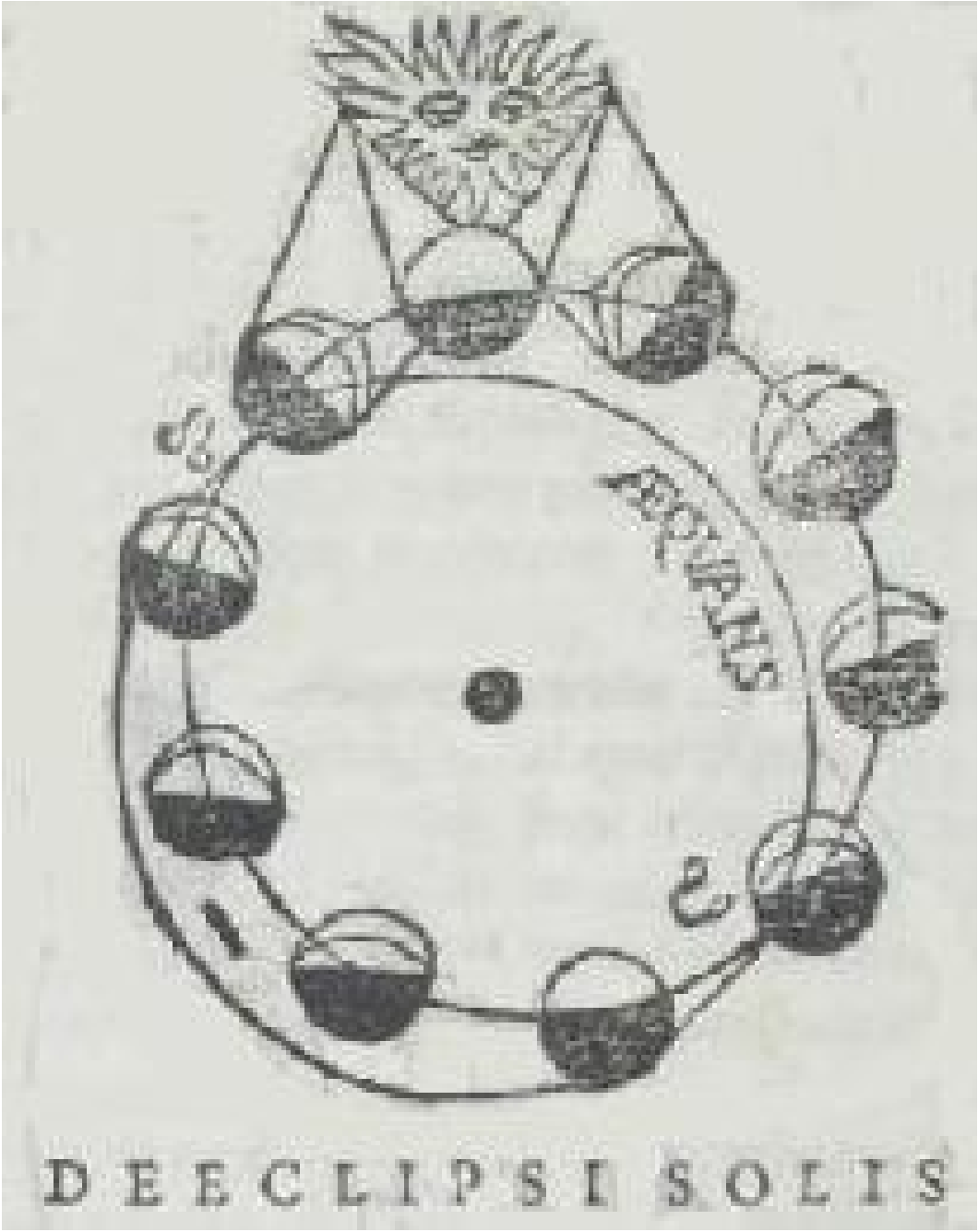

Eclissi e nodi dell’orbita lunare. Quando la Luna è vicino ai nodi si ha un’eclissi. I nodi sono indicati con simboli opportuni: con i riccioletti (nodi) rivolti in alto si rappresenta il nodo ascendente dell’orbita che nella figura è in basso a destra; mentre con i riccioletti in basso si rappresenta il nodo discendente dell’orbita (qui in alto a sinistra). Il moto della Luna è in senso antiorario rispetto alla posizione del Sole. Il sistema rappresentato è geocentrico, che è quello giusto per l’orbita della Luna.
Edizione del 1562 della Sfera del Sacrobosco (Vinetiis, apud Hieronimum Scotum).
Gentile concessione Biblioteca Universitaria Alessandrina, Roma.

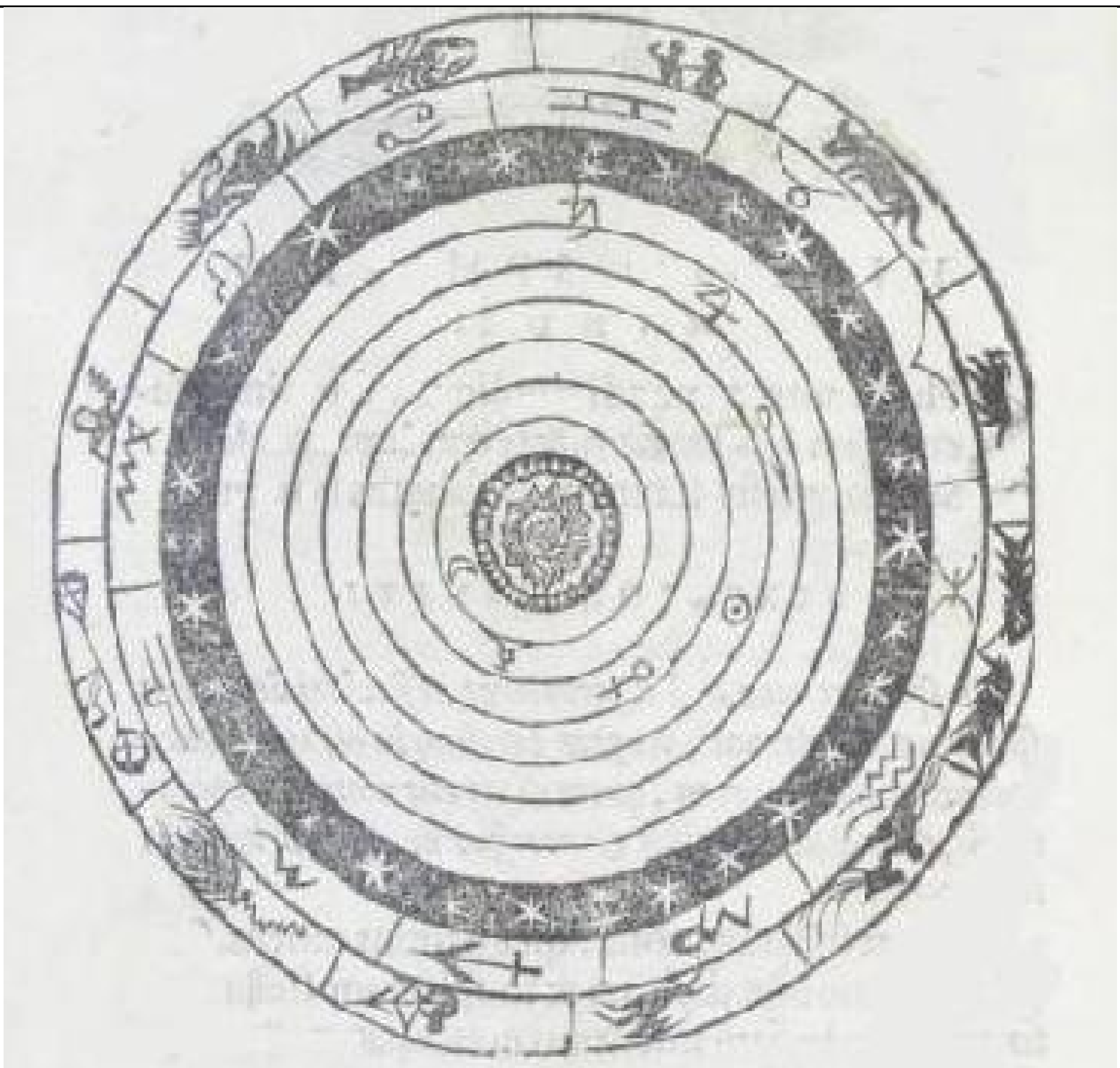

Sfere tolemaiche. Attorno alla Terra ci sono i 4 elementi platonici, con la sfera del fuoco immediatamente adiacente a quella della Luna. All'esterno abbiamo le sfere dei segni zodiacali (indicati da simboli) e delle costellazioni (indicate da disegni). Praticamente i *segni zodiacali* sono delle suddivisioni regolari dell'orbita annuale del Sole di 30° ciascuno, a partire dall'equinozio, che per convenzione cade a 0° nel segno dell'Ariete, sulla linea che separa ✻ e ❀ . I segni si leggono in senso antiorario da Ovest verso Est, secondo il moto annuale del Sole.
Si noti la rotazione della sfera più esterna, la sfera delle costellazioni, rispetto a quella immediatamente interna dei segni zodiacali, dovuta alla precessione degli equinozi. È una rotazione di circa ¼ di un intero segno, che corrisponde a circa 550 anni. La precessione sposta i segni verso la costellazione che li precede. Per questo l'immagine rappresenta una situazione attorno al 400 a. C., 550 anni prima di quando Tolomeo era attivo. Così l'inizio della costellazione dell'Ariete (subito dopo i Pesci), dove era l'equinozio di primavera al tempo della nascita di Cristo, veniva attraversata dal Sole nel 400 a. C. quando si era ancora nel segno dei Pesci, cioè attorno al 13 marzo. Oggi (2005) la sfera esterna dovrebbe essere ruotata in senso orario di più di un intero segno zodiacale, quasi 40°, portando così l'equinozio a cadere tra le costellazioni dei Pesci e dell'Acquario. Edizione del 1562 della Sfera del Sacrobosco (Vinetiis, apud Hieronimum Scotum). Gentile concessione Biblioteca Universitaria Alessandrina, Roma.

**Dùngal EPISTOLA DE DUPLICI SOLIS ECLIPSI ANNO 810. AD CAROLUM MAGNUM. Patrologia Latina vol. 105 col. 447-458**

DUNGALI RECLUSI [nota]
Epistola Dungali. Hujus epistolae auctor est Dungalus Reclusus, hoc est, solitarius, monasticam videlicet agens vitam procul a consortio tum monachorum tum saecularium. Qua in epistola cum salebras aliquot reperissemus, V. C. Ismaelem Bullialdum consuluimus, ut pote in mathematicis rebus apprime versatum, sicut et in omni genere solidae litteraturae, ut de illa quid sentiret, nobis exponere dignaretur. Scripsit itaque observatiunculam, quam hic exscribendam operae pretium duximus. «Haec epistola Dungali, quamvis nec perite, nec clare respondeat quaestioni ab imperatore propositae, neque demonstret quae ostendere oportuit, utilis nihilominus in eo est quod arguat tunc temporis plures, ipsumque Imperatorem dubitasse de eo quod a quibusdam temere et inani jactantia impulsis asserebatur, duas eclipses solis anno Christi 810 factas ac visas fuisse in nostra Europa, qui quidem ratione fulti dubitabant. Cum possibile non fuerit hoc anno 810, VII Id. Jun., solem defecisse, aut defectum illius in Europa adnotatum fuisse. Optime ad hunc annum adnotavit Calvisius, et rectissime judicavit hanc eclipsim solis die VII Id. Jun. ab aliquo, qui Kalendarium scripsit, praedictam fuisse, calculo ex tabulis astronomicis imperfectioribus deducto, sed non factam (addo et non visam Europaeis nostris hominibus). Chronologi nihilominus, Ursamque visam in suis Chronicis scripserunt, opinionem, vel potius errorem vulgi secuti. Addendum mihi videtur, calculatorem anni illius 810 in hoc allucinatum fuisse, quod non monuerit illam eclipsim in locis lineae aequinoctiali subjectis aut meridionalibus apparituram fuisse, et visendam in hemisphaerio australi. Non est enim possibile ut in locis ab aequinoctiali linea paulo remotioribus intra semestre spatium binae eclipses solis cernantur, quod sub linea aequinoctiali, vel in locis subjacentibus parallelis ab ea non longe descriptis accidere potest: intra vero quinquemestre spatium in eodem hemisphaerio boreali vel austrino binae eclipses solares conspici queunt: quae omnia demonstrari possunt, ut pote vera. Sed hujus epistolae auctor Dungalus has differentias ignorasse videtur. Cumque pridie Kal. Decemb. ejusdem anni 810 eclipsis solis, animadversa ab Europaeis, priorem illam VII Id. Jun. factam, ab antoecis nostris et antipodibus, aut saltem in hemisphaerio australi visam fuisse asserene debuit, ut quaestioni plene satisfaceret.» Huc usque Ismael Bullialdus. Dungali epistolam collegit noster Joannes Mabillon e scripto codice Sanremigiano.]

EPISTOLA DE DUPLICI SOLIS ECLIPSI ANNO 810. AD CAROLUM MAGNUM.
(Ex Spicilegio Dacherii, tom. III, pag. 324.)

In nomine Patris, et Filii, et Spiritus sancti. Domino gloriosissimo Carolo serenissimo Augusto, omnium antecedentium Romanorum principum cunctis nobilibus honestisque regalium virtutum donis et exercitiis studiosissimo, vita longaeva, fida salus, continua benevolentia, pax, corona immarcescibilis, gloria sine fine.

Audivi ergo, Domine dilectissime, ego Dungalus vester fidelis famulus et orator, non immemor quod vos Waldoni abbati direxistis epistolam, ut per illam me ipse ex vestris verbis interrogaret de ratione defectus solis, quem anno praeterito ab incarnatione Domini 810 bis evenisse plurium relatu vobis fuisse compertum dixistis, et quem sicut vos legisse memorastis, non solum antiqui gentilium philosophi, sed et quidam Constantinopolitanus episcopus, quasi naturalem concursionis elementorum effectum usitatae et certae explorationis peritia cognitum prius dixere, quam fieret.

Inde vestrae beatissimae et clarissimae serenitati visum est mandare ut de dicta causa ego quasi sectator sapientiae interrogarer quid sentirem, et quid scirem, et quid sentirem proferendo et respondendo faterer, exceptum scriberetur, scriptumque vobis deferretur. Non differam igitur, neque dissimulabo vestro secundum vires sanctissimo et utilissimo parere praecepto; et utinam tam efficax quam voluntarius existerem, ut non solum velle, sed et compote voto assequi cupita valerem, licet apud summum Rectorem pronus et alacris affectus pro re effecta et adimpleta reputatur.

Quia ergo, domine mi, hujus rationis investigatio et peritia ad philosophos, hoc est, physicos proprie et specialiter pertinet, sicut vestri continent apices, quorum libri compositiores et diligentiores quamvis mihi non suppetant, quibus de his rebus et de talibus exercitatiori sermone et enucleatiori expressione tractaverunt, et per quos vobis plenius et eruditius de inquisitis respondere me posse crediderim. Secundum simplices tamen et leves compendiososque libellos qui inter manus sunt, in quantum de ipsis torpor obtunsi cordis et tardus sensus vix lento conamine pigroque nisu reptans et movens, praelibare quiverit, ne vulgari proverbio lupus in fabula, pavido stupidoque silentio reprimi videar, utcunque respondebo; sciens indubitatissime vestram serenissimae et piissimae longanimitatis indulgibilem clementiam, si quid minus aut aliter dixero, facilem mihi veniam donaturam, et paterna correctione me veluti praeter industriam studiumve per fragilitatem infirmitatemque delinquentem modeste castigaturam.

Hujus autem quaestionis origo repetenda est, ut ab initio sicut in caeteris solet disputationibus, per ordinem ratio explicanda procedens, congrua reddatur de interrogatis responsio. Macrobius igitur Ambrosius in expositione Ciceronis inter caetera commemorat de novem circulis, qui aplanen illam ambiunt, hoc est maximam sphaeram in qua duodecim signa videntur infixa, et cui subjectae septem aliae sphaerae per quas duo lumina sol et luna et vaga quinque discurrunt. Orbium autem, hoc est circulorum praedictorum, primus est galactias, quod latine lacteus interpretatur, qui solus subjacet oculis, caeteris circulis magis cogitatione quam visu comprehendendis. Secundus zodiacus circulus, hoc est signifer, signa, id est, stellas et sidera ferendo et continendo dictus. Quinque alii circuli parelleli vocantur, dicti hoc nomine, quod neque in omnibus aequales sunt, neque inaequales; de quibus Virgilius memorat in Georgicis.

Praeter hos alii duo sunt Coluri quibus nomen dedit imperfecta conversio: duo, qui numero praedicto superadduntur, Meridianus et Horizon non scribuntur in sphaera, quia certum locum habere non possunt, sed pro diversitate circumspicientis habitantisve variantur: quae omnia transitive nominata et numerata in hoc loco non est opus exponere. Duo ergo praedicta lumina, hoc est sol et luna, et quinque stellae quae appellantur vagae, septem memoratas sphaeras maximae sphaerae duodecim signa continenti, quae aplanes vocatur, subjectas dispertiverunt, et occupatis regionibus quasi proprias et speciales haereditates singulae singulas obtinuerunt.

In prima autem sphaera de septem illa est stella quae dicitur Saturni, in secunda Jovis, in tertia Martis, in quarta hoc est media, sol, in quinta stella Veneris, in sexta Mercurii; septimam quae est omnium extima et infima, luna tenet. Ita Cicero describit, cui Archimedes et Chaldaeorum ratio consentit. Plato vero a luna sursum secundum, hoc est, inter septem a summo locum sextum solem tenere confirmat, secutus Aegyptios omnium philosophiae disciplinarum parentes, qui ita solem inter lunam et Mercurium locatum volunt. Quamvis autem ista persuasio Tullii et auctorum ejus quibusdam edictis et credibilibus rationibus fulta convaluit, et ab omnibus pene in usum recepta est, perspicacior tamen Platonis observatio veriorem ordinem deprehendisse videtur, quam praeter indaginem visus haec quoque ratio commendat, quod lunam quae luce propria caret, et de sole mutuatur, necesse est fonti luminis sui esse subjectam; haec enim ratio lunam facit non habere lumen proprium, caeteras omnes stellas lucere suo, quod illae supra solem locatae in ipso purissimo aethere sunt, in quo omne quidquid est, ut verbis Philosophi loquar, lux naturalis et sua est, quae tota cum igne suo ita sphaerae solis incumbit, ut coeli zonae quae procul a sole sunt perpetuo frigore oppressae sint; luna vero, quia ipsa sola sub sole est, et caducorum jam regioni luce sua carenti proxima, lucere non potuit. Denique quia totius mundi ima pars terra est, aetheris autem ima pars luna est, lunam autem terram sed aethercam vocaverunt, immobilis autem ut terra esse non potuit, quia in sphaera quae volvitur nihil manet immobile praeter centrum; mundanae autem sphaerae terra centrum est, ideo sola immobilis perseverat: rursus terra accepto solis lumine clarescit tantummodo, non lucet; luna speculi instar lumen quo illustratur emittit, quae quamvis densius corpus sit quam caetera coelestia, ut multum tamen terreno purius, fit acceptae luci penetrabile, adeo ut eam de se rursus emittat, nullum tamen ad nos praeferentem sensum caloris, quia lucis radius cum ad nos de origine sua, id est, de sole pervenit, naturam secum ignis de quo nascitur devehit; cum vero in lunae corpus infunditur, et inde resplendet, solam refundit claritudinem, non calorem: nam et speculum cum splendorem de se vi oppositi eminus ignis immittit, solam ignis similitudinem carentem sensu caloris ostendit.

His illud adjiciendum est, praeter solem et lunam, et stellas quinque quae appellantur vagae, reliquas omnes alias infixas coelo, nec nisi cum coelo moveri. Alii, quorum assertio vero propior est, has quoque dixerunt suo motu, praeter quod cum coeli conversione feruntur, accedere, sed propter immensitatem extimi globi excedentia credibilem numerum saecula multa una cursus sui ambitione consumere, et ideo nullum earum motum ab homine sentiri; cum non sufficiat humanae vitae spatium ad breve saltem punctum tam tardae accessionis deprehendendum. Solem autem ac lunam et stellas quinque, quibus ab errore nomen, praeter quod secum trahit ab ortu in occasum coeli diurna conversio, ipsa suo motu in orientem ab occidente procedere, argumentis ad verum ducentibus comprobatur, moveri enim coeloque non esse infixas et visus et ratio affirmat, dum modo in hac, modo in illa coeli regione visuntur, et saepe cum in unum duae pluresve convenerint, et a loco in quo simul visae sunt, et a se postea separantur, quod infixae stellae non faciunt, sed in iisdem locis semper videntur, nec a sui unquam se copulatione dispergunt, ab ortu vero ad occasum in contrarium motu propiore volvi non solum manifestissima ratione, sed oculis quoque approbantibus demonstratur.

Considerato enim signorum ordine, quibus zodiacum divisum vel distinctum videmus, ab uno signo quolibet ordinis ejus sumam exordium: cum Aries exoritur, post ipsum Taurus emergit; hunc Gemini sequuntur, hos Cancer, et per ordinem reliqua. Si istae ergo in occidentem ab oriente procederent, non ab Ariete in Taurum, qui retro locatus est, nec a Tauro in Geminos signum posterius volverentur: sed a Geminis in Taurum, et a Tauro in Arietem rectae et mundanae volubilitatis consona accessione procederent. Cum vero a primo signo in secundum, a secundo ad tertium, et inde ad reliqua quae posteriora sunt revolvuntur, signa autem infixa coelo feruntur, sine dubio constat has stellas non cum coelo, sed contra coelum moveri. Hoc ut plene luceat, astruam de lunae cursu, quae et claritate sua et velocitate notabilior est. Secundo fere die circa occasum videtur, et quasi vicina soli, quem nuper reliquit, postquam ille demersus est, ipsa coeli marginem tenet antecedenti super occidens; tertio die tardius occidit quam secundo; et ita quotidie longius ab occasu recedit, ut septimo die circa solis occasum in medio coelo ipsa videatur, post alios vero septem cum ille mergit, haec oritur, adeo media parte mensis dimidium coelum, id est, unum hemisphaerium ab occasu in orientem recedendo metitur; rursus post septem alios circa solis occasum latentis hemisphaerii verticem tenet, cujus rei indicium est quod medio noctis exoritur: postremo totidem diebus exactis, additis insuper plus minusve aliis duobus, solem denuo comprehendit, et vicinus videtur ortus amborum, quandiu soli succedens rursus moveatur, et rursus recedens paulatim semper in orientem regrediendo relinquat occasum. Sol quoque ipse non aliter quam ab occasu in orientem movetur, et rursus recedens paulatim semper in orientem regrediendo relinquit occasum; sol quoque ipse naturaliter quam ab occasu in orientem movetur. Et licet tardius recessum suum quam luna conficiat, quippe quod tanto tempore signum unum emetiatur quanto totum zodiacum luna discurrit, manifesta tamen et subjecta oculis motus sui praestat indicia: hunc enim in Ariete esse ponam, quod quia aequinoctiale signum est, pares horas somni et diei facit. In hoc signo cum occidit, Libram, id est, Scorpii chelas mox oriri videmus, et apparet Taurus vicinus occasui: nam Vergilias et Hyadas partes Tauri clariores non multo post solem mergentes videmus.

Sequenti mense sol in signum posterius, id est in Taurum recedit, et ita fit ut neque Vergiliae, neque alia pars Tauri illo mense videatur: signum enim quod et cum sole occidit, semper occulitur, adeo ut et vicina astra solis propinquitate celentur: nam et Canis tunc, quia vicinus Tauro est, non videtur, tectus lucis propinquitate. Et hoc est quod Virgilius ait:

Candidus auratis aperit eum cornibus annum
Taurus, et adverso cedens Canis occidit astro.
Non enim vult intelligi Tauro oriente cum sole mox in occasu fieri Canem, qui proximus Tauro est, sed occidere eum dixit Tauro gestante solem, quia tunc incipit non videri, sole vicino. Tunc tamen occidente sole Libra adeo superior invenitur, ut totus Scorpius ortus appareat; Gemini vero vicini tunc videntur occasui; rursus post Tauri mensem Gemini non videntur, quod in eos solem migrasse significat. Post Geminos redit in Cancrum, et tum cum occidit mox Libra in medio coelo videtur: adeo constat solem tribus signis peractis, id est, Ariete et Tauro et Geminis ad medietatem hemisphaerii recessisse.

Denique post tres menses sequentes, tribus signis quae sequuntur incensis, Cancrum dico, Leonem et Virginem, invenitur in Libra; quae rursus aequat noctem diei, et dum in ipso signo occidit, mox oritur Aries, in quo sol ante sex menses occidere solebat. Ideo autem occasum ejus quam ortum elegimus praeponendum, quia signa posteriora post occasum videntur, et dum ad haec quo sole mergente videri solent, solem redire monstramus, sine dubio eum contrario motu recedere quam coelum movetur, ostendimus. Haec autem quae de sole ac luna diximus, etiam quinque stellarum recessum assignare sufficient: pari enim ratione in posteriora signa migrando, semper mundanae volubilitati contraria recessione versantur, quarum cursus recursusque ipso sole moderari perhibetur; nam certa spatii definitio est, ad quam cum unaquaeque erratica stella recedens ad solem pervenerit, tanquam ultra prohibeatur accedere, agi retro videtur et rursus cum certam partem recedendo contigerit, ad directi cursus consueta revocatur, ita solis vis et potestas motus reliquorum luminum constituta demensione moderatur.

Circus ergo sive circulus intelligitur uniuscujusque stellae una integra et peracta conversio, id est, ab eodem loco post emensum sphaerae per quam movetur ambitum in eumdem locum regressus. Est autem hic linea ambiens sphaeram ac veluti semitam faciens, per quam sol et luna discurrit, et intra quam vagantium stellarum error legitimus coercetur, quas ideo veteres errare dixerunt, quia et cursu suo feruntur, et contra sphaerae maximae, id est, ipsius coeli impetum contrario motu ad orientem ab occidente volvuntur, et omnium quidem par celeritas, motus similis, idem est modus meandi, sed non omnes eodem temporis spatio circos suos orbesque conficiunt. Causam vero sub eadem celeritate diversi spatii inaequalitas sphaerarum efficit, quas singulae stellae perlustrant: a Saturni enim sphaera, quae est prima de septem, usque ad sphaeram Jovis, a summo secundam interjecti spatii tanta distantia est, ut zodiaci ambitum superior triginta annis, duodecim vero subjecta conficiat, rursus tantum a Jove sphaera Martis recedit, ut eumdem cursum biennio peragat. Venus autem tanto est a regione Martis inferior, ut ei annus satis sit ad zodiacum peragrandum.

Jam vero ita Veneri proxima est stella Mercurii, et Mercurio sol propinquus, ut hi tres coelum suum pari temporis spatio, id est, anno plus minusve circumeant; ideo et Cicero hos duos cursus comites solis vocavit, quia in spatio pari longe a se nunquam recedunt: luna autem tantum ab his deorsum recessit, ut quod illi anno, viginti octo diebus conficiat. Sed Cicero cum quartum de septem solem velit, quartus autem inter septem non fere medius, sed omnimodo medius et sit et habeatur, non abrupte medium solem, sed fere medium dixit his verbis: «Deinde de septem mediam fere regionem sol obtinet.» Sed non vacat adjectio qua haec pronuntiatio temperatur; nam sol quartum locum obtinens, mediam regionem tenet numero, sed non spatio; Saturni enim stella, quae summa est,

Zodiacum triginta annis peragrat: sol medius anno uno; luna ultima uno mense non integro. Tantum ergo interest inter solem et Saturnum, quantum inter unum et triginta; tantum inter lunam solemque, quantum inter duodecim et unum; unde apparet totius a summo in imum spatii certam ex media parte divisionem solis regione non fieri. Sed quantum ad numerum pertinet, veluti inter septem quartus medius dicitur, quamvis propter inaequalitatem spatiorum adjectione fere particulae temperatur. Zodiacus ergo circulus est unus ex undecim supradictis, qui solus potuit latitudinem hoc modo quem referimus, adipisci. Natura coelestium circulorum incorporalis est, lineaque ita mente concipitur, ut sola longitudine censeatur, latum habere non possit; in Zodiaco longitudinem signorum capacitas exigebat. Quantum igitur spatii lata dimensio porrectis sideribus occupat, duabus lineis limitatum est, et tertia ducta per medium ecliptica vocatur, quia cum cursum suum in eadem linea pariter sol et luna conficiunt, alterius eorum necesse est evenire defectum; solis, si ei tunc luna succedat; lunae, si tunc adversa sit soli. Ideo nec sol unquam deficit, nisi cum tricesimus lunae dies est; et nisi quinto decimo cursus sui die, nescit luna defectum: sic enim evenit ut aut lunae contra solem positae, ad mutuandum ab eo solidum lumen, sub eadem linea inventus terrae conus obsistat, aut soli ipsa succedens, o jectu suo ab humano aspectu lumen ejus repellat. In defectu autem sol ipse nihil patitur, sed noster fraudatur aspectus, luna vero circa proprium defectum laborat, non accipiendo solis lumen, cujus beneficio noctem colorat; quod sciens Virgilius, disciplinarum omnium peritissimus, ait:

Defectus solis varios, lunaeque labores.
Quamvis igitur trium linearum ductus zodiacum et claudat et dividat, unum tamen circum auctor vocabulorum dici voluit antiquitas, secundum vero quosdam philosophos latitudo zodiaci duodecim lineis mensuratur, ex quibus juxta paris numeri naturam duas haberi medias necesse est, quas tantummodo a sole lustrari confirmantes, lunam per omnes discurrere dicunt. Quapropter ultro citroque vagata eclipsim fieri singulis mensibus non sinit. Omnibus tamen annis utriusque sideris defectus statutis diebus horisque evenire comprobant, quamvis non semper appareant, ideo quia aliquando fiunt subtus terram in parte latentis hemisphaerii, aliquando supra; sed propter nubila et propter globositatem et convexitatem terrae neque ubique, neque eisdem horis ab omnibus cerni possunt; unde certissimum est saepius ista fieri quam videri, nec aequaliter cunctis apparere cum videntur; unde vespertinos solis ac lunae defectus Orientales non sentiunt, neque matutinos Occidentales, obstante cono terrae atque visum arcente. Lunae autem defectum aliquando quinto mense a priori, solis vero septimo ejusdem bis in triginta diebus super terras occuliari, necnon ab aliis visum esse, quondam in duodecim diebus utrumque sidus deficere probabili ratione et traditione cognovimus.

Respondi ergo, ut mihi videtur, beatissime Auguste, secundum vestrarum exactionem litterarum, et dixi ex eorumdem auctoritate quemadmodum antiqui philosophi et scierunt et praescierunt, quomodo fieret defectus solis, et quando fieret illi enim omnium disciplinarum peritissimi, et nullius sectae inscii veteribus approbatae, sagacissima elimatae et defaecatae mentis intentione, et perspicacissima purgatissimaque interni sensus acie praefixa, omnium rerum naturas, rationes, causas et origines subtilissime et instantissime naturali investigatione quaesiverunt, accuratissime et efficacissime illo a quo omne datum optimum est et omne donum perfectum offerente quaesita invenerunt, inventa et deprehensa diligentissime et intentissime observaverunt inde physici astronomiae specialiter studentes, eadem diutissima meditatissimaque diligentia ortus et obitus stellarum intuentes et intuendo experientes, solis et lunae et reliquarum quinque vagantium cursus et recursus, accessus et recessus plenissime exploraverunt, in tantum ut explorando indubitatissime scirent quantas lineas zodiaci circuli unaquaeque stella erratica lustraret, et per quam proprie et specialiter de ipsis lineis in praesenti cursum dirigeret, et in quo signo et in qua parte ipsius signi esset. Qui ergo ita de subtilioribus, licet veris et naturalibus, aliarum stellarum motibus certissime et studiosissime cognoverunt, cur solis et lunae cursus, qui vere notabiliores et faciliores sunt ad cognoscendum, ignorarent, ut eos lateret quomodo vel quando per eamdem zodiaci circuli eclipticam lineam currerent, et illam unam eamdemque lustrantes, in unum signum et in unam partem coirent, et in eamdem partem coeuntes, lunaque in ipsa soli succedente eclipsis fieret solis.

Non solum ergo praedicti philosophi eclipsim, hoc est defectum solis praesciebant, et praescientes praedicebant quando post unum mensem futurus esset; sed quando per annum, aut XX aut C M annos sequeretur, per supradictam sagacem explorationem et diligentem observationem longe ante experti praesignabant. Sed ut plus miremini, usque ad quindecim millia annorum talibus argumentis protenderunt. Inde Cicero visionem Africani referens, ita dicit: «Homines populariter annum tantummodo solis unius astri reditum metiuntur; re autem recta cum ad idem unde semel profecta sunt, cuncta astra redierunt, eademque totius coeli descriptionem longis intervallis retulerunt: tunc ille vere vertens annus appellari potest, in quo vix dicere audeo quam multa hominum saecula teneantur. Namque ut olim deficere sol hominibus exstinguique visus est, cum Romuli animus haec ipsa in templa penetravit, quandoque ab eadem parte sol eodemque tempore iterum defecerit, tum signis omnibus ad idem principium stellisque revocatis, expletum annum habeto; cujus quidem anni nondum vicesimam partem scito esse conversam.»

Quae verba Tullii Ambrosiana expressio aperit hoc modo atque pandit: Annus non is solus est quem nunc communis omnium usus appellat; sed singulorum seu luminum, hoc est solis et lunae, sive stellarum, emenso omni coeli circuitu, a certo loco in eumdem locum reditus, annus suus est: sic mensis lunae annus est intra quem coeli ambitum lustrat: nam et a luna mensis dicitur, quia Graeco nomine luna ìÞíç vocatur.

Virgilius denique ad discretionem lunaris anni, qui brevis est, annum, qui cursu solis efficitur, significare volens ait:

Interea magnum sol circumvolvitur annum.
Annum magnum vocans solis comparatione lunaris; nam cursus quidem Veneris atque Mercurii pene par solis; Martis vero annus fere biennium tenet: tanto enim tempore coelum circuit. Jovis autem stella duodecim, et Saturni triginta annos in eadem circumitione consumit.

Haec de sole et luna, ac vagis, ut ante retulimus, jam nota sunt: annus vero, qui mundanus vocatur, qui vere vertens est, quia conversione plenae universitatis efficitur, largissimis saeculis explicatur, cujus ratio talis est. Stellae omnes et sidera, quae infixa coelo videntur, quorum proprium motum nunquam visus humanus sentire vel deprehendere potest, moventur tamen, et praeter coeli volubilitatem qua semper trahuntur, suo quoque accessu tam sero promovent, ut nullius hominum vita tam longa sit, quae observatione continua factam de loco per mutationem, in quo . . . . . primum viderat, deprehendat.

Mundani ergo anni finis est, cum stellae omnes omniaque sidera, quae aplanes habet, a certo loco ad eumdem locum ita remeaverint, ut ne una quidem coeli stella in alio loco sit quam in quo fuit, cum omnes aliae ex eo loco motae sunt ad quem reversae, anno suo finem dederunt, ita ut sol et luna cum erraticis quinque in iisdem locis et partibus sint, in quibus incipiente mundano anno fuerunt: hoc autem, ut physici volunt, post annorum quindecim millia peracta contingit.

Ergo sicut annus lunae mensis est, et annus solis duodecim menses, et aliarum stellarum hi sunt anni quos supra retulimus; ita mundanum annum quindecim millia annorum, quales nunc computamus efficiunt. Ille ergo vere annus vertens vocandus est, quem non solis, id est unius astri, reditus metitur, sed quem stellarum omnium, quae in quocunque coelo sunt, ad eumdem locum reditus sub eadem coeli totius descriptione concludit; unde mundanus dicitur, quia mundus proprie coelum vocatur. Igitur sicut annum solis non solum a Kalendis Januariis usque ad easdem vocamus, sed et a sequente post Kalendas die usque ad eumdem diem, et a quocunque cujuslibet mensis die usque in diem eumdem reditus, annus vocatur; ita hujus mundani anni initium sibi quisque facit quodcunque decreverit, ut ecce nunc Cicero a defectu solis, qui sub Romuli fine contigit, mundani anni principium sibi ipse constituit, et licet etiam saepissime postea defectus solis evenerit, non dicitur tamen mundanum annum repetita defectio solis implesse, sed tunc implebitur cum sol deficiens in iisdem locis et partibus et ipse erit, et omnes coeli stellas omniaque sidera rursus inveniet in quibus fuerant sub Romulo, cum post annorum quindecim millia sicut asserunt philosophi, sol denuo ita deficiet, ut in eodem signo eademque parte sit, ad idem principium, in quo sub Romulo fuerant, stellis quoque omnibus signisque revocatis.

Anno ergo praeterito 810 ab incarnatione Domini non est mirum eclipsin solis evenisse, sicut vestrae indicant litterae; septimo Idus Junias, prima tunc initiante luna, et rursus in eodem anno pridie Kalendas Decembris, trigesima incipiente luna, et a priore defectu septimo mense, hoc est Decembre inchoante; qui sic defectus solis definitur novissima primave luna fieri, et septimo mense a priore defectu, quamvis aliquando penitus non appareat, cum certe sit factus, aut si apparuerit non semper ubique cernatur, aut si ubique conspiciatur, non eisdem horis omnes aequaliter videant evenisse propter supradictas causas.

Si quis ergo etiam in hoc tempore tanto sensus acumine praeditus, tanta instantiae diuturnitate nisus, tanta explorationis et observationis diligentia intentus, eadem otiositate et curiositate sicut priori aetate geniti, sollicitus tantum studium erga astronomiae aut cujuscunque disciplinae assectationem adhibuerit, nonne idem facile credendus est ad eamdem antiquorum scientiam et praescientiam posse pervenire? Voluntas enim dispar, non natura, quae una et aequalis est, homines tantum a se distare facit, quanquam in primis hominibus propter mundi adolescentiam et vim corporum, et sensuum vigorem magis voluisse comperimus.

Hic ergo nunc de eclipsi solis sit finis dicendi, non quod dixisse forsitan sufficienter arbitrer, sed quia ad praesens proprii ingenioli exiguitas amplius memorare non quiverit: Plinius enim secundus et alii libri, per quos aestimem haec me posse supplere, non habentur nobiscum in his partibus, cum de talibus per me ipsum nihil audeam excogitare neque praesumam. Vos autem, domine piissime Auguste, quibus prae omnibus affluentiam sapientiae, sicut et caeterarum sanctarum virtutum Deus distribuit, rogo suppliciter ut in quo vobis de hac causa ignorare videar, aut aliter aestimare quam rectum est, instruere et dirigere dignemini: Stulta enim mundi elegit Deus: Et, Non est apud eum personarum acceptio, ut non solum vestrae purissimae et clarissimae sapientiae lux his qui prope sunt luceat, sed et his qui longe; et non solum per aperta camporum discurrentes illustret, verum etiam Reclusos licet per rimas et juncturas vestri serenissimi splendoris radius exerens perfundat. Omnibus ergo valde necesse est attentis et assiduis precibus rogare et postulare, ut Dominus et Salvator noster Jesus Christus suo populo donet et tribuat multis annis de tali et tanto principe et magistro gaudere, qui omnibus aequaliter omnium bonorum operum et virtutum et honestarum disciplinarum doctor praecipuus, et perfectum habetur exemplar rectoribus ad suos subjectos bene regendos, militibus ad suam exercendam legitime militiam, clericis ad universalis Christianae religionis ritum recte observandum, philosophis et scholasticis ad honeste de humanis philosophandum et sapiendum, reverenterque atque orthodoxe de divinis sentiendum et credendum. Quid plura de nostri domini Augusti Caroli summis virtutibus et excellentibus dicere nitor, cum licet multum elaborare velim totas referre non potero? Hoc tantum veraciter dicimus, quod omnes uno ore conclamant, quia in ista terra, in qua nunc Deo donante Franci dominantur, ab initio mundi talis rex et talis princeps nunquam visus est, qui sic esset fortis, sapiens, et religiosus sicut noster dominus Augustus Carolus. De caetero autem per sua sancta et sublimia merita, forsitan de suo semine talis oriatur. Hoc solum superest ut nos omnes Christiani altissimis vocibus et devotissimis cordibus unanimiter clamemus ad Dominum et rogemus ut nostri optimi domini Augusti Caroli triumphos multiplicet, imperium dilatet, sacram conservet progeniem, sanitatem confirmet, vitam in multos extendat annorum curriculos. Exaudi, exaudi, exaudi, Christe.

Sicut ergo, domine reverentissime atque dulcissime, Deo et vestro fideli famulo Waldoni abbati mandastis, ut me de talibus ex vestris verbis commonendo interrogaret, et exigendo commoneret, qui sicut vobis fidelis, ita mihi de hac re gravis et importunus exactor quamvis moderate exstitit, ita per illum vobis remitto, ut inde ei gratias referatis, si quid in his bene dixerim, quae per ejus urgentem exactionem volens nolens solvi: si autem aliquid male propter meum proprium neglectum, mihi poenitentiam quam velitis clementer imponatis. Opto vos bene semper valere in Deo, optime domine, et non tantum optime domine, sed et piissime atque amantissime Pater.

## LETTERA SULLA DUPLICE ECLISSI DI SOLE DELL'ANNO 810. A CARLO MAGNO.

### *Saluto e preghiera*

**1** In nome del Padre, del Figlio, e dello Spirito Santo. Al gloriosissimo Signore Carlo Serenissimo Augusto di tutti i principi dei Romani, dotato di tutte le qualità nobili e oneste associate alle virtù regali e assiduamente impegnato nelle amministrazioni, lunga vita, salute durevole, continua benevolenza, pace, corona incorruttibile, gloria senza fine.

### *Occasione e tema della lettera*

**2** O amatissimo signore, io Dúngal[76], Vostro fedele servo e oratore, ho sentito dire e non dimentico che voi inviaste la lettera all'abate Baldo, perché con quella mi interrogasse, facendosi latore del Vostro quesito circa i motivi dell'eclissi di Sole, e diceste di essere stato informato da più parti che nell'anno 810 dall'incarnazione del Signore essa si fosse ripetuta due volte[77], e ricordavate di aver letto che non solo gli antichi filosofi dei pagani, ma anche il vescovo di Costantinopoli predissero che

---

[76] Lettera di Dúngal.

*L'autore di questa lettera è il recluso Dúngal, ossia un solitario che condusse uno stile di vita monastico lontano dalla compagnia tanto dei monaci quanto del clero secolare. Di fronte alle difficoltà che pone la lettera, consultiamo V.C. Ismaele Bullialdo, attento ed abilissimo nelle argomentazioni matematiche come in ogni genere di letteratura scientifica, per trarne un parere qualificato circa la lettera stessa. Riportiamo qui la sua breve postilla che abbiamo ritenuto valesse la pena di citare: "questa lettera di Dúngal, benché non risponda né con perizia né con chiarezza alla questione proposta dall'imperatore, né dimostri ciò che si sarebbe dovuto chiarire, è nondimeno utile circa i dubbi sollevati da molti, compreso l'imperatore su quanto veniva asserito temerariamente da alcuni spinti da vana superbia, ossia che due eclissi di Sole nell'anno 810 d.C. fossero state viste e fossero avvenute nella nostra Europa, i quali ne dubitavano certamente a ragion veduta. Poiché non sarebbe stato possibile in quest 'anno 810, 9 giugno, che il Sole si eclissasse, o la sua eclisse fosse registrata in Europa. Giustamente Calvisio annotò e notò a ragione che questa eclisse di Sole del 9 giugno fosse stata predetta da qualcuno, che scrisse il calendario, con calcolo tratto dai tabulati imprecisi degli astronomi, ma non avvenuta (aggiungo e non vista dai nostri europei). Pur tuttavia i cronisti scrissero nelle loro cronache che l'Orsa fu vista, e l'opinione, o meglio l'errore fu recepito dal popolo. Mi sembra opportuno aggiungere che chi calcolò l'eclissi di quell'anno 810 su ciò si fosse sbagliato, poiché non si era reso conto che quell'eclisse sarebbe stata visibile nelle zone limitrofe o meridionali della linea equinoziale e vista così nell'emisfero australe. Non è infatti possibile che nei luoghi poco più distanti dalla linea equinoziale, a distanza di sei mesi si siano viste due eclissi di Sole, cosa che può accadere sotto la linea dell'equinozio, o in luoghi descritti soggiacenti ai paralleli e non lontano da essa . In verità entro lo spazio di cinque mesi nello stesso emisfero boreale o nell'australe possono essere scorte le eclissi solari a due a due. ( due per volta) : tutte cose che possono essere dimostrate come vere.*

*Ma l'autore di questa lettera, Dúngal, sembra ignorasse questa differenza. "Quando il 30 novembre dello stesso anno 810 dell'eclissi di Sole, fu osservata dagli Europei, avrebbe dovuto dire che la precedente del 9 giugno fu vista nell'emisfero australe dai nostri antenati e agli antipodi". Fin qui Ismaele Bullialdo. Riportiamo il testo trascritto dal nostro Jean Mabillon da un codice di Saint Rémy.*

Così il Migne nella Patrologia Latina vol. 105. Il commento tecnico è dell'astronomo francese Bulliau (1605-1694) al quale dobbiamo il termine "evezione" con cui ha chiamato la perturbazione al moto medio lunare di 32,6 giorni dovuta all'azione combinata dell'attrazione del Sole e della variazione di eccentricità dell'orbita lunare. Gli è stato dedicato un cratere lunare visibile presso il lembo Sud Est, già dal padre Giovanni Battista Riccioli nella carta lunare pubblicata nell' *Almagestum Novum* a Bologna nel 1651.

In questa nota del Migne viene menzionata la linea equinoziale, intendendosi l'equatore. Calvisio era anche corrispondente di Keplero.

[77] Nell'810 ci furono 4 eclissi di Sole e 2 di Luna, ma una sola era visibile da Aquisgrana, quella del 30 novembre.

il fenomeno si sarebbe prodotto per effetto, per così dire, del quasi naturale incontro degli elementi che gli esperti conoscevano per via di una ricerca lunga e rigorosa.

***Le buone intenzioni di Dúngal***

**3** Pertanto sembrò opportuno alla Vostra beatissima e illustrissima serenità di ordinare che sulla detta questione io fossi interrogato come discepolo della sapienza sulle mie impressioni e cognizioni in materia, e dichiarassi tali mie impressioni mettendo poi per iscritto la mia risposta e facendovela recapitare. Pertanto non indugerò, né mi sottrarrò al dovere di obbedire come posso al Vostro santissimo e utilissimo ordine; e voglia il cielo che la mia disponibilità contribuisca a rendere più efficace la mia argomentazione, di modo che le mie buone intenzioni portino buoni risultati. Sono comunque sicuro che il mio affetto devoto e ossequiente nei confronti del sommo Imperatore potrà forse compensare la mancata realizzazione del compito affidatomi.

***Con umiltà alla ricerca delle fonti***

**4** Mio signore, capirete che la ricerca e la conoscenza pratica di quella natura spetta ai filosofi, competendo più specificamente e più propriamente ai fisici, per quanto non mi vengano in aiuto i libri più complessi e dettagliati nei quali trattarono di questi argomenti e attraverso i quali pensavo di potere dare una risposta più compiuta e più scientifica circa quanto da voi richiesto. Secondo gli opuscoli semplici, brevi e succinti che ho tra le mani, per quanto di essi abbia saputo cogliere il torpore di un cuore chiuso e il lento procedere del mio ragionamento, movendomi a fatica in un tentativo lento e pigro, risponderò comunque per non fare la figura del lupo, bloccato, per citare il detto della tradizione favolistica, da un pavido e stupido silenzio. So per certo, infatti, che, se avrò detto qualcosa di meno o di diverso, la Vostra indulgente clemenza con la Sua serenissima e generosissima longanimità, mi concederà un più facile perdono e mi punirà moderatamente con paterna correzione per errori dovuti alla mia fragilità e debolezza che superano ogni mio sforzo.
Occorre, comunque, andare alla radice di questa questione, affinché, partendo dall'inizio, come si fa sempre nelle altre dispute, procedendo per ordine nella spiegazione, si ottenga un responso chiaro sull'argomento.

*EXCERPTA DA MACROBIO 1*
*Il cielo delle stelle fisse, i pianeti, la galassia, lo zodiaco, i paralleli*

**5** Ebbene, Ambrogio Macrobio, nel commento a Cicerone, ricorda tra le altre cose i nove cerchi che circondano l'*aplanes*, cioè la massima sfera celeste nella quale sono viste fissate dodici costellazioni, a cui sono affiancate altre *sette sfere attraverso le quali due astri luminosi , il Sole e la Luna, e cinque stelle "mobili" si muovono in ogni direzione*. Delle sfere, invece, ovvero dei circoli suddetti, il primo è la Galassia[78], che in latino si traduce 'lattea', la quale è la sola che è possibile vedere ad occhio nudo, mentre gli altri circoli si possono intuire più con il pensiero che con la vista. Il secondo circolo zodiacale è il signifero cosiddetto perché porta e contiene le stelle e gli astri.*Cinque altri cerchi sono*

[78] Trovandoci sul piano della Galassia, essa descrive nel cielo un cerchio massimo; le miriadi di stelle determinano la lattiginosità visibile sull'equatore galattico, lasciando la vista più libera verso i poli galattici. È un approccio piuttosto insolito quello di considerare il cerchio massimo dell'equatore galattico come l'unico realmente visibile, ed è tanto più sorprendente trovarlo in chi probabilmente non ha avuto la possibilità di vederlo completamente, essendo per metà nell'emisfero australe a declinazioni decisamente inaccessibili per l'Europa.

*chiamati 'paralleli'*, poichè non sono eguali né diversi sotto tutti gli aspetti. Anche Virgilio stesso li menziona nelle sue *Georgiche*.

**6** *Oltre a questi, due sono i Coluri*[79]*, il cui nome è dovuta a una traduzione imprecisa*: i due, ovvero quelli suddetti, Meridiano e Orizzonte, non sono iscritti nella sfera, poiché non possono occupare un luogo preciso, ma essi variano a seconda degli osservatori e dei componenti. Tutto ciò che è stato citato ed enumerato di passaggio non occorre trattarlo in questo discorso.
I due astri più luminosi, il Sole e la Luna ,e le 5 stelle[80] definite "mobili"[81], si collocarono nelle sette sfere prima ricordate soggette alla sfera più grande detta *aplanes* che contiene i 12 segni (zodiacali) e, una volta occupate quelle regioni[82], essi ottennero ciascuna specifiche e speciali eredità.

***Le sette sfere e i pianeti***

**7** Nella prima delle sette sfere vi è la stella di Saturno, nella seconda quella di Giove, nella terza quella di Marte, nella quarta (a metà) il Sole, nella quinta Venere, nella sesta Mercurio, nella settima, che è l'estrema e la più piccola, la Luna. Così le descrive Cicerone e anche Archimede e i Caldei sono dello stesso parere. In realtà Platone posiziona il Sole tra la settima e la sesta sfera a partire dall'estremità cioè il Sole seguito più in alto dalla Luna, imitando gli Egizi , i padri di tutte le discipline filosofiche, che collocano il Sole tra la Luna e Mercurio.

***Cicerone e Platone sulla luminosità della Luna***

**8** Anche se questo parere di Cicerone e delle sue fonti guadagnò terreno grazie ad alcuni editti e credibili argomentazioni e fu accolta da quasi tutti, tuttavia l'osservazione più penetrante di Platone si è dimostrata deduzione più veritiera, suffragata anche dall'osservazione e dal ragionamento; infatti, la Luna, che è priva di luce propria e la mutua dal Sole, è indispensabile che sia soggetta ad una fonte di luce. Questa argomentazione, infatti, ci spiega che la Luna è un astro che non brilla di luce propria come tutte le altre stelle, le quali si trovano al di sopra del Sole nello stesso luminosissimo etere, nel quale tutto ciò che è, per dirla con il Filosofo, è luce naturale e anche propria, la quale si getta così tutta con il suo fuoco nella sfera del Sole, sicché le zone del cielo lontane dal Sole sono oppresse da un freddo perpetuo. In verità la Luna, poiché si trova essa stessa sotto il Sole, vicina alla regione delle stelle cadenti[83] priva di luce, non sarebbe potuta brillare.

*La Luna, "terra dell'etere"*

**9** Infine, dal momento che la Terra occupa la parte più profonda di tutto il cosmo[84] ma la parte più profonda dell'etere è occupata dalla Luna, che chiamarono ugualmente Terra ma Terra dell'etere, essa non avrebbe tuttavia potuto essere immobile come la Terra poiché nella sfera che gira nulla rimane immobile oltre il centro; in verità la Terra è il centro della sfera mondana, perciò essa sola conserva l'immobilità: ancora la terra brilla soltanto con la sola luce del Sole, non splende; la Luna, come uno specchio, trasmette la luce da cui è colpita e, sebbene sia un corpo più denso degli altri corpi celesti ma molto più liscio alla superficie, diviene penetrabile alla luce accolta, al punto da emetterla di nuovo da sé. Ma tale luce non ci porta nessun senso di calore, poiché il raggio di luce, quando ci arriva dalla sua origine, cioè dal Sole, trasporta con sé la natura del fuoco dal quale nasce; ma quando esso è penetrato nel corpo della Luna ed ivi risplende, riversa fuori solo la luce e non il calore: infatti anche lo specchio, quando introduce in sé lo splendore di un fuoco che gli si oppone a distanza, è solamente simile al fuoco, rimanendo privo del senso di calore[85].

---

[79] Sono cerchi massimi anche questi, solo che non ruotano con le stelle fisse, ma sono riferiti all'orizzonte di ogni osservatore.
[80] Sono i pianeti. Fino a Keplero, che scrisse *Astronomia Nova…Stella Martis* per spiegare l'orbita di Marte, i pianeti erano considerati stelle, anche se già Cicerone, *De Natura Deorum II, XLVI, 117* e Diogene Laerzio, *Vitae philosophorum, VII, 145* ne avevano ipotizzato una struttura sferica come Luna Sole e Terra, ma questa conoscenza si era poi *fossilizzata*, come sostiene L. Russo (Flussi e Riflussi, Feltrinelli, Milano 2003). Ancora con Galileo si discute se la Terra sia o meno una stella. Comunque con il telescopio la natura fisica dei pianeti cominciò a rivelarsi, ed anche ad accertarsi che brillassero effettivamente di luce riflessa dal Sole.
[81] Pianeta in greco vuol dire astro errante.
[82] Tutti i pianeti, Sole e Luna inclusi, in quanto sono anch'essi mobili rispetto alle stelle fisse, orbitano nella fascia zodiacale, più o meno tutti sul cerchio massimo dell'eclittica. È questo il senso della complessa costruzione latina.
[83] Le stelle cadenti si infiammano nell'alta atmosfera. Classicamente si riteneva che fossero fenomeni atmosferici o meteore, così come le comete. Fu Tycho Brahe a dimostrare la natura soralunare delle comete.
[84] Letteralmente *Mundus,* il mondo.

***EXCERPTA DA MACROBIO 2***
***Il Sole e le stelle mobili***

**10** *A ciò bisogna aggiungere che, oltre il Sole e la Luna e le cinque stelle che si chiamano "mobili", alcuni autori hanno affermato che tutte le altre stelle sono fisse nel cielo e non sono mosse se non dal cielo stesso, altri, avvicinandosi maggiormente al vero, hanno detto che anche queste ultime sono dotate di un movimento autonomo, oltre quello che possiedono a causa del movimento del cielo ma, per via dell'immensità della sfera esterna, un[86] loro moto di rivoluzione richiede molti secoli, al di là di un numero immaginabile, e perciò nessun loro movimento può essere percepito dall'uomo, dal momento che non sarà sufficiente lo spazio di una vita umana ad osservare almeno un breve tratto di una progressione tanto lenta.*

**11** *In verità si può provare, tramite argomenti illuminanti, che il Sole, la Luna e le cinque stelle (che così vengono chiamate erroneamente) si muovono da occidente ad oriente con moto proprio*, oltre a essere trasportate dal movimento diurno del cielo, dall'alba al tramonto; sia l'osservazione che il ragionamento confermano, infatti, che si muovono e non sono fisse nel cielo.

Non solo tramite il ragionamento più limpido ma anche col supporto dell'osservazione si dimostra che esse sono viste ora *in una, ora in un'altra regione del cielo e spesso, quando due o più di esse arrivano in un luogo si separano successivamente* dal luogo in cui sono simultaneamente viste e da sé, cosa che le stelle fisse non fanno, *ma sono sempre viste negli stessi luoghi, non si sciolgono dalla loro unione e da oriente ad occidente girano con moto più prossimo in senso contrario*[87].

---

[85] Sappiamo che i fotoni trasportano energia E=hν, il fatto che la luce della Luna non scaldi è perché essa riflette solo il 5% della luce solare incidente. Alla distanza della Terra dalla Luna (61 raggi terrestri=221 raggi lunari) questo 5% è sparpagliato su un'area quasi 50000 volte maggiore di quella della Luna stessa. Di conseguenza ci giunge solo 1 parte su un milione del calore solare di riflesso dalla Luna. Questo corrisponde ad una magnitudine di M=-13, 15 magnitudini pù fioca del Sole che splende ad M=-27.4.

[86] Se si stesse parlando dei moti propri non sarebbe stato indicato un unico moto per tutte. Anche i moti propri stellari sono impercettibili nell'arco di una vita umana: Sirio ad esempio si sposta di 1.3 arcosecondi per anno (D. Hoffleit, Bright Star Catalogne, Yale Observatory, 1982) e per percorrere una distanza pari al diametro angolare della Luna impiega 1400 anni. Ciò a cui allude Dùngal è il fenomeno della precessione degli equinozi, scoperto da Ipparco, per cui le coordinate equatoriali delle stelle cambiano secondo un ciclo di oltre 26000 anni, e riguarda tutte le stelle del cielo. Ma al tempo di Dùngal gli Arabi non avevano ancora ottenuto stime migliori di quella di Tolomeo, che riteneva la velocità di precessione pari ad 1° ogni 128 anni. Con gli Arabi si cominciò a supporre che la velocità di precessione variasse e che le stelle fossero soggette quindi anche al moto di *trepidazione* per la quale anche Copernico dovette scomodare una sfera in più nel suo modello cosmologico. Il valore vero è costante e vale 1° ogni 70 anni circa, ed è dovuto al moto di precessione dell'asse terrestre rispetto al piano dell'orbita.

### *Lo zodiaco*

**12** *Considerato poi l'ordine delle costellazioni in cui vediamo diviso e distinto lo Zodiaco, da un segno zodiacale qualsiasi si può determinare l'ordine dello Zodiaco stesso: quando sorge Ariete, dopo di lui appare Toro, lo segue Gemelli, poi Cancro e gli altri segni secondo il loro ordine.*
*Se dunque le stelle mobili procedono da oriente verso occidente*[88]*, non passeranno da Ariete a Toro, che viene dopo, né da Toro a Gemelli, ma da Gemelli a Toro e da Toro ad Ariete procederanno attraverso l'opportuna prosecuzione del moto circolare del cielo.*
*Quando perciò tornano indietro*[89] *dal primo segno al secondo, dal secondo al terzo e quindi ai restanti successivi (mentre i segni zodiacali sono trasportati fissi dal movimento del cielo) senza dubbio è appurato che le stelle mobili non sono mosse dal cielo ma si muovono contro il cielo.*

### *Il moto Lunare*

*13 Perché ciò appaia pienamente chiaro, prenderò in esame il cammino della Luna che, per la sua lucentezza e velocità, è più notevole.*
*Circa al secondo giorno del mese*[90]*, verso il tramonto, la Luna appare quasi vicina al Sole e lo lascia poco dopo che quello si è immerso (nell'oceano); essa stessa occupa il margine di cielo che sta davanti, sopra occidente; al terzo giorno tramonta più tardi che al secondo e così ogni giorno si allontana di più dal tramonto, tanto che al settimo giorno, al tramonto del Sole, essa apparirà in mezzo al cielo ed effettivamente, dopo altri sette giorni, quando il Sole s'immerge, la Luna sorge, finche a metà mese misura mezzo cielo, cioè un emisfero, recedendo da occidente verso oriente. Viceversa, dopo altri sette giorni, verso il tramonto del Sole, raggiunge il vertice dell'emisfero nascosto, come è indicato dal fatto che essa sorge a mezzanotte. Infine, passati altrettanti giorni, più altri due circa*[91]*, di nuovo si unisce al Sole ed il sorgere di entrambi sembra vicino, fino a quando, seguendo il Sole, si muoverà all'indietro e, recedendo gradualmente, sempre regredendo verso oriente, lascerà l'occidente.*

### *Il moto del Sole da equinozio a equinozio*

**14** *Anche il Sole non diversamente si muove da occidente verso oriente e, recedendo gradualmente, sempre regredendo verso oriente, lascerà l'occidente.*
*Esso deve ritirarsi più lentamente della Luna poiché attraversa una costellazione in un tempo equivalente a quanto ne occorre alla Luna per percorrere tutto lo zodiaco; gli effetti del suo moto sono chiaramente osservabili. Supporrò infatti che il Sole sia in Ariete che,*

---

[87] Contrario al moto medio dei pianeti che è detto diretto, e si svolge da Ovest verso Est, come il moto del Sole nelle costellazioni zodiacali.
[88] Questo si chiama moto retrogrado.
[89] Qui parla del moto diretto, benché questo sia opposto al moto diurno della sfera celeste.
[90] Oggi, ,mediante tecniche particolari, sono state osservate Lune giovani 20.6 ore dopo il novilunio. In genere è possibile vedere la falce di Luna crescente di due giorni. Sull'avvistamento della prima falce di Luna si basava il calendario arabo.
[91] Questi altri 2 giorni circa servono a Dùngal per far quadrare i conti: una lunazione dura 29.53 giorni. 4 settimane sono 28 giorni. Non che la fase dall'ultimo quarto alla Luna nuova sia più lunga delle altre, anche se le quattro fasi novilunio, primo quarto, plenilunio e ultimo quarto non sono affatto equispaziate: al moto medio lunare si sovrappongono infatti più di 400 perturbazioni tra cui l'evezione, con periodo di 32,6 giorni, già scoperta da Tolomeo e la variazione, con periodo di 14,8 giorni, scoperta da Tycho, che rendono asimmetriche le durate delle varie fasi.

*essendo il segno equinoziale, rende le ore*[92] *del giorno pari a quelle della notte. Quando cade in questo segno, vediamo subito sorgere Bilancia, poi le chele di Scorpione ed appare Toro sul far del tramonto: infatti vediamo le stelle Iadi e Vergilie*[93] *(le parti più luminose di Toro) apparire non molto dopo il Sole.*

***Toro***

***15** Nel mese seguente, il Sole recede nel segno successivo, cioè in Toro e pertanto né le Vergilie18 né le altre parti del Toro sono visibili in quel mese: infatti, il segno che tramonta col Sole è sempre occultato, al punto che anche gli astri prossimi al Sole sono nascosti dalla sua vicinanza: infatti allora anche la costellazione del Cane, poiché è vicina a Toro, non è visibile, nascosta dalla vicinanza della luce. Ed è questo che dice Virgilio:*

*"la splendente costellazione del Toro apre l'anno con le corna dorate,*
*e tramonta cedendo il posto alla costellazione del Cane*[94]*"*

*Ad ogni modo i versi del Poeta non vanno letti nel senso che, con il Sole ad oriente, al Toro segua subito al tramonto la costellazione del Cane, che segue il Toro, bensí nel senso che il Sole tramonta in braccio al Toro, perché allora si comincia a non vedere, data la vicinanza del Sole.*

***Gemelli***

***16** Infatti, al tramonto del Sole, quando la Bilancia si trova più in alto, lo Scorpione è totalmente visibile; quindi si vedono tramontare i vicini Gemelli; di nuovo, dopo il mese del Toro, non si vedono i Gemelli, il che significa che il Sole è entrato nei Gemelli.*

***Cancro***

***17** Dopo i Gemelli (il Sole) entra in Cancro e, nel momento in cui tramonta, subito la Bilancia compare in mezzo al cielo: così si constata che il Sole, superati i tre segni, Ariete, Toro e Gemelli, si allontana dal centro dell'emisfero.*

***Estate ed equinozio autunnale***

***18** Infine, dopo i tre mesi successivi, che seguono ai tre segni illuminati (dal Sole), dico Cancro, Leone e Vergine, il Sole entra nella Bilancia, la quale di nuovo uguaglia la notte al giorno, e allora, mentre il Sole tramonta nel segno stesso, nasce subito l'Ariete, segno nel quale il Sole era solito tramontare sei mesi prima. Per questo abbiamo scelto di anteporre il suo tramonto piuttosto che il suo sorgere, perché i segni successivi si vedono dopo il tramonto e, mostrando che il Sole ritorna verso di questi che di norma si vedono al calare del Sole, mostriamo senza ombra di dubbio che il Sole si allontana con un moto contrario a quello del cielo.*

[92] La durata qui si si intende.
[93] Sono le Pleiadi.
[94] Gli Egizi seguivano il ciclo annuale di Sirio (Sothis) che spariva nella luce solare il giorno del suo tramonto eliaco e riappariva nella sua levata eliaca, tra questi due momenti passavano circa 70 giorni in cui la stella non è visibile. L'anno sothiaco durava 365 giorni e fu detto anche anno *vago*, cioè mobile. La mobilità è rispetto alle stagioni che slittano di un giorno ogni 4 anni. Fu l'astronomo alessandrino Sosigene ad introdurre l'anno bisestile di 366 giorni nella riforma giuliana del calendario del 46 a.C.

*Le orbite dei pianeti*

***19** Queste cose che abbiamo detto del Sole e della Luna sono comunque sufficienti a spiegare l'arretramento delle cinque stelle: infatti migrando per lo stesso processo nei segni successivi, si trovano sempre ad avere un moto contrario al movimento universale*, e il loro corso e il ricorso si dimostra regolato dal Sole stesso; infatti è sicura la definizione dello spazio, in direzione del quale, quando ciascuna stella "mobile", retrocedendo, ha raggiunto il Sole, sembra sia ri-sospinta, come se fosse proibito andare oltre; di nuovo, quando recedendo è giunta a toccare una determinata parte[95], essa viene solitamente richiamata all'orbita diritta, così la forza del Sole[96] e quella del suo moto sono regolate dalla dimensione stabilita delle restanti luci.

**20** Con orbita o cerchio di una stella si intende una rivoluzione completa e compiuta, ovvero il ritorno da un medesimo luogo allo stesso luogo, dopo aver misurato il circuito della sfera attraverso cui si muove. *È qui la linea che circonda la sfera e che apre, per cosí dire, la strada, attraverso la quale il Sole e la Luna corrono, e dentro la quale è contenuta l'orbita regolare delle stelle "mobili", (le stelle) che gli antichi dicevano che vagassero, poiché, da un lato, si muovono secondo il proprio andamento, dall'altro gravitano dall'oriente all'occidente con moto contrario al movimento della sfera piú grande, cioè del cielo stesso, e la velocità di tutte è effettivamente identica, il movimento simile, identico il modo di muoversi, eppure non tutte descrivono nello stesso tempo i propri cerchi e le proprie orbite.*

***Tempi di rivoluzione dei pianeti***

**21** Il motivo del diverso spazio percorso alla stessa velocità dalle stelle è determinato dalla diversità delle sfere, che le singole stelle percorrono: infatti dalla sfera di Saturno, che è la prima tra le sette, fino alla sfera di Giove, la seconda dal vertice, è tanta la distanza dello spazio interposto, che il giro più alto dello zodiaco si compie in trenta anni[97] (cui in realtà ne vanno aggiunti altri 12[98]), e di nuovo la sfera di Marte si allontana tanto da Giove da compiere un'orbita completa in due anni[99].

[95] Denominati punti stazionari. Su questi esiste il teorema di Apollonio (200 a.C.) riportato nel XII libro dell'Almagesto, che si riferiva al modello cosmologico basato sugli epicicli e i deferenti. Questo teorema permetteva di calcolare la posizione del pianeta sull'epiciclo quando il suo moto raggiungeva il punto stazionario, ma per la soluzione completa era necessario avere le dimensioni dell'epiciclo e del deferente, a cui Apollonio al pari di Aristarco, suo contemporaneo, non erano interessati (cfr. L'astronomia prima del telescopio, a cura di C. Walzer, Dedalo, Bari 1997).

[96] Ogni pianeta procede con il suo moto medio, che è il moto siderale, quello che nel modello copernicano è il periodo orbitale. Il periodo orbitale è rappresentato dal moto descritto dal deferente. Segue il periodo sinodico invece il moto sull'epiciclo. I periodi sinodici dei pianeti sono legati alla durata dell'anno solare dall'equazione $1/S=1/T\pm1/P$, con S periodo sinodico, T anno solare e P periodo siderale del pianeta; il segno negativo si usa per i pianeti esterni alla Terra. Questa dipendenza dal moto medio solare (che troviamo nell'equazione attraverso il valore dell'anno solare) viene espressa nel testo di Macrobio riportato da Dùngal mediante il concetto di influenza del Sole sui pianeti.

[97] Saturno compie un'orbita in 29.47 anni.

[98] Giove compie la sua orbita attorno al Sole in 11.87 anni.

[99] Marte in 687 giorni.

Venere è tanto più bassa rispetto alla regione di Marte, che quell'anno solare è sufficiente a terminare lo zodiaco[100].

**22** Dato che in realtà la stella più vicina a Venere è la stella di Mercurio, il Sole è vicino a Mercurio, così che questi tre percorrono il suo cielo nello stesso spazio di tempo più o meno all'anno25.
Per questo Cicerone ha chiamato queste due orbite compagne del Sole, poiché non si allontanano mai l'una dall'altra in un medesimo spazio di tempo: la Luna, così, si allontana tanto da queste verso il basso, da compiere essa stessa in vent'otto giorni quello che esse compiono in un anno.

***Il Sole***

**23** *Cicerone, volendo che il Sole fosse il quarto dei sette pianeti, e in quanto quarto non esattamente mediano ma in ogni modo pressoché mediano, affermò che non era perfettamente centrale ma quasi centrale con queste parole:*

*"in seguito il Sole occupa l'orbita quasi al centro tra le sette".*

*Ma non manca un'aggiunta a qualificare quest'affermazione; infatti il Sole, ottenendo il quarto posto, occupa la regione mediana come numero ma non come spazio; infatti la stella di Saturno, che è la* più alta, attraversa lo zodiaco in trent'anni: il Sole, posto alla metà, in un anno, la Luna, per ultima, in meno di un mese.

*Distanza tra il Sole e Saturno*

**24** Tanto corre tra il Sole e Saturno, quanto tra l'uno ed il trenta; tanto tra la Luna ed il Sole, quanto fra il dodici e l'uno; il fatto più chiaro è che dal punto più alto al punto più basso di tutto lo spazio la regione del Sole non produce una divisione netta rispetto alla parte centrale.
Ma per quanto concerne il numero si dice per esempio che il quattro cada a metà tra uno e sette, sebbene a causa di una differenza degli spazi vi si aggiunga una frazione.

Lo zodiaco è l'unico circolo tra gli undici sopraindicati, che da solo è capace di raggiungere questa ampiezza nel modo riferito.

***La natura dei circoli celesti e l'Eclittica***

***25*** *La natura dei circoli celesti è incorporea e la linea nella mente è concepita in modo tale che venga registrata con la sola lunghezza (o "longitudine") e non possa avere il fianco (o "lato"); nello Zodiaco la grandezza (o "capacità") presupponeva la lunghezza (o "longitudine") dei segni. Quindi, la quantità di spazio occupata in larghezza dalle stelle estese, è circoscritta da due linee[101], e la terza tracciata nel mezzo è chiamata eclittica, poiché, quando il Sole e la Luna compiono il proprio*

[100] Venere e Mercurio essendo interni alla Terra non si discostano mai dal Sole più di 48° e 28° rispettivamente. Questo fa sì che nel corso del moto annuale il Sole si porti appresso questi due pianeti facendoli percorrere tutto lo zodiaco con loro che o lo seguono o lo precedono.
[101] Sono i due tropici: settentrionale o del Cancro e meridionale o del Capricorno. Posti a declinazione +23°40' e -23°40' al tempo di Dùngal (oggi sono a ±23°26').

*percorso su quella linea nello stesso momento , è necessario che uno dei due subisca un'eclissi: del Sole, se la Luna lo segue*[102]*; della Luna, se essa è in opposizione al Sole*[103].

## *Le eclissi di Sole e di Luna*

**26 E perciò il Sole non si eclissa**[104]**, se non quando è il trentesimo giorno della Luna. E se l'orbita non è nel suo quindicesimo giorno, la Luna non conosce eclissi. Così, infatti, accade che o il cono**[105] **(d'ombra) tracciato dalla Luna posta contro il Sole si collochi davanti alla Terra, sulla stessa linea (l'eclittica), per riceverne essa la luce come al solito, o la stessa (Luna) si sostituisca al Sole**[106]**, e per la sua interferenza respinga la sua luce dalla nostra vista. Nell'eclissi d'altronde, il Sole non patisce nulla, ma la nostra vista è ingannata, mentre la Luna soffre dell'eclissi, non ricevendo la luce del Sole, grazie alla quale colora la notte. Su questa base il sapiente Virgilio, molto esperto in tutte le discipline, disse:**
**"Varie eclissi di Sole, ed oscuramenti della Luna"**[107]

## *La visibilità delle eclissi*

**27** *Per quanto, quindi, l'orbita delle tre linee chiuda e divida lo zodiaco, tuttavia un autore di vocabolari volle che un giro fosse chiamato "antichità*", ma secondo altri filosofi la lunghezza dello zodiaco è misurata mediante 12 linee, di cui per la natura stessa dei numeri pari, occorre che due siano considerate mediane, ed essi, garantendo che siano percorse soltanto dal Sole, dicono che la Luna le percorra tutte. Non si dà, perciò, che, vagando essa di qua e di là, l'eclissi avvenga ogni mese. Tuttavia gli astronomi dimostrano che tutti gli anni avviene l'eclissi[108] di uno dei due astri[109] nei giorni e nelle ore prestabilite, per quanto non sempre sia visibile, per il fatto che alle volte avviene a Sud della Terra, nella parte dell'emisfero che si nasconde, alle volte sopra; ma a causa del cielo coperto, a causa della sfericità e della concavità della Terra le eclissi non possono essere viste né in tutte le stagioni, né da tutti i luoghi della Terra; quindi è certissimo che di eclissi ne avvengano più spesso di quanto ne vengano viste, e che le eclissi non appaiano tutte uniformemente quando sono viste; quindi gli orientali non percepiscono le eclissi serali del Sole e della Luna, né gli occidentali quelle mattutine, quando il cono della Terra si frappone ostacolando la vista. Infatti, abbiamo constatato che l'eclissi di Luna che talvolta si verifica nel quinto mese dalla precedente[110], e quella stessa di Sole che

---

[102] La Luna è in congiunzione col Sole durante l'eclissi di Sole.

[103] Si chiama eclittica il percorso del Sole nel cielo, ovvero la proiezione dell'orbita terrestre nel cielo, poiché è lì che avvengono le eclissi.

[104] In latino si usa *defectus* per indicare l'eclissi, ma il termine è corretto in senso proprio solo nel caso di eclissi di Luna. Infatti in quel caso è la luce del Sole a mancare e la Luna perde la sua sorgente di luce. Nel caso dell'eclissi di Sole questo non smette di brillare, ma è occultato dalla Luna. Si tratta quindi, tecnicamente, di un'occultazione e non di un *defectus.*

[105] Conus inventus potrebbe essere il cono d'ombra, o meglio di penombra, perché non risulti ridondante la casistica successiva di eclissi totale ( che così descritta include anche le eclissi anulari, anche se questa tipologia di eclissi non era ancora mai stata riportata in letteratura, ,essendo Clavio il primo ad osservarne una a Roma il 9 aprile 1567 e a pubblicarlo nel *Commentarium in Sphaera* dell'edizione del 1589).

[106] Ovvero sia in congiunzione, o sovrapposta al Sole. È una descrizione piuttosto ridondante dell'eclissi di Sole, volendosi qui probabilmente distinguere le fasi parziali, quando l'osservatore si trova nel cono di penombra della Luna, dalla fase di centralità, dove l'eclissi è totale, quando i centri del Sole e della Luna sono coincidenti. Non tutte le eclissi solari sono anche totali.

[107] Perdendo un po' la poesia bisognerebbe tradurre *Varie occultazioni del Sole ed oscuramenti della Luna*.

[108] Avvengono infatti da un minimo di 2 eclissi ad un massimo di 7 eclissi all'anno.

[109] Qui si dice letteralmente di una delle due stelle, poiché anche Luna e Sole sono stelle erranti.

[110] Nell'anno 810 la sequenza delle eclissi fu 9/1 (totale, visibile dall'Antartide) 5/6 (anulare, nel Sud Pacifico) 20/6 totale di Luna (con la Luna al centro dell'ombra ed il Sole al nodo dell'orbita) 5/7 (anulare, al polo Nord) 30/11 (quasi totale ad Aquisgrana) e 14/12 (totale di Luna). Nel quinto mese dalla precedente, è da intendersi eclissi di Luna, se ammettiamo che Dùngal stesse consultando proprio le effemeridi dell'anno 810, ed è l'eclissi del 14 dicembre, rispetto a quella precedente

capita due volte nel corso del settimo mese[111] sulle regioni della Terra non sono visibili se non da diversi luoghi.[112], [e sappiamo] che una volta ogni **12 giorni**[113] entrambi gli astri si possono eclissare secondo un calcolo e una tradizione probabili.

### *Ripresa del dialogo tra Dúngal e Carlomagno sui traguardi dell'astrofisica*

**28** Quindi ti ho risposto, come mi sembra, o grande Imperatore, come richiesto delle Vostre lettere, e ho detto con l'autorità dei medesimi come gli antichi filosofi conoscevano e anticipavano come e quando avvenisse un'eclissi; e quei profondi conoscitori di queste discipline, pur non essendo a conoscenza di nessun principio riconosciuto in tempi antichissimi, con la fermissima volontà di una mente sgombra e purificata, con una chiarissima e perspicace acutezza munita di una sensibilità interiore, hanno ricercato la natura, le norme, le cause, e le origini delle cose naturali; impegnandosi in una ricerca molto acuta e pressante, hanno trovato in modo molto preciso ed accurato delle risposte offerte da Colui da cui proviene ogni splendida grazia e ogni dono perfetto. Gli studiosi dell'astronomia, poi, hanno esaminato tali scoperte e deduzioni con impegno e assiduità e, studiando metodicamente e ampiamente mediante l'osservazione e l'esperienza visiva il sorgere e il cammino (= l'orbita) delle stelle, l'andata e il ritorno del Sole, della Luna e delle altre cinque stelle "mobili", ne hanno osservato completamente l'avvicinamento e l'allontanamento, tanto che con l'osservazione hanno saputo senza ombra di dubbio quante linee dell'orbita dello zodiaco percorra ciascuna stella "mobile", e attraverso quale fra quelle linee e in modo specifico e preponderante essa diriga l'orbita, e in quale costellazione e in quale parte essa appartenga alla stessa costellazione.

### *Occorrenza e previsione di un'eclissi*

**29** Coloro, perciò, che conoscevano con molta sicurezza e accuratezza i più sottili moti delle altre stelle, per quale motivo avrebbero ignorato le orbite del Sole e della Luna, che davvero sono le più notevoli e più facili da conoscere, tanto da non sapere come o quando corrèssero attraverso l'eclittica del circolo dello zodiaco, e come, percorrendo quell'unica e medesima eclittica, si incontrino in un unico segno e in un'unica parte e come, incontrandosi in quell'unica parte, e quando la Luna si sovrappone al Sole in quella, si dia eclissi di Sole?

**30** Non solo, quindi, i filosofi predicevano l'eclissi, cioè prevedevano in anticipo l'eclissi del Sole, e, su questa base, predicevano quanto si sarebbe verificata un mese dopo, ma, da profondi conoscitori della materia quali erano, attraverso la suddetta tenace ricerca e la diligente osservazione, essi stabilivano molto tempo prima quando sarebbe caduta la successiva nell'arco dell'anno, o dei 20 o 900 anni seguenti. Ma rimarrete stupefatti che essi estesero le previsioni sulla base di tali argomentazioni fino a 15.000 anni.

---

del 20 giugno. Tuttavia ambedue queste eclissi lunari erano visibili da Aquisgrana poiché accadevano di sera.

[111] Il settimo mese è 6 mesi dopo, dal 5 giugno al 5 luglio nel caso dell'810, con la prima eclissi accaduta il 9 gennaio.

[112] Letteralmente ab aliis. Si intende che se da un determinato luogo è visibile la prima eclissi di Luna, dallo stesso non sarà visibile quella nel quinto mese successivo, e così se da un luogo è visibile una delle due eclissi di Sole, non sarà visibile quella di 30 giorni più tardi dallo stesso luogo.

[113] Probabilmente è un errore di trascrizione: dovrebbe essere 15 giorni, poiché la Luna deve passare dall'opposizione alla congiunzione o viceversa perché si abbia una nuova eclissi. Di fatto l'eclissi di Sole del 30 novembre, visibile ad Aquisgrana fu seguita da quella di Luna del 14 dicembre, altrettanto visibile da quel luogo con cielo sereno, cosa che probabilmente non fu per Dùngal se si deve rifare ad una tradizione ed un calcolo attendibile.

***Cicerone sull'anno di ciascuna stella***

**31** Quindi Cicerone nel *Somnium Scipionis* afferma:

*"Gli uomini, a dire il vero, misurano di norma l'anno solo con il volgere ciclico del Sole, cioè con il ritorno di un'unica stella*[114]*; quando, invece, tutti quanti gli astri saranno ritornati nell'identico punto da cui sono partiti e avranno nuovamente tracciato, dopo lunghi intervalli di tempo, il disegno di tutta la volta celeste, solo allora lo si potrà definire, a ragione, il volgere di un anno; a stento oserei dire quante generazioni di uomini siano in esso contenute. Come un tempo il Sole sembrò agli uomini venir meno e spegnersi, allorché l'anima di Romolo entrò in questi stessi spazi celesti, così, quando per la seconda volta, dalla stessa parte del cielo e nel medesimo istante, il Sole verrà meno, in quell'istante, una volta che saranno ricondotte al punto di partenza tutte le costellazioni e le stelle, considera compiuto l'anno; sappi, comunque, che non ne è ancora trascorsa nemmeno la ventesima parte."*

**32** Aprendosi con queste parole il commento di Macrobio a Marco Tullio Cicerone egli la chiarisce in questo modo:

*"L'anno non è soltanto quello che comunemente siamo soliti chiamarlo, ma ciascuna sia delle luci, cioè il Sole e la Luna, sia delle stelle che percorrono tutto il cielo, da un certo luogo ritornano nello stesso luogo, e questo costituisce il suo anno: così il mese della Luna è un anno in quella parte di cielo che illumina, infatti il mese prende nome dalla Luna poiché il nome greco della Luna significa proprio mese".*

***Virgilio sulla differenza tra 'anno' lunare e anno solare***

***33** Virgilio, infine, volendo far capire la differenza tra l'anno lunare che è breve e l'anno compiuto dall'orbita del Sole:*

*"Nel frattempo il Sole compie il grande giro dell'anno".*

*L'anno del Sole è chiamato grande rispetto a quello della Luna; ed il corso di Venere e Mercurio è quasi uguale a quello del Sole e l'anno di Marte dura quasi un biennio: infatti è in tale periodo di tempo che esso fa il giro del cielo. Il pianeta Giove compie, invece, lo stesso giro in dodici anni e quella di Saturno in trenta."*

[114] Il Sole per l'appunto.

*L'anno cosmico*

***34** Il termine dell' anno cosmico [115] si raggiunge, quindi, quando tutte le stelle e gli astri, compresi nel cielo delle stelle fisse[116], sono ritornati da un punto stabilito a quel medesimo punto, in modo tale che la singola stella del cielo non si trovi in altra posizione se non quella in cui era prima, essendosi tutte le altre spostate da quel luogo a cui sono tornate, di modo che il Sole e la Luna con i cinque pianeti si trovino nelle medesime posizioni e parti, in cui si trovavano all'inizio dell'anno astronomico: ciò, secondo gli astronomi, si verifica ogni 15.000 anni[117].*

***35** Perciò, come l'anno della Luna equivale a un mese, e l'anno del Sole a dodici mesi, e gli anni delle altre stelle sono quelli citati precedentemente, così, 15.000 anni, secondo i nostri calcoli, compongono l'anno astronomico. Sarebbe effettivamente corretto chiamarlo anno di rivoluzione, misurato non dal ritorno del Sole, ossia di un solo astro, ma compiuto con il ritorno di le tutti i pianeti, che si trovino in qualsiasi cielo, allo stesso luogo nella medesima configurazione astronomica di tutta la volta celeste; per questo motivo si chiama cosmico (mundanus), perché il termine proprio per 'cielo' è mundus ('il cosmo'). Pertanto, come facciamo partire l'anno solare non solo dal 1° gennaio fino allo stesso giorno, ma dal giorno seguente al primo fino al medesimo giorno, e il ritorno da qualsiasi giorno di qualsiasi mese fino allo stesso giorno si chiama anno, cosí ciascuno potrà fissare a piacere l'inizio di tale anno astronomico, come proprio Cicerone stabilisce l'inizio dell'anno astronomico dall'eclissi di Sole che avvenne al momento della caduta di Romolo, e per quanto l'eclissi si sia ripetuta poi spessissimo, non si dice tuttavia che la ripetizione dell'eclissi di Sole[118] abbia segnato la fine dell'anno cosmico, ma che esso si compirà quando il Sole, al momento dell'eclissi, si troverà nel medesimo punto, e troverà tutte le stelle e tutti gli astri di nuovi nelle medesime posizioni in cui si trovavano sotto Romolo: quando, al compiersi dei 15.000 anni come asseriscono i filosofi, il Sole si eclisserà nuovamente, di modo da trovarsi nel medesimo punto, essendo tornate tutte le stelle e costellazioni al medesimo punto iniziale in cui si erano trovate sotto Romolo.*

***Le eclissi di Sole dell'810, la loro cadenza e visibilità***

**36** Non è perciò strano che nell'anno 810 dall'incarnazione del Signore si sia verificata un'eclissi di Sole, come indica la Vostra missiva; il 9 giugno[119], all'inizio allora della prima giornata della lunazione[120], e nuovamente nel stesso anno il 30 novembre, all'inizio della trentesima giornata della lunazione durante il settimo[121] mese dalla precedente eclissi, alle soglie del mese di dicembre. Si stabilisce così che queste eclissi di Sole avvengono per l'interposizione di una Luna assai nuova, e nel settimo mese[122] dalla precedente eclissi, sebbene l'eclissi stessa talvolta sia quasi invisibile, pur essendosi sicuramente verificata, e non sempre viene vista dappertutto, o, pur essendo vista dappertutto, non tutti nelle stesse ore vedono ugualmente ciò che si è verificato[123].

[115] Letteralmente *annus mundanus* cioè relativo al mondo, la sfera delle stelle fisse e tutte le sfere delle stelle mobili. Da non confondere con l'anno platonico, quello in cui si compie il ciclo della precessione degli equinozi.
[116] L'*aplanes*, cioè dove non vi sono moti relativi di una stella rispetto all'altra.
[117] 15000 anni è il minimo comune multiplo tra i periodi sinodici di tutti i pianeti, Sole e Luna inclusi. Si tratta di un valore ottenuto con grandi approssimazioni. Ponendo i periodi sinodici oggi noti, ed approssimandoli alla decina più vicina, salvo Venere e Terra, si ottiene un grande anno di quasi 60000 anni. Ma le frazioni di giorno trascurate non permettono il riallineamento perfetto dopo un tale periodo. Il calcolo ha dunque un senso puramente numerologico. Si noti Dùngal assegna a Marte un periodo sinodico di quasi 2 anni. In realtà Marte ha un periodo sinodico di 50 giorni più lungo, ed uno siderale di 43 più corto dei 2 anni.
Mercurio120(116)Venere584Terra365Marte780Giove400(399)Saturno380(378)Luna30(29.53) Grande Anno59280 anni
[118]Era noto già ai Caldei col nome di *Saros* il periodo di 18 anni e 10 giorni e 1/3 in capo al quale si ripete un'eclissi solare o lunare. L'*Exeligmos* comprende 3 Saros e vale 54 anni e 31 giorni, e dopo tale periodo un'eclissi torna a vedersi da uno stesso luogo quasi nelle medesime circostanze. Di questo dato, però, Dùngal non sembra a conoscenza.
[119] L'eclissi è avvenuta il 6 giugno 810
[120] Letteralmente si parla della prima Luna, e poi della trentesima, ma è chiaro che si tratta di giorni.
[121] Sono passati 6 mesi lunari, cioè 177 giorni, che è metà dell'anno delle eclissi, o anno draconitico.

***Lo studio e la previsione delle eclissi***

**37** Se quindi qualcuno in questa occasione, mosso da profondo interesse, dotato di così grande acutezza di ingegno e sostenuto dalla ricerca e dalla osservazione e da un prolungato impegno, si sarà avvicinato allo studio della astronomia o di qualsiasi disciplina, non se ne può forse dedurre che questi possa ugualmente dedicarsi alla stessa scienza degli antichi o alla prescienza? Diversa è, infatti, la volontà, non la natura (che è una ed uguale), che fa in modo che gli uomini siano diversi, sebbene noi sappiamo che nei primi uomini, a causa dell'immaturità del mondo, la natura abbia fatto prevalere la forza del corpo e l'energia dei sensi.

*Chiusura di un discorso pur incompiuto*

**38** Quindi, a questo punto, terminiamo pure questo discorso sull'eclissi di Sole, non perché io ritenga di averne trattato a sufficienza, ma perché l'esiguità del mio debole ingegno non mi consente di ricordare di più. Non dispongo, infatti, qui né del II libro [della *Naturalis Historia*] di Plinio né degli altri libri che possano colmare le mie carenze su questi temi perché io stesso non oso né dar spazio all'immaginazione né fare supposizioni in materia.

*Supplica finale a Carlomagno*

**39** Ma quanto a voi, o mio piissimo Imperatore, al quale Dio ha elargito abbondanza di sapienza e di altre virtù sante, vi prego supplichevolmente di degnarvi di istruirmi e di indirizzarmi in quanto vi sembri che io non conosca in questa materia, o in ambiti in cui la mia opinione non sia corretta:
"Dio sceglie le cose stolte del mondo, e "non fa favoritismi", affinché non solo la luce della Vostra santissima e celeberrima sapienza illumini costoro che sono vicini, ma anche coloro che sono lontani, e non solo il raggio luminoso del Vostro splendore tocchi coloro che corrono qua e là attraverso gli spazi aperti della campagna, ma esso ricolmi i 'reclusi' in disparte[124] attraverso le fessure e le fenditure.

*Augurio e preghiera per Carlomagno*

**40 a** Quindi è fondamentale per tutti chiedere con frequenti e assidue preghiere che il nostro salvatore Gesù Cristo ci doni e conceda di godere per molti anni di tanto principe e maestro stimato ugualmente per tutti modello straordinario in tutte le buone opere e virtù e in ogni nobile attività: esempio perfetto per i governanti nel guidare i propri sudditi, per i soldati nello svolgere regolarmente il proprio servizio militare, per i chierici nell'osservare rettamente il rito dell'universale religione cristiana, per i rétori e i maestri nello sviluppare argomentazioni e crescere in sapienza e promuovere la rettitudine nelle cose umane e la devozione e l'ortodossia di fede nelle cose divine.

[122] Cioè dopo 6 mesi lunari completi.
[123] In questo caso Dùngal ha consultato delle effemeridi precise, che non fa altro che commentare. L'eclissi è avvenuta, ma le regioni di visibilità di questa eclissi non sono state previste, dunque si limita a commentare che è avvenuta certamente, anche se non è stata riportata alcuna osservazione. La precisione delle effemeridi nel calcolo delle circostanze di un'eclissi è cominciata ad essere soddisfacente solo con la pubblicazione delle *Tabulae Rudolphinae* di Keplero nel 1627. È celebre l'episodio dei padri Gesuiti a Pechino nel 1629 quando predissero le circostanze dell'eclissi parziale in modo più preciso dei loro colleghi astronomi cinesi ed arabi sulla base dei calcoli copernicani. Erano i padri G. Schreck e G. A. Schall von Bell, partiti da Lisbona nel 1618 con un cannochiale. Padre Schreck era stato allievo di Galieleo a Padova nel 1603-4 ed il settimo accademico dei Lincei. Aveva chiesto a Galileo e Keplero un aiuto per la riforma del calendario cinese e Keplero gli diede precise risposte ed una copia delle sue opere (cfr. P. Maffei, La cometa di Halley, Mondatori, Milano, 1984, ,p. 104).
[124] Come Dungal stesso, *reclusus* presso l'abbazia di Saint-Denis.

*Encomio dell'Imperatore*

Perché mai mi affanno a dire di più sulle eccellenti e grandi virtù del Nostro Signore Imperatore Carlo, quando non potrò riferirle tutte, per quanto mi sforzi? Lo diciamo senza ipocrisie, ripetendo ciò che tutti gridano da una sola bocca, poiché in questa terra, in cui ora i Franchi dominano per volontà divina, dall'inizio del mondo non si è mai visto un tale re e un tale principe, un principe, cioè, che fosse altrettanto forte, sapiente e religioso come il nostro signore imperatore Carlo.

**40 b** Del resto per i suoi santi e illustri meriti forse nascerà dalla sua stirpe un uomo così grande. Manca soltanto questo: che tutti noi cristiani con voci squillanti e devotissimi cuori unanimemente ci rivolgiamo a Dio e preghiamo affinché accresca i trionfi del nostro Signore e Imperatore Carlo, gli estenda l'impero, gli conservi la santità della stirpe, gli rafforzi la salute, gli allunghi la vita di molti anni. Esaudisci, esaudisci, esaudisci, o Cristo.

***Finale***

**41** Come quindi, o signore tanto venerabile e caro, avete raccomandato a Dio e al Vostro fedele abate Baldo di interrogarmi su tali argomenti, rimproverandomi con le Vostre parole, e di sollecitarmi con le Vostre richieste, Baldo che, come è a voi fedele, rappresenta per me, sia pure in modo contenuto, un duro e inflessibile ispettore, così mi rimetto al Vostro buon cuore che gli diate poi prova di riconoscenza, se avrò detto qualcosa di buono tra le cose che ho spiegato, rispondendo al Vostro invito; se invece avrò detto qualcosa di male a causa della mia negligenza, imponetemi pazientemente il pentimento. Auspico che voi stiate sempre bene secondo i disegni di Dio, eccellentissimo signore, e pur sempre padre devotissimo e affettuosissimo.

.